\documentclass[english,aps,pra,showpacs]{revtex4}
\usepackage[T1]{fontenc}
\usepackage[latin1]{inputenc}
\usepackage{amsmath}
\usepackage{amssymb}
\usepackage{graphicx}
\usepackage{esint}

\makeatletter

\providecommand{\tabularnewline}{\\}

\@ifundefined{textcolor}{}
{%
 \definecolor{BLACK}{gray}{0}
 \definecolor{WHITE}{gray}{1}
 \definecolor{RED}{rgb}{1,0,0}
 \definecolor{GREEN}{rgb}{0,1,0}
 \definecolor{BLUE}{rgb}{0,0,1}
 \definecolor{CYAN}{cmyk}{1,0,0,0}
 \definecolor{MAGENTA}{cmyk}{0,1,0,0}
 \definecolor{YELLOW}{cmyk}{0,0,1,0}
}

\@ifundefined{definecolor}{\@ifundefined{definecolor}{\@ifundefined{definecolor}{\@ifundefined{definecolor}{\@ifundefined{definecolor}
 {\usepackage{color}}{}
}{}}{}}{}}{}\makeatother

\makeatother

\usepackage{babel}

\makeatother

\usepackage{babel}

\makeatother

\usepackage{babel}

\makeatother

\usepackage{babel}

\makeatother

\usepackage{babel}

\makeatother

\usepackage{babel}

\makeatother

\usepackage{babel}
\begin{document}

\title{Virial expansion for a strongly correlated Fermi system and its application
to ultracold atomic Fermi gases}

\author{Xia-Ji Liu$^{1}$}

\email{xiajiliu@swin.edu.au; Telephone: +61-3-9214-8166; Fax: +61-3-9214-5160 }

\affiliation{$^{1}$\ ARC Centre of Excellence for Quantum-Atom Optics, Centre
for Atom Optics and Ultrafast Spectroscopy, \\
 Swinburne University of Technology, Melbourne 3122, Australia}

\date{\today{}}
\begin{abstract}
Strongly correlated Fermi system plays a fundamental role in very
different areas of physics, from neutron stars, quark-gluon plasmas,
to high temperature superconductors. Despite the broad applicability,
it is notoriously difficult to be understood theoretically because
of the absence of a small interaction parameter. Recent achievements
of ultracold trapped Fermi atoms near a Feshbach resonance have ushered
in enormous changes. The unprecedented control of interaction, geometry
and purity in these novel systems has led to many exciting experimental
results, which are to be urgently understood at both low and finite
temperatures. Here we review the latest developments of virial expansion
for a strongly correlated Fermi gas and their applications on ultracold
trapped Fermi atoms. We show remarkable, quantitative agreements between
virial predictions and various recent experimental measurements at
about the Fermi degenerate temperature. For equation of state, we
discuss a practical way of determining high-order virial coefficients
and use it to calculate accurately the long-sought third-order virial
coefficient, which is now verified firmly in experiments at ENS and
MIT. We discuss also virial expansion of a new many-body paramter
- Tan's contact. We then turn to less widely discussed issues of dynamical
properties. For dynamic structure factor, the virial prediction agrees
well with the measurement at the Swinburne University of Technology.
For single-particle spectral function, we show that the expansion
up to the second order accounts for the main feature of momentum-resolved
rf-spectroscopy for a resonantly interacting Fermi gas, as recently
reported by JILA. In the near future, more practical applications
with virial expansion are possible, owing to the ever-growing power
in computation. 
\end{abstract}


\maketitle
\tableofcontents{}

\section{Introduction}

\subsection{Universal strongly correlated Fermi systems: From dilute neutron
matter to ultracold trapped Fermi atoms}

\begin{figure}[htp]
\begin{centering}
\includegraphics[clip,width=0.8\textwidth]{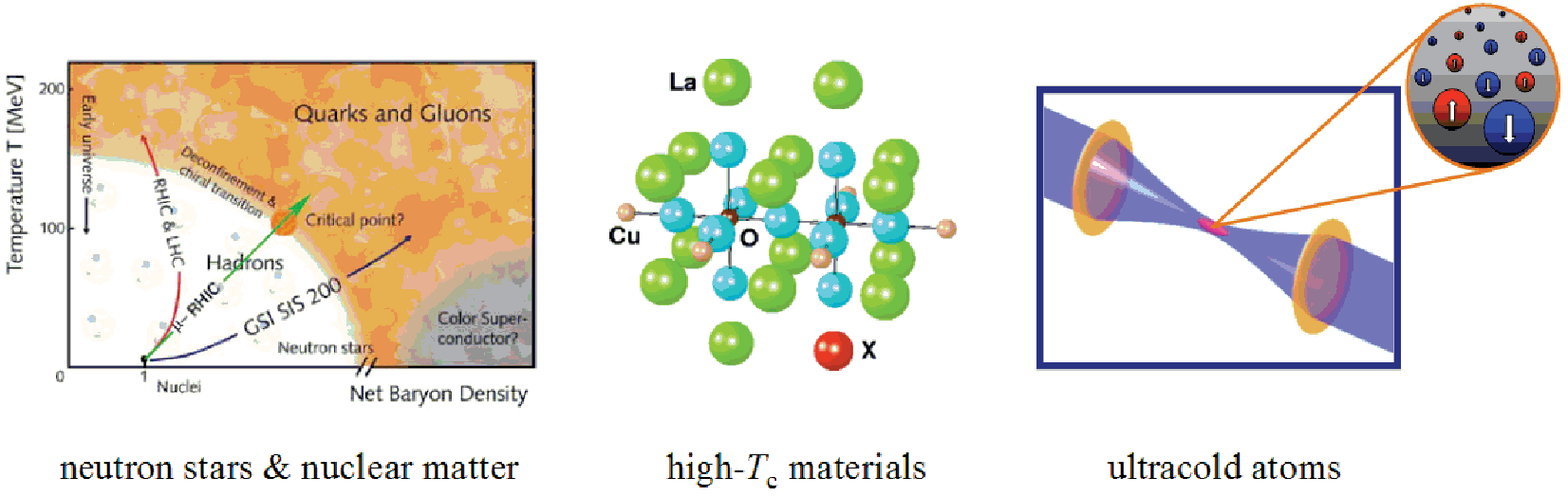} 
\par\end{centering}

\caption{(color online) Ubiquitous strongly correlated Fermi gases in nature.}

\label{fig:FermiGas} 
\end{figure}

The strongly correlated Fermi gas is a ubiquitous system in nature
\cite{ThomasPhysicsToday2010}. It appears in the quark-gluon plasmas
in the early Universe \cite{QuarkGluonPlasma}, neutron stars \cite{PethickARNPS1995,LeePRC2006},
high-temperature superconductors \cite{LeeRMP2006}, and most recently
in ultracold atoms \cite{BlochRMP2008,GiorginiRMP2008,KetterleCourse}
(see Fig. \ref{fig:FermiGas}). The strong correlation is a result
of a large separation of length scales and the Fermi system is close
to an interesting universal limit with {\em infinitely} large scattering
length and {\em zero} effective range of interaction \cite{HeiselbergPRA2001,HoUniversality}.
The absence of length scale implies that the type and detail of interactions
are not important. It is anticipated that universal behaviors in both
static and dynamic properties would emerge \cite{HoUniversality,HDLNaturePhysics,HLPreprint}.

Dilute neutron matter is a good example of strongly-correlated Fermi
systems \cite{PethickARNPS1995,LeePRC2006}. The neutron scattering
length is about $a_{s}\simeq-18$ fm and the effective range is $r_{0}\simeq2.8$
fm $\ll a_{s}$. For typical neutron densities $0.1\rho_{N}>\rho>10^{-4}\rho_{N}$,
where $\rho_{N}\simeq0.16$ fm$^{-3}$ is the saturation density of
nuclear matter, the dimensionless interaction parameter $k_{F}\left|a_{s}\right|\gg1$
while $k_{F}\left|r_{0}\right|$ is small. Here, $k_{F}=\left(3\pi^{2}\rho\right)^{1/3}$
is the Fermi wave-vector. Therefore, the neutron matter is close to
the unitary limit, with which the \textit{s}-wave scattering amplitude
becomes saturated at a zero-energy resonance. Understanding the nuclear
matter appears to be a challenging many-body theoretical problem \cite{BertschChallenge,BakerPRC1999}.

In this context, a unitary atomic Fermi gas realized recently in ultracold
atom laboratory attracts particular attention \cite{BlochRMP2008,GiorginiRMP2008,KetterleCourse}.
It serves as a new paradigm for studying strong-correlations because
of its unprecedented controllability and purity. By tuning an external
magnetic field across a collisional Feshbach resonance \cite{ChinRMP2010},
the interatomic attractions in a two-component Fermi gas can be changed
precisely from weak to infinitely strong, leading to the observation
of crossover from Bardeen-Cooper-Schrieffer (BCS) superfluids to Bose-Einstein
condensates (BEC) \cite{JILASuperfluidity,MITSuperfluidity}, which
was anticipated long time ago \cite{NSR,SadeMeloPRL1993,OhashiPRL2002}.
At the resonance, the \textit{s}-wave scattering length $a_{s}$ is
exactly infinity and the effective range of interaction is negligible.
This unitary limit can now be routinely achieved in laboratories with
fermionic potassium-40 ($^{40}$K) \cite{JILASuperfluidity} and lithium-6
($^{6}$Li) atoms \cite{MITSuperfluidity}.

Experimentally, the near resonance regime was first approached by
O'Hara \textit{et al.} in 2002 with $^{6}$Li atoms \cite{OharaScience2002}.
The stability of atomic Fermi gases under strong attractions was observed
and the ground state energy was found to reduce significantly with
respect to that of an ideal, non-interacting Fermi gas. Since then,
a number of different aspects of a unitary Fermi gas have been characterized
after substantial experimental efforts. Hydrodynamic ballistic expansion
and collective excitations due to strong correlations were confirmed
\cite{KinastPRL2004,BartensteinPRL2004,HuPRL2004}; superfluidity
at BEC-BCS crossover was unambiguously verified by generating quantized
vortices \cite{MITVortices}; universal thermodynamics was evidenced
by heat capacity measurement \cite{HDLNaturePhysics,BourdelPRL2003,KinastScience2005,StewardPRL2006,LuoPRL2007,NascimbeneNature2010,HorikoshiScience2010,EoSMIT};
nearly ideal hydrodynamic flow and universal viscosity was observed
\cite{CaoScience2011}. Recently, these studies have been extended
to Fermi gases with unequal densities for the spin-up and spin-down
components \cite{ZwierleinScience2006,PartridgeScience2006} and to
Fermi gases in low-dimensions \cite{MartiyanovPRL2010,DykePRL2011,FrohlichPRL2011},
giving the prospects of realizing exotic inhomogeneous Fulde-Ferrell-Larkin-Ovchinnikov
(FFLO) superfluidity \cite{FFLO,OrsoPRL2007,HLDFFLO,RiceFFLO} and
Berezinskii-Kosterlitz-Thouless (BKT) transition \cite{BKT,ZhangPRA2008,TemperePRA2009}.
Along with rapid experimental progress, new measurement techniques
have been developed to characterize strongly correlated Fermi gases.
These include the momentum-resolved rf-spectroscopy for measuring
single-particle spectral function \cite{rfJILANature,rfJILANaturePhysics}
and the two-photon Bragg spectroscopy for dynamic and static structure
factors \cite{BraggSwinburne}.

In contrast, the parallel theoretical development is much slower.
There are numerous activities on developing better strong-coupling
theories \cite{OhashiPRL2002,HaussmannPRB1994,EngelbrechtPRB1997,PeraliPRL2004,ChenPhysicsReport2005,HLDEPL2006,LHEPL2006,HaussmannPRA2007,DienerPRA2008,CombescotPRA2009,GubbelsPRA2011}
or ab-initio quantum Monte Carlo (QMC) methods \cite{AstrakharchikPRL2004,BulgacPRL2006,AkkineniPRB2007,BurovskiPRL2008,CarlsonPRL2008,HouckePreprint}.
However, a deep understanding of strongly correlated Fermi gases is
prohibited because of the absence of a controllable small interaction
parameter. The use of standard strong-coupling theories requires infinite
order expansions and the truncation to a particular order can not
be fully justified a priori \cite{HLDPRA2008,HLDNJP2010}. At this
stage, numerically exact QMC simulations are less accurate than one
might expect, suffering from either the notorious sign problem for
fermions \cite{AkkineniPRB2007} or the finite size effects in small
samples used in the simulation \cite{BulgacPRL2006,BurovskiPRL2008}.

In this respect, exact results of strongly correlated Fermi gases
in some non-trivial limits are very valuable. Two recent efforts are
notable. In the limit of short-distance, large-momentum, and/or large-frequency,
Tan derived a set of exact universal relations \cite{TanRelations,BraatenReview}.
It was shown that all the limiting behaviors are governed by a many-body
parameter called the \emph{contact}, which measures the density of
pairs within short distances. Tan's relations can be conveniently
understood using the short-distance and/or short-time operator product
expansion (OPE) method \cite{BraatenPRL2008}, which separates in
a natural way the few-body physics from many-body physics. In another
limit of high temperature, quantum virial expansion \cite{HoVE,OhkumaPRA2006,OurVE,OurVELongPRA,OurVELongPRB,OurVirialDSF,OurVirialAkw,OurVirialContact,BlumeVE1,ViewpointPhysics}
provides another rigorous means to bridge few-body and many-body physics.
The properties of a strongly correlated Fermi gas, either static \cite{HoVE,OurVE,OurVELongPRA,OurVirialContact,BlumeVE1}
or dynamic \cite{OurVirialDSF,OurVirialAkw}, can be expanded non-perturbatively
using some exact expansion coefficients or expansion functions, which
are calculable from few-fermion solutions \cite{BuschFP1998,WernerPRL2006,KestnerPRA2007,StecherPRA2008,BlumePRA2009,DailyPRA2010,RittenhouseJPB2011}.
Both Tan relations and virial expansion give useful insights into
the challenging many-body problem, though in the different perspective.

In this paper, we review the recent theoretical development on quantum
virial expansion, and show that virial expansion gives a complete
solution of strongly-correlated Fermi gas above the Fermi degenerate
temperature. We focus our attention on ultracold atomic Fermi gases,
and compare in a quantitative way the virial expansion predictions
with available experimental measurements for various fundamental properties.
We note that the virial expansion has also been used frequently to
study the equation of state of neutron matter \cite{RopkeNPA1982,HorowitzNPA2006,MekjianPRC2009,TypelPRC2010,ShenPRC2010,NatowitzPRL2010}.

\subsection{Overview of virial expansion picture}

Quantum virial expansion, alternatively referred to as quantum cluster
expansion, is a standard method in quantum statistical mechanics \cite{KahnPhysica1938,HuangBook}.
It is practically useful for a dilute quantum gas. The basic idea
of virial expansion is simple. Though we have a strongly correlated
system at low temperatures, with increasing temperature the correlation
between particles would become {\em increasingly} weak. At sufficiently
high temperatures, the scattering cross section is of the order the
square of the thermal de Broglie wavelength, which becomes much smaller
than the average inter-atomic distance. As a result, the inclusion
of few-body correlations is already sufficient to describe the underlying
properties of the system. These few-body correlations can be exactly
taken into account using few-particle solutions and virial expansion.

As a concrete example, let us consider the thermodynamic potential
$\Omega$ for a given Hamiltonian ${\cal H}$, which in the grand
canonical ensemble is given by \cite{FetterBook}, 
\begin{equation}
\Omega=-k_{B}T\ln{\cal Z},
\end{equation}
 where $k_{B}$ is the Boltzmann constant, 
\begin{equation}
{\cal Z}=\text{Tr}\exp\left[-\left({\cal H}-\mu{\cal N}\right)/k_{B}T\right]
\end{equation}
 is the grand partition function, and ${\cal N}$ is the field operator
of total number of particles. The thermodynamic potential can be written
in terms of the partition function of clusters, 
\begin{equation}
Q_{n}=\text{Tr}_{n}\left[\exp\left(-{\cal H}/k_{B}T\right)\right],
\end{equation}
 where the integer $n$ denotes the number of particles in the cluster
and the trace Tr$_{n}$ is taken over $n$-particle states with a
proper symmetry. $Q_{n}$ is calculable using the complete solutions
of a $n$-particle system. The grand partition function then takes
the form 
\begin{equation}
{\cal Z}=1+zQ_{1}+z^{2}Q_{2}+z^{3}Q_{3}\cdots,
\end{equation}
 where $z=\exp\left(\mu/k_{B}T\right)$ is the fugacity \cite{FetterBook}.
At large temperatures, it is well-known that the chemical potential
$\mu$ diverges to $-\infty$, so the fugacity would be very small,
$z\ll1$. By Taylor-expanding $\ln{\cal Z}$ in powers of the small
fugacity, it is obvious that 
\begin{equation}
\Omega=z\tilde{\Omega}^{(1)}+z^{2}\tilde{\Omega}^{(2)}+z^{3}\tilde{\Omega}^{(3)}+\cdots,\label{veOmega}
\end{equation}
 where $\tilde{\Omega}^{(n)}$ can be expressed in terms of $Q_{i}$
($i\leq n$) and therefore contains the contribution from few-body
physics up to $n$-particles.

In principle, all the properties of a quantum gas could be cluster
expanded in powers of fugacity, no matter how strong the interactions.
The fugacity is a natural small parameter at large temperatures. Na\"{i}vely,
virial expansion is applicable when $z<1$. For a two-component spin-1/2
Fermi gas, using the fact that the fugacity is roughly equal to the
phase-space density $\rho\lambda_{dB}^{3}/2$, where $\rho$ is the
density, $\lambda_{dB}\equiv[2\pi\hbar^{2}/(mk_{B}T)]^{1/2}$ is the
thermal de Broglie wavelength and $m$ the mass of atoms, one can
estimate that virial expansion is useful at temperature $T>T_{F}$.
Here $T_{F}=\hbar^{2}k_{F}^{2}/(2mk_{B})$ is the Fermi degenerate
temperature.

\begin{figure}[htp]

\begin{centering}
\includegraphics[clip,width=0.5\textwidth]{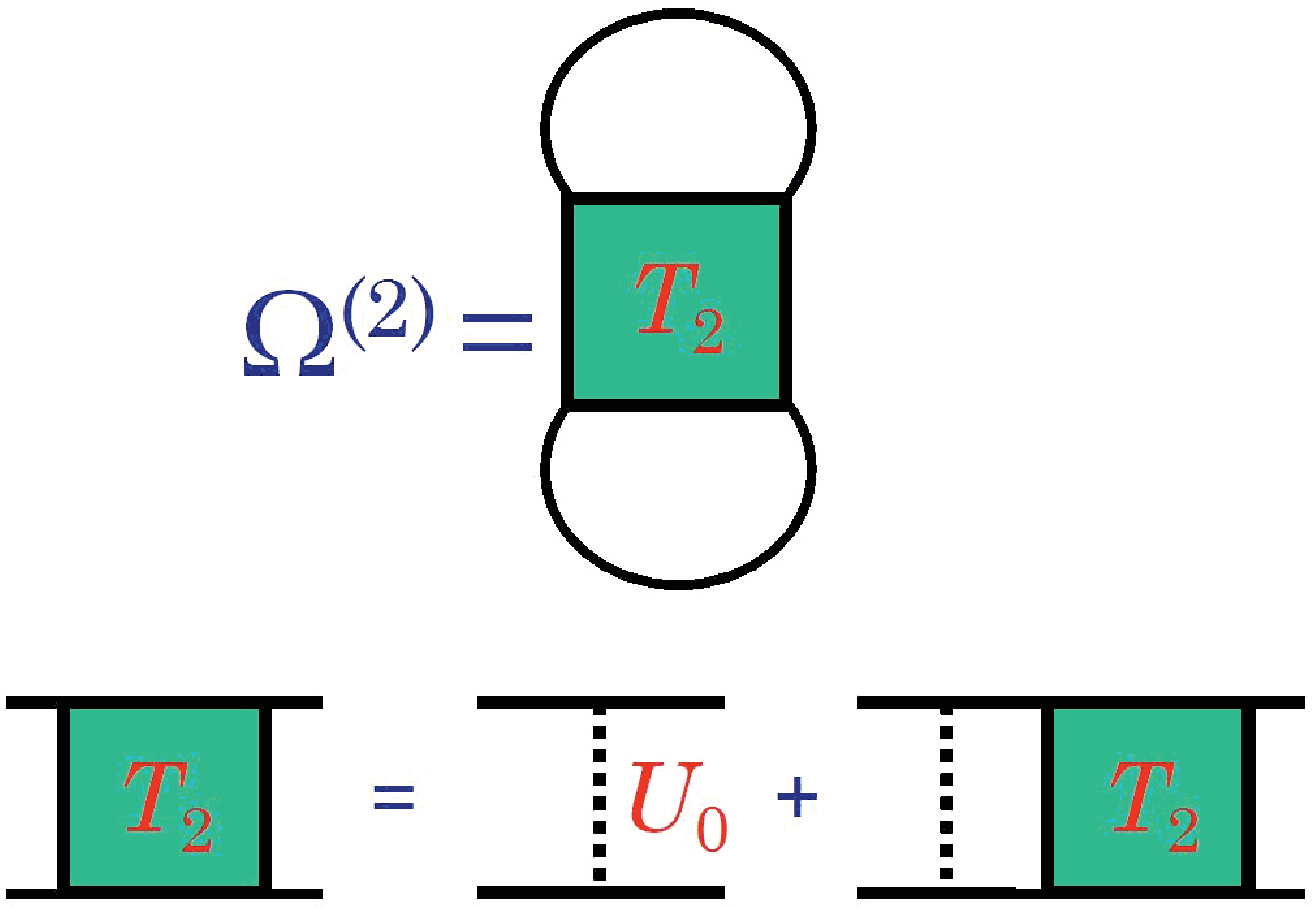} 
\par\end{centering}

\caption{(color online) Diagrammatic representation of the contribution of
two-particle scattering process to the thermodynamic potential. Here
$T_{2}$ is the two-particle vertex function, obtained by summing
all the ladder-type diagrams. The dashed line and solid line represent
the bare contact interaction $U_{0}$ and the single-particle Green
function, respectively. For details, see refs. \cite{NSR} and \cite{HLDEPL2006}.}

\label{fig:T2} 
\end{figure}

\begin{figure}[htp]
\begin{centering}
\includegraphics[clip,width=0.8\textwidth]{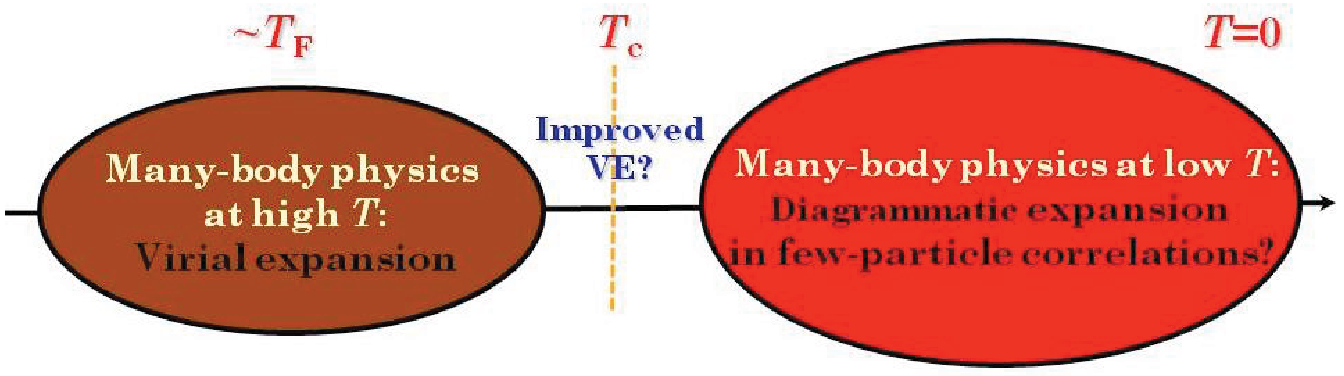} 
\par\end{centering}

\caption{(color online) Schematic illustration of virial expansion and diagrammatic
expansion, both of which are expansion in few-particle correlations.
It is desirable to find an improved theory, which connects smoothly
these two expansion theories.}

\label{fig:diagram} 
\end{figure}

Though virial expansion is a large-temperature expansion, it has an
intrinsic relation with low-temperature strong-coupling diagrammatic
theory. In the absence of a small interaction parameter, we may conjecture
that a {\em reliable} strong-coupling theory of strongly correlated
systems should be developed by successively including few-particle
scattering process. Thus, we may write 
\begin{equation}
\Omega=\Omega^{(1)}+\Omega^{(2)}+\Omega^{(3)}+\cdots,\label{deOmega}
\end{equation}
 where $\Omega^{(1)}$ is the thermodynamic potential of a non-interacting
system, and $\Omega^{(n)}$ ($n\geq2$) is the contribution from the
$n$-particle scattering process, which is to be calculated at {\em
all} temperatures by summing a series of diagrams to infinite order
(i.e., the $n$-particle vertex function $T_{n}$). We shall refer
to such an expansion as the \emph{diagrammatic} expansion. As an example,
in Fig. \ref{fig:T2} we show the diagrammatic representation of $\Omega^{(2)}$.
It sums up all the two-particle scatterings via the two-particle vertex
function $T_{2}$ \cite{NSR}. In the language of functional path-integral
method, $\Omega^{(2)}$ corresponds to the gaussian fluctuations around
the mean-field saddle point \cite{NSR,SadeMeloPRL1993,OhashiPRL2002,HLDEPL2006}.
It is obvious that at large temperatures, by expanding $\Omega^{(i)}$
($i\leq n$) in powers of $z$, we can calculate directly $\tilde{\Omega}^{(n)}$
appeared in the virial expansion. In this respect, virial expansion
and diagrammatic expansion are closely related. Both of them are the
expansion in few-particle correlations. The advantage of the diagrammatic
expansion is that it takes into account the few-particle scatterings
in the medium, and therefore is applicable at all temperatures. It
is a natural generalization of virial expansion to the low-temperature
regime. These two expansion theories are sketched in Fig. \ref{fig:diagram}.

Ideally, for a strongly correlated system, we anticipate that $\Omega^{(n)}$
becomes less important with increasing $n$ and the diagrammatic expansion
thus converges. Indeed, for a unitary Fermi gas at the BEC-BCS crossover,
the theoretical calculation of $\Omega^{(2)}$ at zero temperature
gives fairly accurate equation of state \cite{HLDEPL2006}, as confirmed
by the latest experimental measurement \cite{NavonScience2010}. Others
terms of $\Omega^{(n)}$ with $n\geq3$ are notoriously difficult
to obtain, but are conjectured to be small at low temperatures. Virial
expansion provides systematic determinations of $\Omega^{(n)}$ at
high temperatures and may shed light on their low temperature behavior.

At this point, we may anticipate that the applicability of virial
expansion is not limited to small fugacity $z<1$. The expansion could
be meaningful in the deep quantum degenerate regime through an analytic
continuation across the point $z=1$, and therefore is applicable
down to the superfluid phase transition temperature $T_{c}$. The
pursuit of such an improved virial expansion theory is a theoretical
challenge.

\subsection{Key technical issues in the latest development of virial expansion}

Despite the usefulness of virial expansion, its application to strongly
correlated quantum gases is less documented in the literature. There
are two severe technical difficulties in applying virial expansion
to a {\em homogeneous} system: (i) insufficient knowledge on the
exact few-particle solutions and (ii) continuous energy spectrum.
As a result, it seems impossible to calculate the essential few-particle
cluster partition function $Q_{n}$ when $n\geq3$, as the calculation
requires infinitely large number of energy levels. Therefore, previous
applications of virial expansion have been restricted to the second
order, where $Q_{2}$ can be calculated using an elegant phase-shift
formalism derived by Beth and Uhlenbeck in 1937 \cite{BethPhysica1937,LandauBook}.

The latest development of virial expansion, to be reviewed in this
paper, relies on the recent theoretical progress on the exact few-particle
solutions of {\em trapped} strongly interacting fermions \cite{BuschFP1998,WernerPRL2006,KestnerPRA2007,StecherPRA2008,BlumePRA2009,DailyPRA2010,RittenhouseJPB2011}.
Due to the trapping potential, the energy spectrum becomes discrete.
As the thermal energy $k_{B}T$ provides a natural high-energy scale
in the cluster partition function, the number of energy levels required
by the calculation is finite. In principle, we can always calculate
numerically these energy levels using the ever-growing computation
power, if few-particle solutions are not known analytically. In addition,
virial expansion of dynamic properties becomes possible, based on
the calculated few-particle wave-functions \cite{OurVirialDSF,OurVirialAkw}.

To get back to the homogeneous system, one can utilize the so-called
local density approximation, which treats the trapped Fermi gas as
a collection of many locally uniform blocks. In the unitary limit,
it is found that the virial expansion results for trapped and homogeneous
systems are convertible by some universal relations \cite{OurVE,HLPreprint}.
The details will be discussed later.

\subsection{Model Hamiltonian}

Throughout this Review, we focus on the strongly-interacting Fermi
gases with {\em zero-range} interactions in \emph{three} dimensions,
which have emerged as the simplest strongly-correlated model system
that has been accessed experimentally with ultracold atoms of $^{6}$Li
and $^{40}$K \cite{BlochRMP2008,GiorginiRMP2008,KetterleCourse}.
They also serve as a ``bare bone'' description of nuclear and neutron
matter. The generalization of virial expansion in low dimensional
systems is straightforward \cite{OurVELongPRB}.

\begin{figure}[htp]
\begin{centering}
\includegraphics[clip,width=0.5\textwidth,angle=-90]{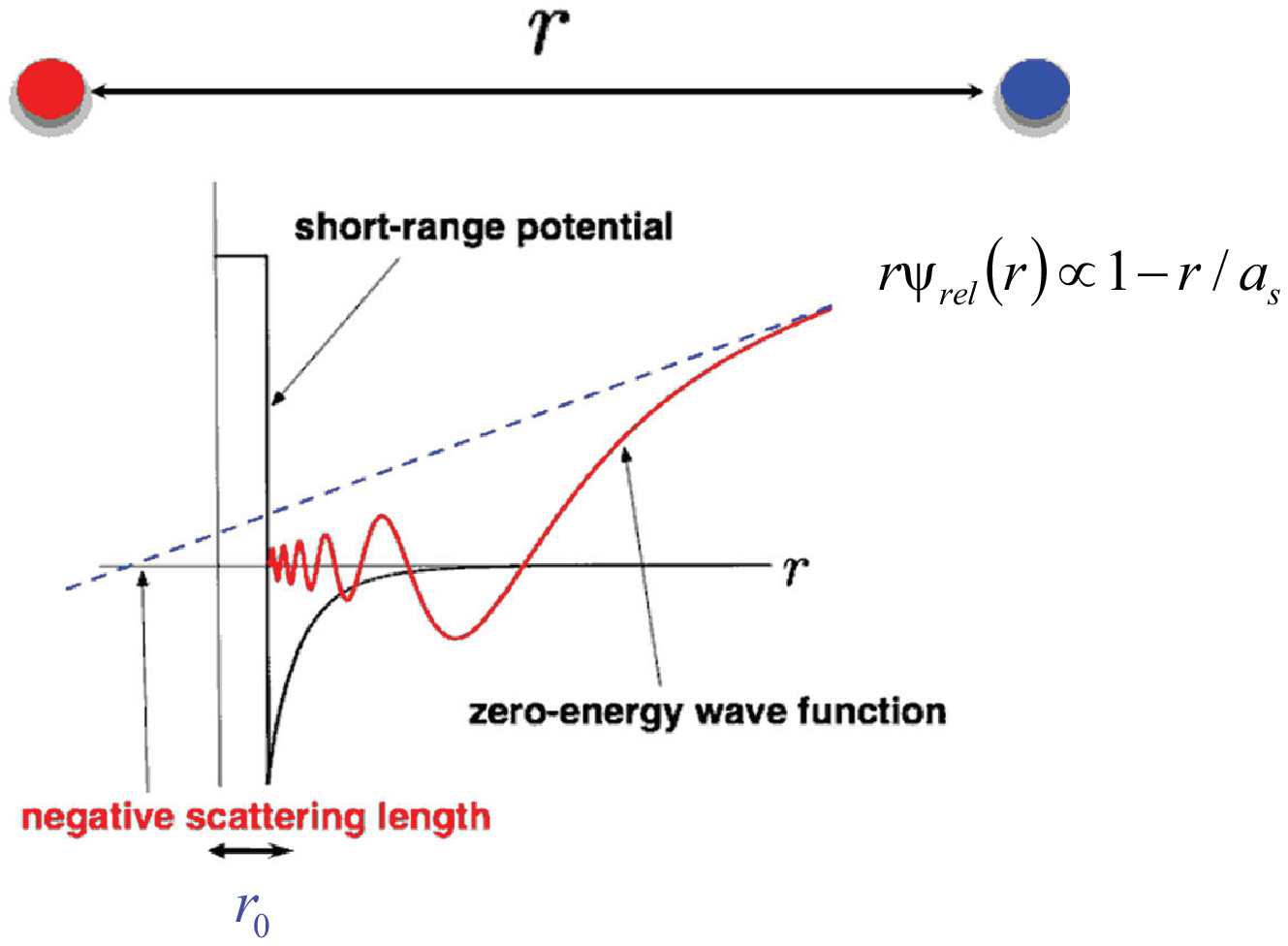} 
\par\end{centering}

\caption{(color online) Explanation on the use of zero-range interactions.
For a dilute quantum gas at low temperatures, the short-range behavior
of zero-energy scattering wave-function between spin-up and spin-down
fermions (represented by different colors) can not be accessed. It
can therefore be safely replaced by a simple asymptotic form, $r\psi_{rel}(r)\propto1-r/a_{s}$,
which defines an $s$-wave scattering length, $a_{s}$. The interatomic
interactions are characterized by the single parameter $a_{s}$ only.}

\label{fig:zeroRangeInteraction} 
\end{figure}

The use of zero-range interactions can be understood from Fig. \ref{fig:zeroRangeInteraction},
which plots the short-range behavior of the zero-energy scattering
wave-function $\psi_{rel}(r)$ for real interatomic interactions with
a range of interactions $r_{0}$. For an ultracold \textit{dilute}
Fermi gas, $r_{0}$ is typically at the order of $10^{-9}$ m, much
smaller than the mean inter-particle distance $\rho^{-1/3}\sim10^{-6}$
m. As a result, the complicated short-range behavior of the wave-function,
which corresponds to high-energy physics, becomes irrelevant for typical
atomic collisions. Therefore, we may set effectively $r_{0}=0$ and
approximate $\psi_{rel}(r)\propto1/r-1/a_{s}$ at $r\sim r_{0}=0$
\cite{HuangBook}. Here $a_{s}$ is the $s$-wave scattering length
\cite{HuangBook}. This is the so-called Bethe-Peierls (BP) boundary
condition, which is equivalent to the $s$-wave zero-range interactions
(or the so-called pseudopotential \cite{HuangBook}). We note that,
when the $s$-wave scattering length becomes positive, the interaction
potential will support a two-body bound state with energy $E_{B}=-\hbar^{2}/(ma_{s}^{2})$.
Therefore, in the unitary limit, where $a_{s}\rightarrow\pm\infty$,
a shallow two-body bound state with infinitely small energy emerges.
For more details, see ref. \cite{ChinRMP2010}.

In the ultracold atom experiments, a harmonic trap is necessary to
prevent atoms from escaping. We thus consider $N$ fermions in a three-dimensional
isotropic harmonic trap $V_{T}({\bf x})=m\omega_{T}^{2}(x^{2}+y^{2}+z^{2})/2$
with the same mass $m$ and trapping frequency $\omega_{T}$, occupying
two different hyperfine states or two spin states. The zero-range
$s$-wave interaction between fermions with {\em unlike} spins
is replaced by the BP boundary condition. That is, when any particles
$i$ and $j$ with unlike spins are close to each other, $r_{ij}=\left|{\bf x}_{i}-{\bf x}_{j}\right|\rightarrow0$,
the many--body wave function $\psi\left({\bf x}_{1},{\bf x}_{2},...,{\bf x}_{N}\right)$
with proper symmetry should satisfy \cite{WernerPRA2006,WernerPhDThesis},
\begin{equation}
\psi={\cal A}_{ij}({\bf X}_{ij}=\frac{{\bf x}_{i}+{\bf x}_{j}}{2},\{{\bf x}_{k\neq i,j}\})\left(\frac{1}{r_{ij}}-\frac{1}{a_{s}}\right),\label{ManyBodyWaveFunctionBP}
\end{equation}
 where ${\cal A}_{ij}({\bf X}_{ij},\{{\bf x}_{k\neq i,j}\})$ is a
function independent of $r_{ij}$. This BP boundary condition can
be equivalently written as, 
\begin{equation}
\lim_{r_{ij}\rightarrow0}\frac{\partial\left(r_{ij}\psi\right)}{\partial r_{ij}}=-\frac{r_{ij}\psi}{a_{s}}.\label{BP}
\end{equation}
 Otherwise, the wave function $\psi$ obeys a non-interacting Schrödinger
equation, 
\begin{equation}
\sum_{i=1}^{N}\left[-\frac{\hbar^{2}}{2m}{\bf \nabla}_{{\bf x}_{i}}^{2}+\frac{1}{2}m\omega_{T}^{2}\left(x_{i}^{2}+y_{i}^{2}+z_{i}^{2}\right)\right]\psi=E\psi.
\end{equation}

In the second quantization, the system can be instead described by
the model Hamiltonian,

\begin{equation}
{\cal H}=\sum_{\sigma=\uparrow,\downarrow}\int d{\bf x}\psi_{\sigma}^{\dagger}({\bf x})\left[-\frac{\hbar^{2}\nabla^{2}}{2m}+V_{T}({\bf x})-\mu_{\sigma}\right]\psi_{\sigma}({\bf x})+U_{0}\int d{\bf x}\psi_{\uparrow}^{\dagger}({\bf x})\psi_{\downarrow}^{\dagger}({\bf x})\psi_{\downarrow}({\bf x})\psi_{\uparrow}({\bf x}),\label{ManyBodyHami}
\end{equation}
 where the chemical potentials $\mu_{\uparrow}$ and $\mu_{\downarrow}$
could be different due to unequal spin-populations. The zero-range
interaction is given by a contact potential $U_{0}\delta({\bf x-x}^{\prime})$.
The bare interaction strength $U_{0}$ is to be renormalized by two-particle
vertex function $T_{2}$ in the vacuum \cite{FetterBook}, using 
\begin{equation}
\frac{1}{U_{0}}=\frac{m}{4\pi\hbar^{2}a_{s}}-\frac{1}{V}\sum_{{\bf k}}\frac{m}{\hbar^{2}{\bf k}^{2}},\label{Renormalization}
\end{equation}
 where the momentum ${\bf k}$ has a high-energy cut-off, $k<\Lambda=r_{0}^{-1}$,
in accord with the use of zero-range interactions. $U_{0}$ scales
to zero as the cut-off momentum $\Lambda\rightarrow\infty$.

The (single-channel) model Hamiltonian shown above provide the simplest
description of ultracold $^{6}$Li and $^{40}$K atoms near {\em
broad} Feshbach resonances \cite{BlochRMP2008,GiorginiRMP2008,KetterleCourse}.
In case of narrow Feshbach resonances, it is necessary to use a two-channel
model and to include molecules in the closed channel \cite{DienerPreprint,LiuPRA2005}.

\subsection{Brief introduction to Tan relations}

Here we introduce briefly the exact Tan relations, which were derived
by Shina Tan in 2005 \cite{TanRelations,BraatenReview}. These Tan
relations link the asymptotic behavior of many-body systems at short-range,
large-momentum, and high-frequency to their thermodynamic properties.
For instance, the momentum distribution $\rho_{\sigma}(q)$ falls
off as ${\cal I}/q^{4}$ at large momentum, the pair correlation function
$g_{\uparrow\downarrow}({\bf x}_{i}-{\bf x}_{j})\equiv\int d{\bf X}_{ij}\left\langle \hat{\rho}_{\uparrow}({\bf x}_{i})\hat{\rho}_{\downarrow}({\bf x}_{j})\right\rangle $
diverges like 
\begin{equation}
g_{\uparrow\downarrow}(r_{ij}=\left|{\bf x}_{i}-{\bf x}_{j}\right|\rightarrow0)\simeq\frac{{\cal I}}{16\pi^{2}}\left(\frac{1}{r_{ij}^{2}}-\frac{2}{a_{s}r_{ij}}\right),
\end{equation}
 and the rf-spectroscopy has the tail of ${\cal I}/\omega^{5/2}$
at large frequency \cite{PunkPRL2007,BaymPRL2007}. All the Tan relations
are related to each other by a single{\em \ }coefficient ${\cal I}$,
referred to as the integrated contact density or contact. The contact
measures the probability of two fermions with unlike spins being close
together \cite{BraatenPRL2008}. It also links the short-range behavior
to thermodynamics via the adiabatic relation, 
\begin{equation}
\left[\frac{\partial E}{\partial\left(-1/a_{s}\right)}\right]_{S,N}=\frac{\hbar^{2}}{4\pi m}{\cal I},\label{AdiabaticTanRelation}
\end{equation}
 which gives the change in the total energy $E$ due to adiabatic
changes in the scattering length. The fundamental importance of the
Tan relations arises from their wide applicability. They are useful
at both zero or finite temperature, superfluid or normal phase, homogeneous
or trapped, few-body or many-body systems.

While the original rigorous derivation by Shina Tan is difficult to
follow, the underlying physics of Tan relations can be easily understood
from several points of view \cite{BraatenReview}. The simplest way
is from the two-body wave function under the BP boundary condition,
$\psi_{rel}(r)\propto1/r-1/a_{s}$. Na\"{i}vely, the momentum distribution
$\rho_{\sigma}(q)$ is simply the square of the Fourier transform
of $\psi_{rel}(r)$ and the pair correlation function $g_{\uparrow\downarrow}(r)\propto\left|\psi_{rel}(r)\right|^{2}$.
The asymptotic behavior of $\rho_{\sigma}(q)\propto q^{-4}$ and $g_{\uparrow\downarrow}(r)\propto r^{-2}$
is then straightforward to check.

At the many-body level, Tan's relations can be elegantly proved by
using the short-distance and/or short-time operator product expansion
(OPE) method \cite{BraatenPRL2008}, in which the contact ${\cal I}$
is identified as 
\begin{equation}
{\cal I}=U_{0}^{2}\int d{\bf x}\psi_{\uparrow}^{\dagger}({\bf x})\psi_{\downarrow}^{\dagger}({\bf x})\psi_{\downarrow}({\bf x})\psi_{\uparrow}({\bf x})\text{.}
\end{equation}
 For example, the adiabatic relation Eq. (\ref{AdiabaticTanRelation})
can be obtained directly by applying Hellmann-Feynman theorem to the
model Hamiltonian \cite{BraatenPRL2008}, 
\begin{equation}
\left[\frac{\partial E}{\partial\left(-1/a_{s}\right)}\right]_{S,N}=\left\langle \frac{\partial{\cal H}}{\partial\left(-1/a_{s}\right)}\right\rangle =\frac{\partial U_{0}}{\partial\left(-1/a_{s}\right)}\left\langle \int d{\bf x}\psi_{\uparrow}^{\dagger}\psi_{\downarrow}^{\dagger}\psi_{\downarrow}\psi_{\uparrow}\right\rangle =\frac{\hbar^{2}}{4\pi m}{\cal I}.
\end{equation}
 The last step follows the renormalization for the bare interaction
strength Eq. (\ref{Renormalization}).

The contact is a fundamental parameter that characterizes the many-body
properties of strongly correlated Fermi gases. Recently, its measurement
receives considerable attentions \cite{TanJILA,TanSwinburne1,TanSwinburne2,TanSwinburne3}.
It turns out that the most accurate way is through the Tan relation
for spin-antiparallel static structure factor \cite{ourTanRelation},
which is obtained by a direct Fourier transform of pair correlation
function, 
\begin{equation}
S_{\uparrow\downarrow}\left(q\gg k_{F}\right)\simeq\frac{{\cal I}}{4Nq}\left[1-\frac{4}{\pi a_{s}q}\right].\label{StructureFactorTanRelation}
\end{equation}
 The simple power-law tail of $1/q$ in the structure factor relation
is more amenable for experimental measurement than the $q^{-4}$ or
$\omega^{-5/2}$ tail in the momentum distribution or in rf-spectroscopy.
In the latter two cases, the fast decay due to the higher-order power
law imposes more stringent signal-to-noise requirements at a given
momentum or frequency. Experimentally, the static structure factor
can be measured by two-photon Bragg spectroscopy \cite{BraggSwinburne,TanSwinburne1}.

\subsection{Brief summary of virial expansion results}

We now summarize briefly the main results of the latest development
in virial expansion. In general, thermodynamic properties such as
the thermodynamic potential can be expanded in terms of some virial
coefficients, while dynamic properties, i.e., the single-particle
spectral function and dynamic structure factor, can be expanded in
terms of some virial expansion functions. The latest developments
of virial expansion include:

(i) The third virial coefficient $b_{3}$ for thermodynamic potential
has been precisely determined \cite{OurVE,BlumeVE1}. For a homogeneous
Fermi gas in the unitary limit, it is found that $\Delta b_{3}=b_{3}-b_{3}^{(1)}=-0.3551030264897$
\cite{BlumeVE1}. Here $b_{3}^{(1)}$ is the third virial coefficient
of an ideal, non-interacting Fermi gas. This theoretical prediction
has been confirmed experimentally with the experimental value $\Delta b_{3,extp}=-0.35\pm0.02$
\cite{NascimbeneNature2010} and has been independently checked by
a field theoretic approach \cite{LeyronasPreprint}, which gives $\Delta b_{3}=-0.3551\pm0.0001$.
The fourth virial coefficient in the unitary limit has also been calculated
\cite{BlumeVE1}, but with much less accuracy, $\Delta b_{4}=-0.016\pm0.004$.
The virial coefficients of a trapped system $b_{n,T}$ and a homogeneous
system $b_{n}$ are related by \cite{OurVE}, $b_{n,T}=n^{-3/2}b_{n}$
($n=1,2,3,\cdots$). These virial coefficients predict an accurate
equation of state for a trapped Fermi gas in the unitary limit, for
temperature down to $0.5T_{F}$ \cite{HLDNJP2010}. We review the
virial expansion of thermodynamics in Sec. II.

(ii) The contact parameter ${\cal I}$ can be virial expanded in terms
of contact coefficients $c_{n}$ \cite{OurVirialContact}. For a homogeneous
Fermi gas in the unitary limit, it is predicted that $c_{2}=1/\pi$
and $c_{3}=-0.1399\pm0.0001$ \cite{LeyronasPreprint}. In analogy
with the virial coefficient, the contact coefficients of a trapped
system $c_{n,T}$ and a homogeneous system $c_{n}$ are related by,
$c_{n,T}=n^{-3/2}c_{n}$. Likewise, for a trapped Fermi gas in the
unitary limit, the virial expansion of contact provides an excellent
explanation for the experimental measurement at $T>0.5T_{F}$ \cite{TanSwinburne2}.
This part will be reviewed in Sec. III.

(iii) The virial expansion functions for the single-particle spectral
function and dynamic structure factor have been determined \cite{OurVirialDSF,OurVirialAkw},
to the second order in fugacity. These results enable an important
qualitative understanding of recent experimental measurements on momentum-resolved
rf-spectroscopy \cite{rfJILANature,rfJILANaturePhysics} and two-photon
Bragg spectroscopy \cite{TanSwinburne2,TanSwinburne3}, for a trapped
Fermi gas in the unitary limit at temperature down to the superfluid
transition. The determination of the third virial expansion functions
is straightforward, but involves much heavier numerical efforts. The
virial expansion of dynamic structure factor and of single-particle
spectral function will be reviewed in Secs. IV and V, respectively.

\section{Virial expansion of equation of state}

Let us consider the virial expansion of the thermodynamic potential
$\Omega$, for a balanced spin-1/2 Fermi gas with equal spin populations
($\mu_{\uparrow}=\mu_{\downarrow}=\mu$). The spin-population imbalanced
case with $\mu_{\uparrow}\neq\mu_{\downarrow}$ will be discussed
at the end of the section. All the equations of state can be derived
from the thermodynamic potential. By Taylor-expanding $\Omega=-k_{B}T\ln{\cal Z}$
in the fugacity, where ${\cal Z}=1+zQ_{1}+z^{2}Q_{2}+\cdots$, the
thermodynamic potential takes the form, 
\begin{equation}
\Omega=-k_{B}TQ_{1}\left[z+b_{2}z^{2}+\cdots+b_{n}z^{n}+\cdots\right],
\end{equation}
 where $b_{n}$ is referred to as the $n$-th (virial) expansion coefficient.
Note that, by definition $\tilde{\Omega}^{(3)}$ in Eq. (\ref{veOmega})
is given by $\tilde{\Omega}^{(3)}$ $=-k_{B}TQ_{1}b_{n}$. It is readily
seen that, 
\begin{eqnarray}
b_{2} & = & \left(Q_{2}-Q_{1}^{2}/2\right)/Q_{1},\\
b_{3} & = & \left(Q_{3}-Q_{1}Q_{2}+Q_{1}^{3}/3\right)/Q_{1},\ etc.
\end{eqnarray}
 These equations give a general definition of virial expansion, which
is applicable to both homogeneous and trapped systems. The calculation
of the $n$-th virial coefficient requires the input of cluster partition
function $Q_{i}$ ($i\leq n$), and hence requires the solutions of
up to the $n$-particle problem. In practice, it is convenient to
concentrate on the interaction effects only. We therefore consider
the difference $\Delta b_{n}\equiv b_{n}-b_{n}^{(1)}$ and $\Delta Q_{n}\equiv Q_{n}-Q_{n}^{(1)}$,
where the superscript ``$1$'' denotes the non-interacting systems
\footnote{We note that there are varying definitions of virial coefficients
in the literature. In some works, e.g. ref. \cite{NascimbeneNature2010},
the ideal gas contribution $b_{n}^{(1)}$ is excluded from the definition
of virial coefficients. That is, $\Delta b_{n}$ defined in the present
work is treated as the virial coefficient $b_{n}$.%
}. For the second and third virial coefficients, one shall calculate
respectively 
\begin{equation}
\Delta b_{2}=\Delta Q_{2}/Q_{1}\label{db2eq}
\end{equation}
 and 
\begin{equation}
\Delta b_{3}=\Delta Q_{3}/Q_{1}-\Delta Q_{2}.\label{db3eq}
\end{equation}

As we mentioned earlier, the calculation of virial coefficients in
the strongly correlated regime is a subtle theoretical problem. The
second virial coefficient was known long time ago through the elegant
Beth-Uhlenbeck formalism, which relates in a simple manner the second
virial coefficient to the two-body $S$-matrix or the two-body scattering
phase shift \cite{BethPhysica1937,LandauBook,ServadioPRA1971}. A
connection between the virial series and the scattering matrix has
been suspected since then. However, the computation of the third virial
coefficient, along the line of Beth and Uhlenbeck's original work,
met the very difficulties of the three-particle problem \cite{PaisPR1959,LarsenPRA1970}.

Until very recently, there is renewed interest in calculating higher-order
virial coefficients, largely due to the creation of ultracold atomic
Fermi gases. Initial attempt was based on the field theoretic method,
by calculating the contribution of three-particle scattering process
to the thermodynamic potential $\Omega^{(3)}$ \cite{BedaquePRB2003}.
It was shown by Rupak in 2007 that in the unitary limit, $\Delta b_{3}\simeq+1.05$
\cite{RupakPRL2007}. However, it was soon realized by Liu, Hu and
Drummond \cite{OurVE} that this value does not agree with the high-temperature
heat-capacity measurements reported by Thomas's group at Duke University
\cite{LuoPRL2007}. Using an entire different strategy based on Eq.
(\ref{db3eq}) and the exact three-particle solution in harmonic traps,
they predicted $\Delta b_{3}=-0.35510298$. The numerical accuracy
of $\Delta b_{3}$ can be improved by including more three-particle
energy levels. The latest calculation by Rakshit, Daily and Blume,
along the line of Liu, Hu and Drummond's work, gave $\Delta b_{3}=-0.3551030264897$
and $\Delta b_{4}=-0.016\pm0.004$ \cite{BlumeVE1}. In parallel,
new field theoretic calculations for the third virial coefficient
have been performed. It was shown by Kaplan and Sun $\Delta b_{3}=-0.3573\pm0.0005$
\cite{KaplanPRL2011} and by Leyronas $\Delta b_{3}=-0.3551\pm0.0001$
\cite{LeyronasPreprint}. At this stage, more complete field theoretic
calculation is desirable, in order to confirm independently the fourth
virial coefficient and to predict new virial coefficients.

On the other hand, the experimental accuracy in measuring the equation
of state of a unitary Fermi gas is improved very rapidly. The measurement
by Salomon's group at École Normale Supérieure (ENS) gave $\Delta b_{3,expt}=-0.35\pm0.02$
and $\Delta b_{4,expt}=0.096\pm0.015$ \cite{NascimbeneNature2010}.
The latest measurement by Zwierlein's group at MIT reported $\Delta b_{4,expt}=0.096\pm0.010$
\cite{EoSMIT}. We anticipate that the new predictions on the virial
coefficients, improved continuously by many theorists, will play an
important role in deepening our understanding of the equation of state
of strongly correlated Fermi systems.

\subsection{Virial coefficients of non-interacting Fermi gases}

The background non-interacting virial coefficients can be conveniently
determined by the non-interacting thermodynamic potential. For a {\em
homogeneous} two-component Fermi gas, it takes the form \cite{FetterBook},
\begin{equation}
\Omega^{(1)}=-2k_{B}T\sum_{{\bf k}}\ln\left[1+e^{-\left(\epsilon_{{\bf k}}-\mu\right)/\left(k_{B}T\right)}\right],
\end{equation}
 where $\epsilon_{{\bf k}}=\hbar^{2}k^{2}/(2m)$ is the single-particle
energy, the factor of $2$ accounts for the spin degree of freedom.
By using $\sum_{{\bf k}}=V\int_{0}^{\infty}4\pi k^{2}dk/(2\pi)^{3}$
and introducing a new variable $t=\epsilon_{{\bf k}}/(k_{B}T)$, the
ideal thermodynamic potential becomes, 
\begin{equation}
\Omega^{(1)}=-V\frac{2k_{B}T}{\lambda_{dB}^{3}}\frac{2}{\sqrt{\pi}}\int\limits _{0}^{\infty}t^{1/2}\ln\left(1+ze^{-t}\right)dt,\label{idealOmegaHomo}
\end{equation}
 where $\lambda_{dB}\equiv[2\pi\hbar^{2}/(mk_{B}T)]^{1/2}$ is the
thermal de Broglie wavelength. It is easy to identify $Q_{1}=2V/\lambda_{dB}^{3}$.
Therefore, by Taylor-expanding $\ln\left(1+ze^{-t}\right)$ in fugacity
$z$ and integrating out $t$ term by term, we obtain the non-interacting
virial coefficients in free space, 
\begin{equation}
b_{n}^{(1)}=\frac{\left(-1\right)^{n+1}}{n^{5/2}}.
\end{equation}

For a Fermi gas in a harmonic trapping potential $V_{T}({\bf x})=m\omega_{T}^{2}(x^{2}+y^{2}+z^{2})/2$,
it is convenient to use the semiclassical approximation, or the so-called
local density approximation. In the non-interacting limit, this amounts
to setting, 
\begin{equation}
\Omega_{T}^{(1)}=\int d{\bf x}\frac{\Omega^{(1)}\left({\bf x}\right)}{V}=-2k_{B}T\int d{\bf x}\left\{ \frac{1}{V}\sum_{{\bf k}}\ln\left[1+e^{-\left[\epsilon_{{\bf k}}+V_{T}\left({\bf x}\right)-\mu\right]/\left(k_{B}T\right)}\right]\right\} ,
\end{equation}
 where locally the single-particle energy is given by $\epsilon_{{\bf k}}+V_{T}({\bf x})$.
Hereafter, we take the subscript ``$T$'' to denote the quantity
in the trapped system, otherwise, by default we refer to a homogeneous
system. As before, the integrations over ${\bf x}$ and ${\bf k}$
can be done by introducing a new variable $t=[\epsilon_{{\bf k}}+V_{T}({\bf x})]/(k_{B}T)$
. This leads to, 
\begin{equation}
\Omega_{T}^{(1)}=-\frac{2\left(k_{B}T\right)^{4}}{\left(\hbar\omega_{T}\right)^{3}}\left[\frac{1}{2}\int\limits _{0}^{\infty}t^{2}\ln\left(1+ze^{-t}\right)dt\right],\label{idealOmegaTrap}
\end{equation}
 where $Q_{1,T}=2\left(k_{B}T\right)^{3}/\left(\hbar\omega_{T}\right)^{3}$.
By Taylor-expanding the log-term in fugacity $z$, we find the non-interacting
virial coefficients in harmonic traps, 
\begin{equation}
b_{n,T}^{(1)}=\frac{\left(-1\right)^{n+1}}{n^{4}}.
\end{equation}
 We note that the use of semi-classical approximation means to neglect
the discreteness of the energy spectrum in traps. Mathematically,
this is equivalent to take a small parameter $\tilde{\omega}_{T}=\hbar\omega_{T}/(k_{B}T)$
and to keep in the results the leading term in $\tilde{\omega}_{T}$.
We note also that the non-interacting virial coefficients in the homogeneous
case and trapped case are related by, $b_{n,T}^{(1)}=n^{-3/2}b_{n}^{(1)}$
($n=1,2,3,\cdots$).

\subsection{Universal relation between homogeneous and trapped virial coefficients}

The correspondence relation discussed in Sec. IIA holds for a strongly
interacting Fermi gas as well. Here, the crucial point is that the
virial coefficients become temperature {\em independent}. To understand
this, we note that in general the coefficients should be a function
of the ratio $\lambda_{dB}/a_{s}$, between the only two length scales
$\lambda_{dB}$ and $a_{s}$. The temperature dependence enters through
the thermal de Broglie wavelength. However, for a unitary Fermi gas
where $\lambda_{dB}/a_{s}=0$, this dependence disappears. This is
indeed a manifestation of fermionic universality, shared by many quantum
systems with strong short-range interactions.

In the thermodynamic limit, let us consider the thermodynamic potential
of a harmonically trapped Fermi gas in the local density approximation,
\begin{equation}
\Omega_{T}{\bf =}-\frac{2\left(k_{B}T\right)^{4}}{\left(\hbar\omega_{T}\right)^{3}}\left[z+b_{2,T}z^{2}+b_{3,T}z^{3}+\cdots\right]=\int d{\bf x}\frac{\Omega({\bf x})}{V},
\end{equation}
 where the trapped virial coefficients $b_{n,T}$ are to be determined
and $\Omega({\bf x})$ is the local thermodynamic potential, 
\begin{equation}
\Omega({\bf x})=-V\frac{2k_{B}T}{\lambda_{dB}^{3}}\left[z\left({\bf x}\right)+b_{2}z^{2}\left({\bf x}\right)+b_{3}z^{3}\left({\bf x}\right)+\cdots\right].
\end{equation}
 Here, the local fugacity $z\left({\bf x}\right)\equiv\exp[\mu\left({\bf x}\right)/(k_{B}T)]=z\exp[-V_{T}({\bf x})/(k_{B}T)]$
is given by the local chemical potential $\mu({\bf x})=\mu-V_{T}({\bf x})$.
Because of the temperature-independent (constant) virial coefficients
$b_{n}$, the spatial integration can be done explicitly. This immediately
leads to the universal relation for the virial coefficients of a unitary
Fermi gas, 
\begin{equation}
b_{n,T}=\frac{b_{n}}{n^{3/2}}.\label{UniversalRelationBn}
\end{equation}

\subsection{Second virial coefficient of interacting Fermi gases}

\subsubsection{Beth-Uhlenbeck formalism}

As shown by Beth and Uhlenbeck in 1937 \cite{BethPhysica1937}, the
second virial coefficient can be expressed in terms of the phase shifts
of a two-body scattering problem. For a spin-1/2 Fermi gas, it takes
the form, 
\begin{equation}
\frac{\Delta b_{2}}{\sqrt{2}}=\sum_{i}e^{-E_{B}^{i}/\left(k_{B}T\right)}+\sum_{l}\frac{\left(2l+1\right)}{\pi}\int\limits _{0}^{\infty}dk\frac{d\delta_{l}}{dk}e^{-\frac{\lambda_{dB}^{2}k^{2}}{2\pi}},\label{BethUhlenbeck}
\end{equation}
 where the first summation is over all the two-body bound states (with
the energy $E_{B}^{i}$) and $\delta_{l}\left(k\right)$ is the phase
shift of the $l$-th partial wave. The second virial coefficient therefore
can be determined for {\em arbitrary} interatomic interactions.
For a pedagogical explanation of the elegant Beth-Ulenbeck formalism,
we refer to the classical book by Kerson Huang \cite{HuangBook}.
This formalism has been applied by Ho and Mueller to explore the universal
properties of atomic gases near a Feshbach resonance at high temperatures
\cite{HoVE}. It has also been used extensively to study the equation
of state of nuclear and neutron matter \cite{RopkeNPA1982,HorowitzNPA2006,MekjianPRC2009,TypelPRC2010,ShenPRC2010,NatowitzPRL2010}.

For a $s$-wave Feshbach resonance that is of interest in ultracold
atoms, the general expression for the $s$-wave phase shift $\delta_{l=0}\left(k\right)$
is \cite{HuangBook,LandauBook}, 
\begin{equation}
k\cot\delta_{0}\left(k\right)=-\frac{1}{a_{s}}+\frac{1}{2}r_{0}k^{2}+\cdots,
\end{equation}
 which leads to, 
\begin{equation}
\frac{d\delta_{0}}{dk}=-\left(\frac{1}{a_{s}}+\frac{r_{0}k^{2}}{2}\right)/\left[\left(\frac{1}{a_{s}}-\frac{r_{0}k^{2}}{2}\right)^{2}+k^{2}\right].
\end{equation}
 Here for general discussion we have kept a nonzero range of interactions,
$r_{0}$. It is easy to see that the main contribution to the integral
in Eq. (\ref{BethUhlenbeck}), $I=\int\nolimits _{0}^{\infty}dk(d\delta_{0}/dk)\exp(-\lambda_{dB}^{2}k^{2}/2\pi)$,
comes from the region $k\sim1/\left|a_{s}\right|$. Thus, after introducing
a new variable $y=k\left|a_{s}\right|$, the integral becomes, 
\begin{equation}
I=-\text{sgn}(a_{s})\int\limits _{0}^{\infty}dy\frac{1+y^{2}r_{0}/\left(2a_{s}\right)}{\left[1-y^{2}r_{0}/\left(2a_{s}\right)\right]^{2}+y^{2}}\exp\left(-\frac{\lambda_{dB}^{2}}{2\pi a_{s}^{2}}y^{2}\right).\label{integral}
\end{equation}
 In the case of zero-range approximation ($r_{0}=0$), we obtain \cite{HoVE},
\begin{equation}
I^{(0)}=-\text{sgn}(a_{s})\frac{\pi}{2}\left[1-\mathop{\rm erf}\left(x\right)\right]e^{x^{2}},
\end{equation}
 where $x=\lambda_{dB}/\left(\sqrt{2\pi}\left|a_{s}\right|\right)$
and $\mathop{\rm erf}\left(x\right)$ is the error function. The correction
due to a finite range of interactions $r_{0}$ can be taken into account
by Taylor-expanding the function in Eq. (\ref{integral}) in a series
of $r_{0}/\left(2a_{s}\right)$. To the leading order, we find that,
\begin{equation}
I^{\left(1\right)}=-\frac{r_{0}\sqrt{\pi}}{4\left|a_{s}\right|}\left\{ \frac{1-2x^{2}}{x}+2\sqrt{\pi}x^{2}\left[1-\mathop{\rm erf}\left(x\right)\right]e^{x^{2}}\right\} .
\end{equation}
 Near a Feshbach resonance where $x\ll1$ and $r_{0}\ll\left|a_{s}\right|$,
we have, 
\begin{equation}
I=I^{(0)}+I^{\left(1\right)}=-\text{sgn}(a_{s})\frac{\pi}{2}+\frac{\lambda_{dB}}{\sqrt{2}a_{s}}-\frac{\pi r_{0}}{2\sqrt{2}\lambda_{dB}}+\cdots.
\end{equation}
 In terms of the small dimensionless interaction strength ($1/(k_{F}a_{s})\ll1$)
and the range of interactions ($k_{F}r_{0}\ll1$), the second virial
coefficient can be written as, 
\begin{equation}
\Delta b_{2}=\sqrt{2}e^{-\beta E_{B}}-\frac{\text{sgn}(a_{s})}{\sqrt{2}}+\frac{2}{\sqrt{\pi}}\sqrt{\frac{T_{F}}{T}}\frac{1}{k_{F}a_{s}}-\frac{1}{4\sqrt{\pi}}\sqrt{\frac{T}{T_{F}}}k_{F}r_{0},
\end{equation}
 where the single bound state exists only for a positive scattering
length with its energy $E_{B}$ depending on both $a_{s}$ and $r_{0}$.
In the unitary limit, where $1/(k_{F}a_{s})=0$, $k_{F}r_{0}=0$,
and $E_{B}=0$, we obtain the well-known result \cite{HuangBook,LandauBook},
\begin{equation}
\Delta b_{2}=\frac{1}{\sqrt{2}}.
\end{equation}

Concerning the experimental measurement, as an example, we estimate
the second virial coefficient for $^{6}$Li atoms using {\em realistic}
experimental parameters. Let us consider the negative scattering length
(BCS) side of the Feshbach resonance, for which the second virial
coefficient takes the form, 
\begin{equation}
\Delta b_{2}\left(a_{s}<0\right)=\frac{1}{\sqrt{2}}+\frac{2}{\sqrt{\pi}}\sqrt{\frac{T_{F}}{T}}\frac{1}{k_{F}a_{s}}-\frac{1}{4\sqrt{\pi}}\sqrt{\frac{T}{T_{F}}}k_{F}r_{0}.
\end{equation}
 The second and third terms on the right-hand-side of the above equation
are non-universal since both of them depend on the temperature. These
non-universal corrections are caused by a finite scattering length
or a finite range of interactions.

For $^{6}$Li atoms, the finite scattering length near the Feshbach
resonance $B_{0}\simeq834$ G can be conveniently calculated using
\cite{BartensteinPRL2005}, 
\begin{equation}
a_{s}=a_{bg}\left(1-\frac{\Delta B}{B-B_{0}}\right),
\end{equation}
where $a_{bg}\simeq-1405a_{B}$ in units of the Bohr radius $a_{B}\simeq0.529\times10^{-10}$
m, and $\Delta B\simeq300$ G. At the typical experimental density,
where $1/k_{F}\sim400$ nm, we find that $k_{F}a_{s}\simeq\pm100$,
if the magnetic field is tuned away from the resonance by one Gauss.
This leads to about a percent correction to the second virial coefficient
at the degenerate temperature $T_{F}$. On the other hand, the finite
range of interactions near the resonance can be modeled as \cite{WernerEPJB2009},
\begin{equation}
r_{0}=-2R_{*}\left(1-\frac{a_{bg}}{a_{s}}\right)^{2}+\frac{4b}{\sqrt{\pi}}-\frac{2b^{2}}{a_{s}},
\end{equation}
where $R_{*}\simeq0.0269$ nm and $b\simeq2.1$ nm is essentially
the Van der Waals length. As $R_{*}\ll b\ll\left|a_{s}\right|$ across
the Feshbach resonance, the finite range of interactions is reduced
to a constant $r_{0}\simeq4.7$ nm. Thus, we obtain the dimensionless
range of interactions $k_{F}r_{0}\sim0.012$, for the typical Fermi
wavelength. It gives about 0.1\% correction to the second virial coefficient
at $T_{F}$.

\subsubsection{Field theoretic method}

The second virial coefficient can also be conveniently calculated
using the field theoretic method \cite{VedenovJETP1959}, i.e., the
diagrammatic expansion method we mentioned earlier in Sec.IB. This
provide a simple example to illustrate the close relation between
the virial expansion and the diagrammatic expansion. In the following,
we introduce briefly the procedure. To obtain the virial coefficients,
the basic idea of field theoretic method is to calculate $\Omega^{(l)}$,
which is the contribution of $l$-particle scattering process to the
thermodynamic potential \cite{RupakPRL2007}. At large temperatures,
we expand $\Omega^{(l)}$ in fugacity, 
\begin{equation}
\Omega^{(l)}=-V\frac{2k_{B}T}{\lambda_{dB}^{3}}\sum_{n=l}^{\infty}b_{n}^{(l)}z^{n},
\end{equation}
 where $b_{n}^{(l)}$ is the $n$-th virial coefficient from $l$-particle
interaction. The total $n$-th virial coefficient $b_{n}=b_{n}^{(1)}+b_{n}^{(2)}+\cdots+b_{n}^{(n)}$,
where $b_{n}^{(1)}=\left(-1\right)^{n+1}n^{-5/2}$ is the virial coefficient
of an ideal Fermi gas. In the case of $l=2$, we have $\Delta b_{2}=b_{2}-b_{2}^{(1)}=b_{2}^{(2)}$.

We start from a path-integral functional action \cite{SadeMeloPRL1993,HLDEPL2006},
using the single-channel fermionic model with zero-range interactions
Eq. (\ref{ManyBodyHami}). By performing a Hubbard-Stratonovich transformation
to decouple the interaction term, the original fermionic partition
function ${\cal Z}=\int{\cal D}[\psi({\bf x}),\bar{\psi}({\bf x})]e^{-S}$
can be expressed as ${\cal Z}=\int{\cal D}[\Delta({\bf x}),\Delta^{*}({\bf x})]e^{-S_{eff}}$,
in terms of bosonic variables $\Delta({\bf x})$. The ``effective''
bosonic action can be written in a series expansion: $S_{eff}=\sum_{l=2}^{\infty}S_{eff}^{(l)}$.
In the normal state, the first term in the expansion reads \cite{NSR,SadeMeloPRL1993,HLDEPL2006},
\begin{equation}
S_{eff}^{(2)}=\sum_{q}\left[-\chi\left(q\right)\right]\Delta(q)\Delta^{*}(q),
\end{equation}
 where 
\begin{equation}
\chi\left(q\right)=\frac{m}{4\pi\hbar^{2}a_{s}}+\frac{1}{V}\sum_{{\bf k}}\left[\frac{f_{F}(\xi_{{\bf q}/2+{\bf k}})+f_{F}(\xi_{{\bf q}/2-{\bf k}})-1}{i\nu_{n}-\xi_{{\bf q}/2+{\bf k}}-\xi_{{\bf q}/2-{\bf k}}}-\frac{1}{2\epsilon_{{\bf k}}}\right]
\end{equation}
 is the two-particle propagator. Here we have used the abbreviation
$q=({\bf q},i\nu_{n})$, the bosonic (fermionic) Matsubara frequency
$\nu_{n}=2n\pi k_{B}T$ ($\omega_{m}=(2m+1)\pi k_{B}T$), $\xi_{{\bf k}}=\epsilon_{{\bf k}}-\mu=\hbar^{2}{\bf k}^{2}/(2m)-\mu$,
and the Fermi distribution function $f_{F}\left(x\right)=1/[\exp(x/k_{B}T)+1]$.
The action $S_{eff}^{(2)}$ accounts for the scatterings between two-particles
in the presence of other particles (i.e., medium), and thus includes
the two-body contribution to {\em all} the virial coefficients
of $b_{n}^{(2)}$ ($n\geq2$). It gives rise to the following thermodynamic
potential given by Nozières and Schmitt-Rink (NSR) in 1985 \cite{NSR},
\begin{equation}
\Omega^{(2)}=k_{B}T\sum_{{\bf q},i\nu_{n}}\ln\left[-\chi\left(q\right)\right]\exp\left(i\nu_{n}0^{+}\right)=-\frac{1}{\pi}\sum_{{\bf q}}\int_{-\infty}^{+\infty}d\Omega f_{B}\left(\Omega\right)\delta\left({\bf q},\Omega\right),\label{NSROmega}
\end{equation}
 where the summation over the Matsubara frequency has been converted
into an integral using a phase shift, 
\begin{equation}
\delta\left({\bf q},\Omega\right)=-\mathop{\rm Im}\ln\left[-\chi\left({\bf q},i\nu_{n}\rightarrow\Omega^{+}\right)\right],
\end{equation}
 and $f_{B}\left(x\right)=1/[\exp(x/k_{B}T)-1]$ is the Bose-Einstein
distribution function. The diagrammatic representation of $\Omega^{(2)}$
has been illustrated earlier in Fig. 1.

In the high-temperature limit, where the fugacity $z=\exp(\mu/k_{B}T)\ll1$
and $f_{F}\left(\xi_{{\bf k}}\right)\simeq z\exp(-\xi_{{\bf k}}/k_{B}T)$,
we may Taylor-expand the phase shift in powers of $z$. Focusing on
the unitary limit, we approximate the two-particle propagator 
\begin{equation}
\chi\left({\bf q},\Omega^{+}\right)=\chi^{(0)}\left({\bf q},\Omega^{+}\right)+z\chi^{(1)}\left({\bf q},\Omega^{+}\right)+O(z^{2}),
\end{equation}
 where 
\begin{eqnarray}
\chi^{(0)} & = & \frac{i}{4\pi}\left(\frac{m}{\hbar^{2}}\right)^{3/2}\left(\Omega^{+}-\frac{\epsilon_{{\bf q}}}{2}+2\mu\right)^{1/2},\\
\chi^{(1)} & = & \sum_{{\bf k}}\frac{\exp\left(-\xi_{{\bf q/2}+{\bf k}}/k_{B}T\right)+\exp\left(-\xi_{{\bf q}/2-{\bf k}}/k_{B}T\right)}{\Omega^{+}-\epsilon_{{\bf q}}/2-2\epsilon_{{\bf k}}+2\mu},
\end{eqnarray}
 and $\Omega^{+}\equiv\Omega+i0^{+}$. To the leading order of $\chi^{(0)}\left({\bf q},\Omega^{+}\right)$,
the phase shift is exactly a step function, 
\begin{equation}
\delta^{(0)}\left({\bf q},\Omega\right)=\frac{\pi}{2}\Theta\left(\Omega-\frac{\epsilon_{{\bf q}}}{2}+2\mu\right).
\end{equation}
 Thus, to the leading order of fugacity we have, 
\begin{equation}
\Omega^{(2)}=-\frac{1}{\pi}\sum_{{\bf q}}\int_{\epsilon_{{\bf q}}/2-2\mu}^{+\infty}d\Omega f_{B}\left(\Omega\right)\frac{\pi}{2}=\frac{k_{B}T}{2}\sum_{{\bf q}}\ln\left[1-z^{2}\exp\left(-\frac{\epsilon_{{\bf q}}}{2k_{B}T}\right)\right]=\left(-V\frac{2k_{B}T}{\lambda_{dB}^{3}}\right)\left(\frac{1}{\sqrt{2}}\right)z^{2}.
\end{equation}
 This gives rise to the second virial coefficient $b_{2}^{(2)}=\Delta b_{2}=1/\sqrt{2}$.
Away from the unitary limit, it is straightforward to show that we
can recover the Beth-Uhlenbeck formalism from Eq. (\ref{NSROmega}),
by taking the phase shift $\delta({\bf q},\Omega)$ in vacuum.

We note that the higher-order contribution of $b_{n}^{(2)}$ ($n\geq3$)
can be obtained by successively calculating the $z^{n}$ term in Eq.
(\ref{NSROmega}).

\subsection{Virial coefficients from exact few-body solutions in harmonic traps}

We now turn to calculate the third virial coefficient, by using an
entirely different method \cite{OurVELongPRA}. We solve first the
two-particle and three-particle problems in an \emph{isotropic} 3D
harmonic trap $V_{T}({\bf x})=m\omega_{T}^{2}x^{2}/2$, and then use
the solutions to obtain the second and third virial coefficients.
In the end, we discuss the possibility of calculating the fourth virial
coefficient.

We note that in cold-atom experiments the harmonic trap is often highly
anisotropic. The three-particle problem at unitaritiy in an anisotropic
trap can hardly be solved exactly. Fortunately, for a large number
of particles, for which the local density approximation is valid,
we are free to use harmonic traps of any aspect ratio to calculate
the virial coefficients, by using the universal relation Eq. (\ref{UniversalRelationBn}).

\subsubsection{Relative Hamiltonian of few-particle systems}

In a harmonic trap, it is useful to separate the center-of-mass motion
and relative motion. We thus define the following center-of-mass coordinate
${\bf R}$ and relative coordinates ${\bf r}_{i}$ ($i\geq2$) for
$N$ fermions in a harmonic trap \cite{WernerPRA2006,WernerPhDThesis},
\begin{equation}
{\bf R}=\left({\bf x}_{1}+\cdots+{\bf x}_{N}\right)/N,
\end{equation}
 and 
\begin{equation}
{\bf r}_{i}=\sqrt{\frac{i-1}{i}}\left({\bf x}_{i}-\frac{1}{i-1}\sum_{k=1}^{i-1}{\bf x}_{k}\right),
\end{equation}
 respectively. In this Jacobi coordinate, the Hamiltonian of the non-interacting
Schrödinger equation takes the form ${\cal H}_{0}={\cal H}_{cm}+{\cal H}_{rel}$,
where, 
\begin{equation}
{\cal H}_{cm}=-\frac{\hbar^{2}}{2M}{\bf \nabla}_{{\bf R}}^{2}+\frac{1}{2}M\omega_{T}^{2}R^{2},
\end{equation}
 and 
\begin{equation}
{\cal H}_{rel}=\sum_{i=2}^{N}\left[-\frac{\hbar^{2}}{2m}{\bf \nabla}_{{\bf r}_{i}}^{2}+\frac{1}{2}m\omega_{T}^{2}r_{i}^{2}\right].
\end{equation}
 The center-of-mass motion is simply that of a harmonically trapped
particle of mass $M=Nm$, with well-known wave functions and spectrum
$E_{cm}=(n_{cm}+3/2)\hbar\omega_{T}$, where $n_{cm}=0,1,2...$ is
a non-negative integer. In the presence of interactions, the relative
Hamiltonian should be solved in conjunction with the Bethe-Peierls
boundary condition, Eq. (\ref{BP}).

\subsubsection{Two fermions in a 3D harmonic trap}

Let us consider the two-fermion problem in a harmonic trap, where
the relative Schrödinger equation becomes 
\begin{equation}
\left[-\frac{\hbar^{2}}{2\mu}{\bf \nabla}_{{\bf r}}^{2}+\frac{1}{2}\mu\omega_{T}^{2}r^{2}\right]\psi_{2b}^{rel}({\bf r})=E_{rel}\psi_{2b}^{rel}({\bf r}),\label{hamiRel2e}
\end{equation}
 where two fermions with unlike spins do not stay at the same position
($r>0$). Here, we have re-defined ${\bf r}=\sqrt{2}{\bf r}_{2}$
and without confusing with the chemical potential we have used a reduced
mass $\mu=m/2$. It is clear that only the $l=0$ subspace of the
relative wave function is affected by the $s$-wave contact interaction.
According to the Bethe-Peierls boundary condition, as $r\rightarrow0$
the relative wave function should take the form, $\psi_{2b}^{rel}(r)\rightarrow(1/r-1/a_{s})$,
or satisfy, $\partial\left(r\psi_{2b}^{rel}\right)/\partial r=-\left(r\psi_{2b}^{rel}\right)/a_{s}$.
The two-fermion problem in a harmonic trap was first solved by Busch
and coworkers \cite{BuschFP1998}. In the following, we present a
simple physical interpretation of the solution.

The key point is that, regardless of the boundary condition, there
are {\em two} types of general solutions of the relative Schrödinger
equation (\ref{hamiRel2e}) in the $l=0$ subspace, $\psi_{2b}^{rel}(r)\propto\exp(-r^{2}/2d^{2})f(r/d)$.
Here the function $f(x)$ can either be the first kind of Kummer confluent
hypergeometric function $_{1}F_{1}$ or the second kind of Kummer
confluent hypergeometric function $U$. We have taken $d=\sqrt{\hbar/(\mu\omega_{T})}$
as the characteristic length scale of the trap. In the absence of
interactions, the first Kummer function gives rise to the standard
wave function of 3D harmonic oscillators. With interactions, however,
we have to choose the second Kummer function $U$, since it diverges
as $1/r$ at origin and thus satisfies the Bethe-Peierls boundary
condition.

\begin{figure}[htp]
\begin{centering}
\includegraphics[clip,width=0.5\textwidth]{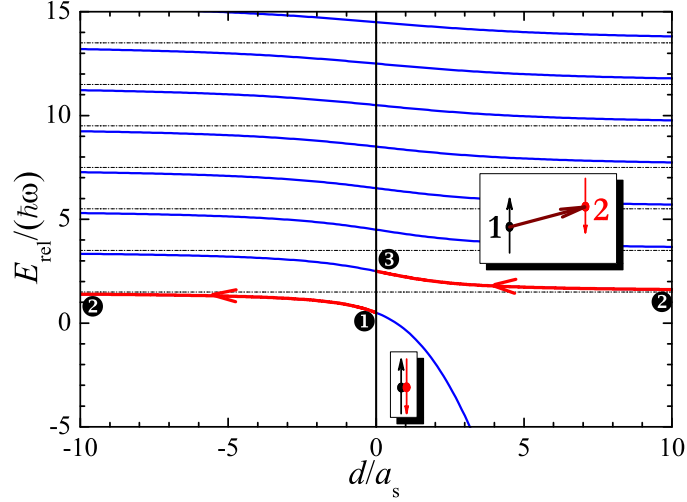} 
\par\end{centering}

\caption{(color online) Energy spectrum of the relative motion of a trapped
two-fermion system near a Feshbach resonance (i.e, $d/a_{s}=0$, where
$d$ is the characteristic harmonic oscillator length). For a positive
scattering length $a_{s}>0$ in the right part of the figure, the
ground state is a molecule with size $a_{s}$, whose energy diverges
as $E_{rel}\simeq-\hbar^{2}/(ma_{s}^{2})$. The excited states or
the upper branch of the resonance may be viewed as the Hilbert space
of a ``repulsive'' Fermi gas with the same scattering length $a_{s}$.
In this two-body picture, the level from the point 2 to 3 is the ground
state energy level of the repulsive two-fermion sub-space, whose energy
initially increases linearly with increasing $a_{s}$ from $1.5\hbar\omega_{T}$
at the point 2 and finally saturates towards $2.5\hbar\omega_{T}$
at the resonance point 3. For comparison, we illustrate as well the
ground state energy level in the case of a negative scattering length
and show how the energy increases with increased scattering length
from point 1 to 2. From ref. \cite{OurVELongPRA}; copyright (2010)
by APS.}

\label{fig:2eSpectrum} 
\end{figure}

Therefore, the (un-normalized) relative wave function and relative
energy should be rewritten as, 
\begin{equation}
\psi_{2b}^{rel}(r;\nu)=\Gamma(-\nu)U(-\nu,\frac{3}{2},\frac{r^{2}}{d^{2}})\exp(-\frac{r^{2}}{2d^{2}}),
\end{equation}
 and 
\begin{equation}
E_{rel}=(2\nu+\frac{3}{2})\hbar\omega_{T},\label{spectrumRel2e}
\end{equation}
 respectively. Here, $\Gamma$ is the Gamma function, the real number
$\nu$ plays the role of a quantum number and should be determined
by the boundary condition, $\lim_{r\rightarrow0}\partial\left(r\psi_{2b}^{rel}\right)/\partial r=-\left(r\psi_{2b}^{rel}\right)/a$.
By examining the short range behavior of the second Kummer function
$U(-\nu,3/2,x)$, this leads to the familiar equation for energy levels
\cite{BuschFP1998}, 
\begin{equation}
\frac{2\Gamma(-\nu)}{\Gamma(-\nu-1/2)}=\frac{d}{a_{s}}.\label{seRel2e}
\end{equation}
 In Fig. \ref{fig:2eSpectrum}, we give the resulting energy spectrum
as a function of the dimensionless interaction strength $d/a_{s}$.

The spectrum is easy to understand. At infinitely small scattering
length $a_{s}\rightarrow0^{-}$, $\nu(a_{s}=0^{-})=n_{rel}$ ($n_{rel}=0,1,2...$),
which recovers the spectrum in the non-interacting limit. With increasingly
attractive interactions, the energies decrease. In the unitarity (resonance)
limit where the scattering length diverges, $a_{s}\rightarrow\pm\infty$,
we find that $\nu(a=\pm\infty)=n_{rel}-1/2$. As the attraction increases
further, the scattering length becomes positive and decreases in magnitude.
We then observe two distinct types of behavior: the ground state is
a {\em molecule} of size $a$, whose energy diverges asymptotically
as $-\hbar^{2}/ma_{s}^{2}$ as $a_{s}\rightarrow0^{+}$, while the
excited states may be viewed as two {\em repulsively} interacting
fermions with the same scattering length $a_{s}$. Their energies
decrease to the non-interacting values as $a_{s}\rightarrow0^{+}$.

In this two-body picture, a universal {\em repulsively} interacting
Fermi gas with zero-range interaction potentials may be realized on
the positive scattering length side of a Feshbach resonance for an
attractive interaction potential, provided that all two fermions with
unlike spins occupy the exited states or the upper branch of the two-body
energy spectrum.

\begin{figure}[htp]
\begin{centering}
\includegraphics[clip,width=0.4\textwidth]{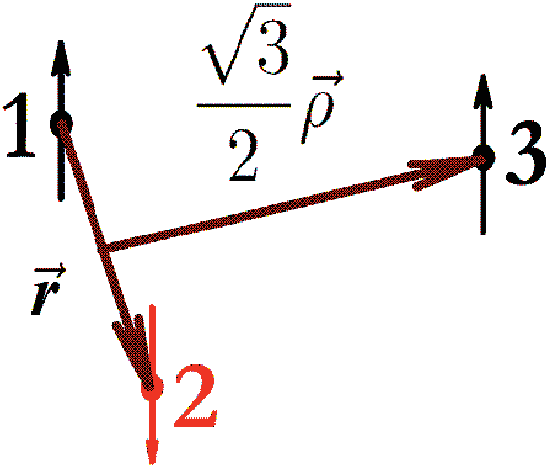} 
\par\end{centering}

\caption{(color online) Configuration of three interacting fermions, two spin-up
and one spin-down. From ref. \cite{OurVELongPRA}; copyright (2010)
by APS.}

\label{fig:3eConfiguration} 
\end{figure}

\subsubsection{Three fermions in a 3D harmonic trap: General exact solutions}

Let us turn to the three fermion case by considering two spin-up fermions
and one spin-down fermion, i.e., the $\uparrow\downarrow\uparrow$
configuration shown in Fig. \ref{fig:3eConfiguration}. The relative
Hamiltonian can be written as \cite{WernerPRA2006,WernerPhDThesis},
\begin{equation}
{\cal H}_{rel}=\frac{\hbar^{2}}{2\mu}{\bf \left(\nabla_{r}^{2}+\nabla_{\rho}^{2}\right)}+\frac{1}{2}\mu\omega_{T}^{2}\left(r^{2}+\rho{\bf ^{2}}\right),\label{hamiRel3e}
\end{equation}
 where we have redefined the Jacobi coordinates ${\bf r}=\sqrt{2}{\bf r_{2}}$
and ${\bf \rho}=\sqrt{2}{\bf r_{3}}$, which measure the distance
between the particle 1 and 2 (i.e., pair), and the distance from the
particle 3 to the center-of-mass of the pair, respectively.

Inspired by the two-fermion solution, it is readily seen that the
relative wave function of the Hamiltonian (\ref{hamiRel3e}) may be
expanded into products of two Kummer confluent hypergeometric functions.
Intuitively, we may write down the following ansatz \cite{OurVE},
\begin{equation}
\psi_{3b}^{rel}\left({\bf r},{\bf \rho}\right)=\left(1-{\cal P}_{13}\right)\chi\left({\bf r},{\bf \rho}\right),\label{wfRel3e}
\end{equation}
 where, 
\begin{equation}
\chi\left({\bf r},{\bf \rho}\right)=\sum\limits _{n}a_{n}\psi_{2b}^{rel}(r;\nu_{l,n})R_{nl}\left(\rho\right)Y_{l}^{m}\left(\hat{\rho}\right).
\end{equation}
 The two-body relative wave function $\psi_{2b}^{rel}(r;\nu_{l,n})$
with energy $(2\nu_{l,n}+3/2)\hbar\omega_{T}$ describes the motion
of the paired particles 1 and 2, and the wave function $R_{nl}\left(\rho\right)Y_{l}^{m}\left(\hat{\rho}\right)$
with energy $(2n+l+3/2)\hbar\omega_{T}$ gives the motion of particle
3 relative to the pair. Here, $R_{nl}\left(\rho\right)$ is the standard
radial wave function of a 3D harmonic oscillator and $Y_{l}^{m}\left(\hat{\rho}\right)$
is the spherical harmonic. Owing to the rotational symmetry of the
relative Hamiltonian (\ref{hamiRel3e}), it is easy to see that the
relative angular momenta $l$ and $m$ are good quantum numbers. The
value of $\nu_{l,n}$ is uniquely determined from energy conservation,
\begin{equation}
E_{rel}=\left[(2\nu_{l,n}+3/2)+(2n+l+3/2)\right]\hbar\omega_{T},
\end{equation}
 for a given relative energy $E_{rel}$. It varies with the index
{}$n$ at a given angular momentum $l$. Finally, ${\cal P}_{13}$
is an exchange operator for particles 1 and 3, which ensures the correct
exchange symmetry of the relative wave function due to Fermi exclusion
principle, i.e., ${\cal P}_{13}\chi\left({\bf r},{\bf \rho}\right)=\chi\left({\bf r}/2+\sqrt{3}{\bf \rho}/2,{\bf \sqrt{3}r}/2-{\bf \rho}/2\right)$.
The relative energy $E_{rel}$ together with the expansion coefficient
$a_{n}$ should be determined by the Bethe-Peierls boundary condition,
i.e., $\lim_{r\rightarrow0}[\partial r\psi_{3b}^{rel}\left({\bf r},{\bf \rho}\right)]/\partial r=-[r\psi_{3b}^{rel}\left({\bf r},{\bf \rho}\right)]/a_{s}$.
We note that the second Bethe-Peierls boundary condition in case of
particle 2 approaching particle 3 is satisfied automatically due to
the exchange operator acting on the relative wave function.

By writing $\chi\left({\bf r},{\bf \rho}\right)=\phi(r,\rho)Y_{l}^{m}\left(\hat{\rho}\right)$,
the Bethe-Peierls boundary condition takes the form ($r\rightarrow0$),
\begin{equation}
-\frac{1}{a_{s}}\left[r\phi(r,\rho)\right]=\frac{\partial\left[r\phi(r,\rho)\right]}{\partial r}-\left(-1\right)^{l}\phi(\frac{\sqrt{3}\rho}{2},\frac{\rho}{2}).\label{BP3e}
\end{equation}
 Using the asymptotic behavior of the second kind of Kummer function,
$\lim_{x\rightarrow0}\Gamma\left(-\nu_{l,n}\right)U(-\nu_{l,n},3/2,x^{2})=\sqrt{\pi}/x-2\sqrt{\pi}\Gamma\left(-\nu_{l,n}\right)/\Gamma\left(-\nu_{l,n}-1/2\right)$,
it is easy to show that in the limit of $r\rightarrow0$, 
\begin{equation}
-\frac{1}{a_{s}}\left[r\phi(r,\rho)\right]=-\frac{\sqrt{\pi}}{a_{s}}\sum\limits _{n}a_{n}R_{nl}\left(\rho\right),\label{BP3e1}
\end{equation}
 and 
\begin{equation}
\frac{\partial\left[r\phi(r,\rho)\right]}{\partial r}=-\sqrt{\pi}\sum\limits _{n}a_{n}R_{nl}\left(\rho\right)\frac{2\Gamma\left(-\nu_{l,n}\right)}{\Gamma\left(-\nu_{l,n}-1/2\right)}.\label{BP3e2}
\end{equation}

Thus, the Bethe-Peierls boundary condition becomes, 
\begin{equation}
\sum\limits _{n}a_{n}\left[B_{n}R_{nl}\left(\rho\right)-R_{nl}\left(\frac{\rho}{2}\right)\psi_{2b}^{rel}(\frac{\sqrt{3}\rho}{2};\nu_{l,n})\right]=0,
\end{equation}
 where 
\begin{equation}
B_{n}=\left(-1\right)^{l}\sqrt{\pi}\left[\frac{d}{a_{s}}-\frac{2\Gamma\left(-\nu_{l,n}\right)}{\Gamma\left(-\nu_{l,n}-1/2\right)}\right].
\end{equation}
 Projecting onto the orthogonal and complete set of basis functions
$R_{nl}\left(\rho\right)$, we find that a secular equation, 
\begin{equation}
\frac{2\Gamma(-\nu_{l,n})}{\Gamma(-\nu_{l,n}-1/2)}a_{n}+\frac{(-1)^{l}}{\sqrt{\pi}}\sum\limits _{n^{\prime}}C_{nn^{\prime}}a_{n^{\prime}}=\left(\frac{d}{a_{s}}\right)a_{n},\label{seRel3e}
\end{equation}
 where we have defined the matrix coefficient, 
\begin{equation}
C_{nn^{\prime}}\equiv\int\limits _{0}^{\infty}\rho^{2}d\rho R_{nl}\left(\rho\right)R_{n^{\prime}l}\left(\frac{\rho}{2}\right)\psi_{2b}^{rel}(\frac{\sqrt{3}\rho}{2};\nu_{l,n^{\prime}}),
\end{equation}
 which arises from the exchange effect due to the operator ${\cal P}_{13}$.
In the absence of $C_{nn^{\prime}}$, the above secular equation describes
a three-fermion problem of a pair and a single particle, {\em un-correlated}
to each other. It then simply reduces to Eq. (\ref{seRel2e}), as
expected.

The secular equation (\ref{seRel3e}) was first obtained by Kestner
and Duan by solving the three-particle scattering problem using Green
function \cite{KestnerPRA2007}. To solve it, for a given scattering
length we may try different values of relative energy $E_{rel}$,
implicit via $\nu_{l,n}$. However, it turns out to be more convenient
to diagonalize the matrix ${\bf A}=\{A_{nn^{\prime}}\}$ for a given
relative energy, where 
\begin{equation}
A_{nn^{\prime}}=\frac{2\Gamma(-\nu_{l,n})}{\Gamma(-\nu_{l,n}-1/2)}\delta_{nn^{\prime}}+\frac{(-1)^{l}}{\sqrt{\pi}}C_{nn^{\prime}}.\label{amat}
\end{equation}
 The eigenvalues of the matrix ${\bf A}$ then gives all the possible
values of $d/a_{s}$ for a particular relative energy. We finally
invert $a(E_{rel})$ to obtain the relative energy as a function of
the scattering length. Numerically, we find that the matrix ${\bf A}$
is symmetric and thus the standard diagonalization algorithm can be
used. We outline the details of the numerical calculation of Eq. (\ref{amat})
in the Appendix A.

\subsubsection{Three fermions in a 3D harmonic trap: Exact solutions in the unitarity
limit}

In the unitarity limit with infinitely large scattering length, $a_{s}\rightarrow\infty$,
we may obtain more physical solutions using hyperspherical coordinates,
as shown by Werner and Castin \cite{WernerPRL2006,WernerPhDThesis}.
By defining a hyperradius $R=\sqrt{(r^{2}+\rho^{2})/2}$ and hyperangles
$\vec{\Omega}=(\alpha,\hat{r},\hat{\rho})$, where $\alpha=\arctan(r/\rho)$
and $\hat{r}$ and $\hat{\rho}$ are respectively the unit vector
along ${\bf r}$ and ${\bf \rho}$, we may write \cite{WernerPRL2006,WernerPhDThesis},
\begin{equation}
\psi_{3b}^{rel}\left(R,\vec{\Omega}\right)=\frac{F\left(R\right)}{R}\left(1-{\cal P}_{13}\right)\frac{\varphi\left(\alpha\right)}{\sin\left(2\alpha\right)}Y_{l}^{m}\left(\hat{\rho}\right),\label{wfRel3eHyper}
\end{equation}
 to decouple the motion in the hyperradius and hyperangles for given
relative angular momenta $l$ and $m$ . It leads to the following
decoupled Schrödinger equations \cite{WernerPhDThesis}, 
\begin{equation}
-F^{\prime\prime}-\frac{1}{R}F^{\prime}+\left(\frac{s_{l,n}^{2}}{R^{2}}+\omega_{T}^{2}R^{2}\right)F=2E_{rel}F,\label{hyperradiusEq}
\end{equation}
 and 
\begin{equation}
-\varphi^{\prime\prime}\left(\alpha\right)+\frac{l\left(l+1\right)}{\cos^{2}\alpha}\varphi\left(\alpha\right)=s_{l,n}^{2}\varphi\left(\alpha\right),\label{hyperangleEq}
\end{equation}
 where $s_{l,n}^{2}$ is the eigenvalue for the $n$-th wave function
of the hyperangle equation.

For three-fermions, $s_{l,n}^{2}$ is always positive. Therefore,
the hyperradius equation (\ref{hyperradiusEq}) can be interpreted
as a Schrödinger equation for a fictitious particle of mass unity
moving in two dimensions in an effective potential $(s_{l,n}^{2}/R^{2}+\omega_{T}^{2}R^{2})$
with a bounded wave function $F(R)$. The resulting spectrum is \cite{WernerPRL2006,WernerPhDThesis}
\begin{equation}
E_{rel}=\left(2q+s_{l,n}+1\right)\hbar\omega_{T},\label{energyRel3eHyper}
\end{equation}
 where the good quantum number $q$ labels the number of nodes in
the hyperradius wave function.

The eigenvalue $s_{l,n}$ should be determined by the Bethe-Peierls
boundary condition, which in hyperspherical coordinates takes the
from \cite{WernerPRL2006,WernerPhDThesis}, 
\begin{equation}
\varphi^{\prime}\left(0\right)-(-1)^{l}\frac{4}{\sqrt{3}}\varphi\left(\frac{\pi}{3}\right)=0.\label{BP3eHyper}
\end{equation}
 In addition, we need to impose the boundary condition $\varphi\left(\pi/2\right)=0$,
since the relative wave function (\ref{wfRel3eHyper}) should not
be singular at $\alpha=\pi/2$. The general solution of the hyperangle
equation (\ref{hyperangleEq}) satisfying $\varphi\left(\pi/2\right)=0$
is given by, 
\begin{equation}
\varphi\propto x^{l+1}{}_{2}F_{1}\left(\frac{l+1-s_{l,n}}{2},\frac{l+1+s_{l,n}}{2},l+\frac{3}{2};x^{2}\right),\label{wfHyperangleEq}
\end{equation}
 where $x=\cos(\alpha)$ and $_{2}F_{1}$ is the hypergeometric function.
In the absence of interactions, the Bethe-Peierls boundary condition
(\ref{BP3eHyper}) should be replaced by $\varphi\left(0\right)=0$,
since the relative wave function (\ref{wfRel3eHyper}) should not
be singular at $\alpha=0$ either. As $\varphi\left(0\right)=\Gamma(l+3/2)\Gamma(1/2)/[\Gamma((l+2+s_{l,n})/2)\Gamma((l+2-s_{l,n})/2)]$,
this boundary condition leads to $[l+2-s_{l,n}^{(1)}]/2=-n$, or $s_{l,n}^{(1)}=2n+l+2$,
where $n=0,1,2,...$ is a non-negative integer and we have used the
superscript ``$1$'' to denote a non-interacting system. However,
a spurious solution occurs when $l=0$ and $n=0$, for which $s_{l,n}^{(1)}=2$,
$\varphi(\alpha)=\sin(2\alpha)/2$ and thus, the symmetry operator
$(1-{\cal P}_{13})$ gives a vanishing relative wave function in Eq.
(\ref{wfRel3eHyper}) that should be discarded \cite{WernerPhDThesis}.
We conclude that for three non-interacting fermions, 
\begin{equation}
s_{l,n}^{(1)}=\left\{ \begin{array}{ll}
2n+4, & l=0\\
2n+l+2, & l>0
\end{array}\right..\label{non_interacting_sln}
\end{equation}
 For three interacting fermions, we need to determine $s_{l,n}$ by
substituting the general solution (\ref{wfHyperangleEq}) into the
Bethe-Peierls boundary condition (\ref{BP3eHyper}). In the Appendix
B, we describe how to accurately calculate $s_{l,n}$. In the boundary
condition Eq. (\ref{BP3eHyper}), the leading effect of interactions
is carried by $\varphi^{\prime}\left(0\right)$ and therefore, $\varphi^{\prime}\left(0\right)=0$
determines the asymptotic values of $s_{l,n}$ at large momentum $l$
or $n$. This gives rise to $(l+1-\bar{s}_{l,n})/2=-n$, or, 
\begin{equation}
\bar{s}_{l,n}=\left\{ \begin{array}{ll}
2n+3, & l=0\\
2n+l+1, & l>0
\end{array}\right.,\label{asymptotic_sln}
\end{equation}
 where we have used a bar to indicate the asymptotic results. By comparing
Eqs. (\ref{non_interacting_sln}) and (\ref{asymptotic_sln}), asymptotically
the attractive interaction will reduce $s_{l,n}$ by a unity.

\subsubsection{Three fermions in a 3D harmonic trap: Energy spectrum}

We can numerically solve both the general exact solution (\ref{wfRel3e})
along the BEC-BCS crossover and the exact solution (\ref{wfRel3eHyper})
in the unitarity limit. In the latter unitary case, the accuracy of
results can be improved to arbitrary precision by using suitable mathematical
software, described in Appendix B. Fig. \ref{fig:3eSpectrum} reports
the energy spectrum of three interacting fermions with increasingly
attractive interaction strength at the ground state angular momentum,
$l=1$. For a given scattering length, we typically calculate several
ten thousand energy levels (i.e., $E_{rel}<(l+256)\hbar\omega_{T}$)
in different subspace. To construct the matrix ${\bf A}$, Eq. (\ref{amat}),
we have kept a maximum value of $n_{\max}=128$ in the functions $R_{nl}\left(\rho\right)$.
Using the accurate spectrum in the unitarity limit as a benchmark,
we estimate that the typical relative numerical error of energy levels
is less than $10^{-6}$. We have found a number of nontrivial features
in the energy spectrum.

\begin{figure}[htp]
\begin{centering}
\includegraphics[clip,width=0.5\textwidth]{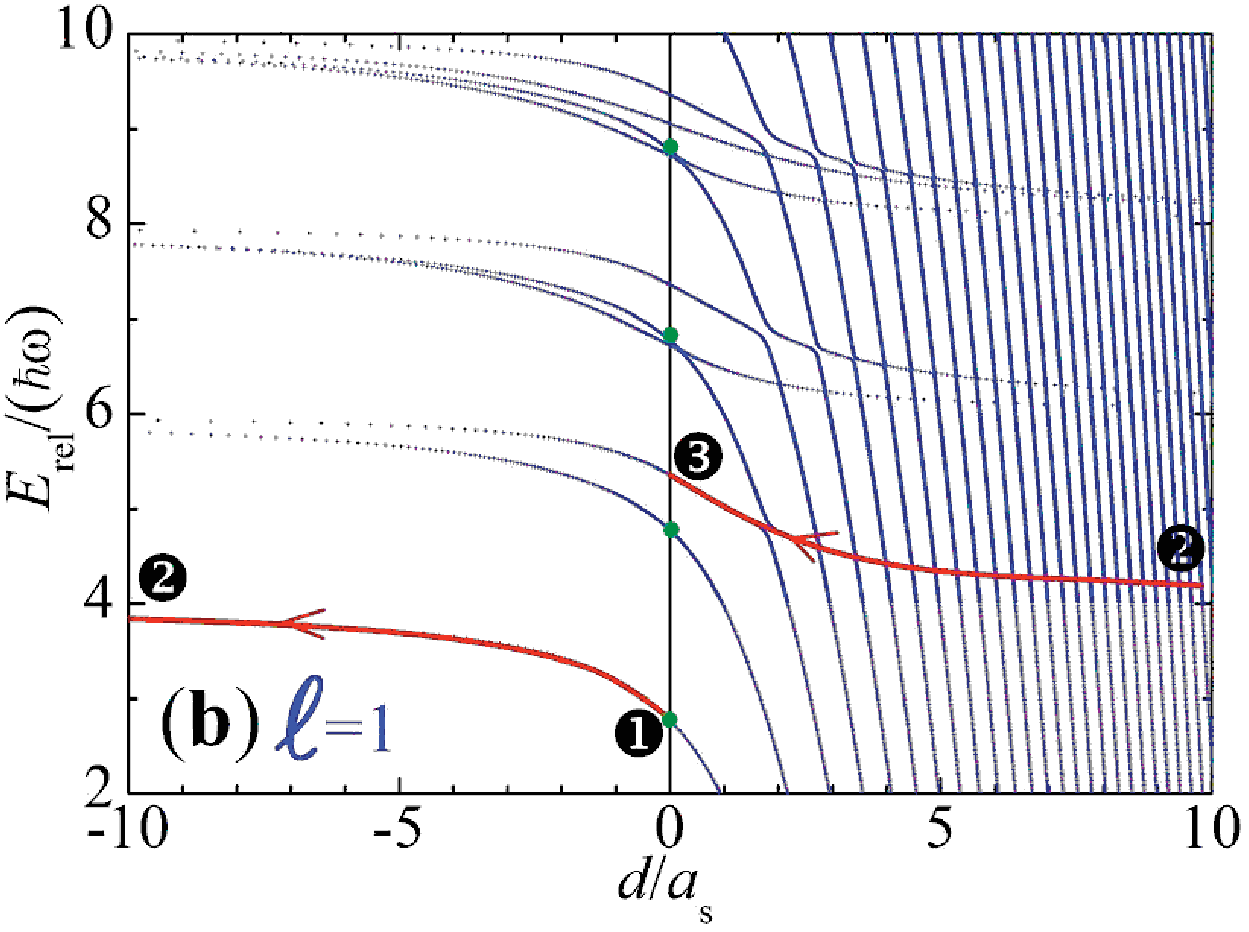} 
\par\end{centering}

\caption{(color online) Relative energy spectrum of three interacting fermions
at the ground state subspace $l=1$. On the positive scattering length
(BEC) side of the resonance, there are two types of energy levels:
one (is vertical and) diverges with decreasing the scattering length
$a_{s}$ and the other (is horizontal) converges to the non-interacting
spectrum. The latter may be viewed as the energy spectrum of three
repulsively interacting fermions. In analogy with the two-fermion
case, we show the ground state energy level of the repulsive three-fermion
system (i.e, from point 2 to 3), as well as the ground state energy
level of the attractive three-fermion system for $a_{s}<0$ (i.e.,
from the point 1 to 2). In the unitarity limit, we show by the circles
the energy levels that should be excluded when we identify the energy
spectrum for infinitely large repulsive interactions. Adapted from
ref. \cite{OurVELongPRA}; copyright (2010) by APS.}

\label{fig:3eSpectrum} 
\end{figure}

The spectrum on the BCS side is relatively simple. It can be understood
as a non-interacting spectrum at $d/a_{s}\rightarrow-\infty$, in
which $E_{rel}=(2Q+3)\hbar\omega_{T}$ at $l=0$ and $E_{rel}=(2Q+l+1)\hbar\omega_{T}$
at $l\geq1$, with a positive integer $Q=1,2,3,...$ that denotes
also the degeneracy of the energy levels. The attractive interactions
reduce the energies and at the same time lift the degeneracy. Above
the resonance or unitary point of $d/a_{s}=0$, however, the spectrum
becomes much more complicated.

There are a group of nearly {\em vertical} energy levels that diverge
towards the BEC limit of $d/a_{s}\rightarrow+\infty$. From the two-body
relative energy spectrum in Fig. \ref{fig:2eSpectrum}, we may identify
these as energy states containing a molecule of size $a_{s}$ and
a fermion. For a given scattering length, these nearly vertical energy
level differ by about $2\hbar\omega_{T}$, resulting from the motion
of the fermion relative to the molecule. In addition to the nearly
vertical energy levels, most interestingly, we observe also some nearly
{\em horizontal} energy levels, which converge to the non-interacting
spectrum in the BEC limit. In analogy with the two-body case, we may
identify these horizontal levels as the energy spectrum of three {\em
repulsively} interacting fermions. We show explicitly in the figure
the ground state level of three repulsively interacting fermions,
which increases in energy from the point 2 to 3 with increasing scattering
length from $a_{s}=0^{+}$ to $a_{s}=+\infty$. For comparison, we
also show the ground state level of three attractively interacting
fermions at a negative scattering length, which decreases in energy
from the point 2 to 1 with increasing absolute value of $a_{s}$.

This identification of energy spectrum for repulsive interactions,
however, is not as rigorous as in the two-body case. There are many
apparent avoided crossings between the vertical and horizontal energy
levels. Therefore, by changing a positive scattering length from the
BEC limit to the unitarity limit, three fermions initially at the
horizontal level may finally transition into a vertical level, provided
that the sweep of scattering length is sufficiently slow and adiabatic.
This leads to the conversion of fermionic atoms to bosonic molecules.
A detailed analysis of the loss rate of fermionic atoms as a function
of sweep rate may be straightforward obtained by applying the Landau-Zener
tunnelling model.

Let us now focus on the resonance case of most significant interest.
In Fig. \ref{fig:3eSpectrum}, we show explicitly by green dots the
vertical energy levels in the unitarity limit. These levels should
be excluded if we are interested in the spectrum of repulsively interacting
fermions. Amazingly, for each given angular momentum, these energy
levels form a regular ladder with an exact energy spacing $2\hbar\omega_{T}$
\cite{WernerPRA2006}. Using the exact solution in the unitarity limit,
Eq. (\ref{energyRel3eHyper}), we may identify unambiguously that
the energy ladder is given by, 
\begin{equation}
E_{rel}=\left(2q+s_{l,0}+1\right)\hbar\omega_{T}.\label{energyRel3eLowest}
\end{equation}
 Therefore, in the unitarity limit the lowest-order solution of the
hyperangle equation gives rise to the relative wave function of a
molecule and a fermion. Thus, it should be discarded when considering
three resonantly interacting fermions with an effective repulsive
interaction.

\subsubsection{Second virial coefficient}

We now calculate the virial coefficients of a trapped {\em attractively}
interacting Fermi gas. In a harmonic trap, the oscillator length $d$
provides a large length scale, compared to the thermal wavelength
$\lambda_{dB}$. Alternatively, we may use $\tilde{\omega}_{T}=\hbar\omega_{T}/(k_{B}T)\ll1$
to characterize the intrinsic length scale relative to the trap. All
the virial coefficients and cluster partition functions in harmonic
traps therefore depend on the small parameter $\tilde{\omega}_{T}$.
We shall be interested in a universal regime with vanishing $\tilde{\omega}_{T}$,
in accord with the large number of atoms in a real experiment.

To obtain $\Delta b_{2,T}$, we consider separately $\Delta Q_{2,T}$
and $Q_{1,T}$. The single-particle partition function $Q_{1,T}$
is determined by the single-particle spectrum of a 3D harmonic oscillator,
$E_{nl}=(2n+l+3/2)\hbar\omega_{T}$. We find that $Q_{1,T}=2/[\exp(+\tilde{\omega}_{T}/2)-\exp(-\tilde{\omega}_{T}/2)]^{3}\simeq2\left(k_{B}T\right)^{3}/\left(\hbar\omega_{T}\right)^{3}$,
in agreement with the previous result based on the local density approximation
(see Eq. \ref{idealOmegaTrap}). The pre-factor of two accounts for
the two possible spin states of a single fermion. In the calculation
of $\Delta Q_{2,T}$, it is easy to see that the summation over the
center-of-mass energy gives exactly $Q_{1,T}/2$. Using Eq. (\ref{spectrumRel2e}),
we find that, 
\begin{equation}
\Delta b_{2,T}=\frac{1}{2}\sum_{\nu_{n}}\left[e^{-\left(2\nu_{n}+3/2\right)\tilde{\omega}_{T}}-e^{-\left(2\nu_{n}^{\left(1\right)}+3/2\right)\tilde{\omega}_{T}}\right],\label{db2}
\end{equation}
 where the non-interacting $\nu_{n}^{\left(1\right)}=n$ ($n=0,1,2,...$).

At resonance with an infinitely large scattering length, the spectrum
is known exactly: $\nu_{n}=n-1/2$, giving rise to, 
\begin{equation}
\Delta b_{2,T}=\frac{1}{2}\frac{\exp\left(-\tilde{\omega}_{T}/2\right)}{\left[1+\exp\left(-\tilde{\omega}_{T}\right)\right]}=+\frac{1}{4}-\frac{1}{32}\tilde{\omega}_{T}^{2}+\cdots.\label{db2attTrap}
\end{equation}
 The term $\tilde{\omega}_{T}^{2}$ in Eq. (\ref{db2attTrap}) is
{\em nonuniversal} and is negligibly small for a cloud with a large
number of atoms. We therefore obtain the universal second virial coefficient:
$\Delta b_{2,T}=1/4$, which are temperature independent.

\begin{figure}[htp]
\begin{centering}
\includegraphics[clip,width=0.5\textwidth]{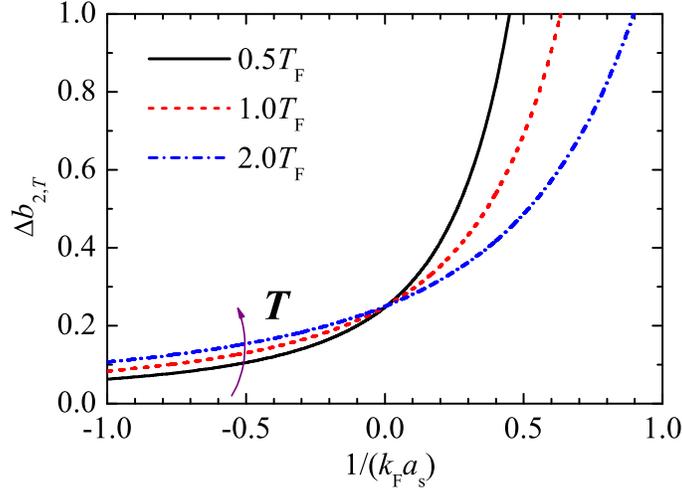} 
\par\end{centering}

\caption{(color online) Second virial coefficient of a trapped attractive Fermi
gas as a function of the interaction parameter $1/(k_{F}a_{s})$.
We have used a total number of atoms $N=100$, leading to $\tilde{\omega}_{T}=(3N)^{-1/3}\approx0.15$
at $T=T_{F}$. Adapted from ref. \cite{OurVE}; copyright (2009) by
APS.}

\label{fig:trappedB2} 
\end{figure}

In Fig. \ref{fig:trappedB2}, we show the second virial coefficient
through the BEC-BCS crossover at three typical temperatures. Here
we consider a gas with $N=100$ atoms and scale the inverse scattering
length using the Fermi vector at the trap center, $k_{F}=(24N)^{1/6}/(d/\sqrt{2})$.
The temperature is given in units of Fermi temperature $T_{F}=E_{F}/k_{B}=(3N)^{1/3}(\hbar\omega_{T}/k_{B})$.
All the curves with distinct temperatures cross at $a_{s}\rightarrow\pm\infty$.
This is the manifestation of universal behavior anticipated if there
is no any intrinsic length scale. However, the characteristic length
scale $d$ of harmonic traps brings a small (non-universal) temperature
dependence that decreases as $N^{-2/3}$, shown by the terms $\tilde{\omega}_{T}^{2}$
in Eq. (\ref{db2attTrap}).

According to the universal relation between trapped and homogeneous
virial coefficients, Eq. (\ref{UniversalRelationBn}), we obtain immediately
the homogeneous second virial coefficient in the unitarity limit,
$\Delta b_{2}=1/\sqrt{2}$, which is in agreement with the result
obtained from the Beth-Uhlenbeck formalism and from the field theoretic
calculation.

\subsubsection{Third virial coefficient}

The calculation of the third virial coefficient, which is given by
$\Delta b_{3,T}=\Delta Q_{3,T}/Q_{1,T}-\Delta Q_{2,T}$, is more complicated.
Either the term $\Delta Q_{3,T}/Q_{1,T}$ or $\Delta Q_{2,T}$ diverges
as $\tilde{\omega}_{T}\rightarrow0$, but the leading divergences
cancel with each other. In the numerical calculation, we have to carefully
separate the leading divergent term and calculate them analytically.
It is readily seen that the spin states of $\uparrow\downarrow\uparrow$
and $\downarrow\uparrow\downarrow$ configurations contribute equally
to $Q_{3,T}$. The term $Q_{1,T}$ in the denominators is canceled
exactly by the summation over the center-of-mass energy. We thus have
\begin{equation}
\Delta Q_{3,T}/Q_{1,T}=\sum\exp(-E_{rel}/k_{B}T)-\sum\exp(-E_{rel}^{(1)}/k_{B}T)\,.
\end{equation}

To proceed, it is important to analyze analytically the behavior of
$E_{rel}$ at high energies. For this purpose, we introduce a relative
energy $\bar{E}_{rel}$, which is the solution of Eq. (\ref{amat})
in the absence of the exchange term $C_{nm}$, and can be constructed
directly from the two-body relative energy. In the subspace with a
total relative momentum $l$, it takes the form, 
\begin{equation}
\bar{E}_{rel}=\left(2n+l+3/2\right)\hbar\omega_{T}+(2\nu+3/2)\hbar\omega_{T},\label{ebar3e}
\end{equation}
 where $\nu$ is the solution of the two-body spectrum of Eq. (\ref{seRel2e}).
At high energies the full spectrum $E_{rel}$ approaches asymptotically
to $\bar{E}_{rel}$ as the exchange effect becomes increasingly insignificant.
There is an important exception, however, occurring at zero total
relative momentum $l=0$. As mentioned earlier, the solution of $\bar{E}_{rel}$
at $n=0$ and $l=0$ is spurious and does not match any solution of
$E_{rel}$. Therefore, for the $l=0$ subspace, we require $n\geq1$
in Eq. (\ref{ebar3e}).

\begin{figure}[htp]
\begin{centering}
\includegraphics[clip,width=0.5\textwidth]{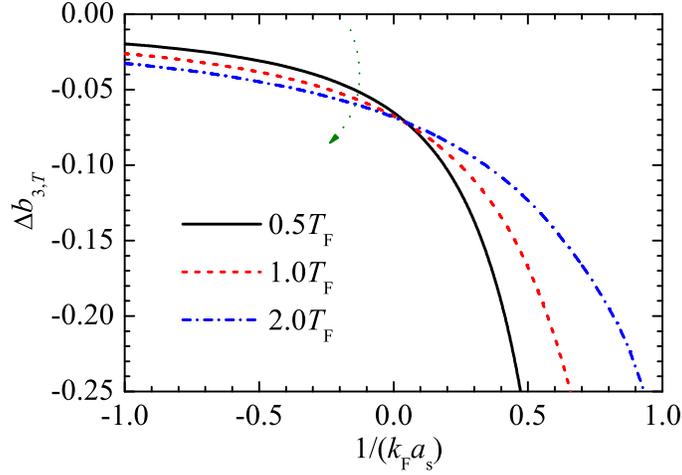} 
\par\end{centering}

\caption{(color online) Third virial coefficient of a trapped attractive Fermi
gas as a function of the interaction parameter $1/(k_{F}a_{s})$.
As in Fig. \ref{fig:trappedB2}, we have used a total number of atoms
$N=100$. Adapted from ref. \cite{OurVE}; copyright (2009) by APS.}

\label{fig:trappedB3} 
\end{figure}

It is easy to see that if we keep the spurious solution in the $l=0$
subspace, the difference $[\sum\exp(-\bar{E}_{rel}/k_{B}T)-\sum\exp(-E_{rel}^{(1)}/k_{B}T)]$
is exactly equal to $\Delta Q_{2,T}$, since in Eq. (\ref{ebar3e})
the first part of spectrum is exactly identical to the spectrum of
center-of-mass motion. The spurious solution gives a contribution,
\begin{equation}
\sum_{\nu_{n}}\left[e^{-\left(2\nu_{n}+3\right)\tilde{\omega}_{T}}-e^{-\left(2\nu_{n}^{\left(1\right)}+3\right)\tilde{\omega}_{T}}\right]\equiv2e^{-3\tilde{\omega}_{T}/2}\Delta b_{2,T},
\end{equation}
 which should be subtracted. Keeping this in mind, we finally arrive
at the following expression for the third virial coefficient of a
trapped Fermi gas with attractive interactions: 
\begin{equation}
\Delta b_{3,T}=\sum\left[e^{-\frac{E_{rel}}{k_{B}T}}-e^{-\frac{\bar{E}_{rel}}{k_{B}T}}\right]-2e^{-3\tilde{\omega}_{T}/2}\Delta b_{2,T}.
\end{equation}
 The summation is over all the possible relative energy levels $E_{rel}$
and their asymptotic values $\bar{E}_{rel}$. It is well-behaved and
converges at any scattering length. The third virial coefficient of
a trapped attractive Fermi gas in the BEC-BCS crossover was shown
in Fig. \ref{fig:trappedB3}.

In the unitarity limit, it is more convenient to use the exact spectrum
given by Eq. (\ref{energyRel3eHyper}), where $s_{l,n}$ can be obtained
numerically to arbitrary accuracy and the non-interacting $s_{l,n}^{\left(1\right)}$
is given by Eq. (\ref{non_interacting_sln}). To control the divergence
problem, we shall use the same strategy as before and to approach
$s_{l,n}$ by using its asymptotic value $\bar{s}_{l,n}$ given in
Eq. (\ref{asymptotic_sln}).

Integrating out the $q$ degree of freedom and using Eq. (\ref{db2attTrap})
to calculate $\Delta Q_{2,T}$, we find that, 
\begin{equation}
\Delta b_{3,T}=\frac{e^{-\tilde{\omega}_{T}}}{1-e^{-2\tilde{\omega}_{T}}}\left[\sum_{l,n}\left(e^{-\tilde{\omega}_{T}s_{l,n}}-e^{-\tilde{\omega}_{T}\bar{s}_{l,n}}\right)+A\right],
\end{equation}
 where $A$ is given by 
\begin{equation}
A=\sum_{l,n}\left(e^{-\tilde{\omega}_{T}\bar{s}_{l,n}}-e^{-\tilde{\omega}_{T}s_{l,n}^{\left(1\right)}}\right)-\frac{e^{-\tilde{\omega}_{T}}}{\left(1-e^{-\tilde{\omega}_{T}}\right)^{2}}.
\end{equation}
 We note that for the summation, implicitly there is a pre-factor
$\left(2l+1\right)$, accounting for the degeneracy of each subspace.
The value of $A$ can then be calculated analytically, leading to,
\begin{equation}
A=-e^{-\tilde{\omega}_{T}}\left(1-e^{-\tilde{\omega}_{T}}\right).
\end{equation}
 We have calculated numerically $\sum_{l,n}(e^{-\tilde{\omega}_{T}s_{l,n}}-e^{-\tilde{\omega}_{T}\bar{s}_{l,n}})$
by imposing the cut-offs of $n<n_{\max}=512$ and $l<l_{\max}=512$.
We find that, 
\begin{equation}
\Delta b_{3,T}\simeq-0.06833960+0.038867\tilde{\omega}_{T}^{2}+\cdots.
\end{equation}
 The numerical accuracy can be further improved by suitably enlarging
$n_{\max}$ and $l_{\max}$. By neglecting the dependence on $\tilde{\omega}$
in the thermodynamic limit, we obtain the universal third virial coefficient:
$\Delta b_{3,T}\simeq-0.06833960$. Using the universal relation between
trapped and homogeneous virial coefficients, Eq. (\ref{UniversalRelationBn}),
we obtain immediately the homogeneous third virial coefficient in
the unitarity limit, $\Delta b_{3}\simeq-0.35510298$.

In a recent study by Rakshit, Daily and Blume \cite{BlumeVE1}, much
more energy levels are included in the calculation of the third virial
coefficient in the unitary limit. As a result, the accuracy is much
improved. It was shown that \cite{BlumeVE1} $\Delta b_{3,T}=-0.068339609311287$
and $\Delta b_{3}=-0.3551030264897$.

\subsubsection{Fourth virial coefficient}

The calculation of the fourth virial coefficient could follow the
same strategy. However, the determination of $\Delta Q_{4,T}$ appears
to be a daunting task, since so far the problem of four interacting
fermions in harmonic traps has no exact solutions.

This difficulty was overcome by Rakshit, Daily and Blume \cite{BlumeVE1},
by using a scheme that allows to extrapolate the high temperature
behavior of the virial coefficients from the low-lying portion of
the excitation spectra only. This scheme is largely due to the weak
$\tilde{\omega}_{T}$ dependence of the trapped virial coefficient
$\Delta b_{n,T}$: because of the peculiarity of the harmonic trapping
potential, $\Delta b_{n,T}$ is a function of $\tilde{\omega}_{T}^{2}$.
As a result, one can determine $\Delta b_{n,T}$ at relatively large
$\tilde{\omega}_{T}$ (i.e., $\tilde{\omega}_{T}\sim1$) and then
extrapolate it to the zero-$\tilde{\omega_{T}}$ limit. This procedure
requires a small portion of the excitation spectra, which can be calculated
using the stochastic variational approach \cite{DailyPRA2010}, with
moderate computational resources. It was predicted that $\Delta b_{4,T}=-0.0020\pm0.0005$
and $\Delta b_{4}=-0.016\pm0.004$.

By using the same token, Rakshit, Daily and Blume estimated the fifth
virial coefficient, $0.0017\leq\Delta b_{5}\leq0.101$, and conjectured
the sign of the higher-order virial coefficients is $+,-,-,+,+,-,\cdots$
for $n=6,7,8,9,10,11,\cdots$.

\subsection{Third virial coefficient from field theoretic method}

\begin{figure}[htp]
\begin{centering}
\includegraphics[clip,width=0.5\textwidth]{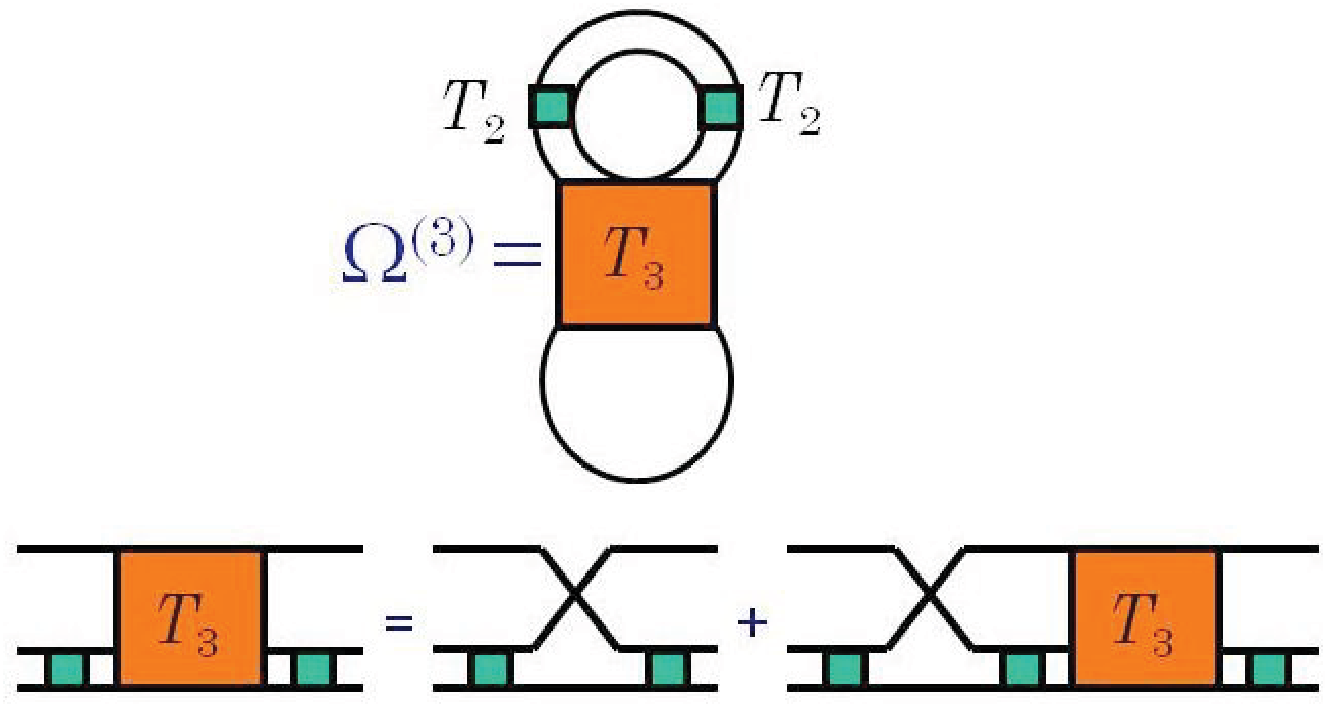} 
\par\end{centering}

\caption{(color online) Diagrammatic representation of the contribution of
three-particle scattering process to the thermodynamic potential.
Here $T_{2}$ and $T_{3}$ are respectively the two- and three-particle
vertex functions. For details, see refs. \cite{BrodskyJETP2005} and
\cite{RupakPRL2007}.}

\label{fig:T3} 
\end{figure}

Here we review briefly the diagrammatic calculation of the third virial
coefficient. The basic idea is to calculate $\Omega^{(3)}$ or $n^{(3)}=-\partial\Omega^{(3)}/\partial\mu$,
which involves the contribution from the three-particle scattering
process. As the three-particle vertex function is solved \cite{SkorniakovJETP1957,BrodskyJETP2005},
in principle the third virial coefficient could be determined. However,
as we shall see, the calculation turns out to be subtle. The diagrammatic
representation of $\Omega^{(3)}$ is shown in Fig. \ref{fig:T3}.
The two- and three-particle vertex functions are indicated by $T_{2}$
and $T_{3}$, respectively.

The calculation of $\Omega^{(3)}$ at large temperatures was pioneered
by Rupak \cite{RupakPRL2007}, by using a two-channel model for the
description of Feshbach resonances. A dimer field is introduced, designed
to reproduce the continuum two-body phase shift. In the unitary limit,
it was predicted that $\Delta b_{3}\simeq1.05$. This calculation
was recently improved by Kaplan and Sun \cite{KaplanPRL2011}, with
the development of a new diagrammatic method for $-\partial\Omega^{(3)}/\partial\mu$.
The sum over discrete Matsubara frequencies is converted to a Possion
resummation. This leads to $\Delta b_{3}=-0.3573\pm0.0005$.

\begin{figure}[htp]
\begin{centering}
\includegraphics[clip,width=0.5\textwidth]{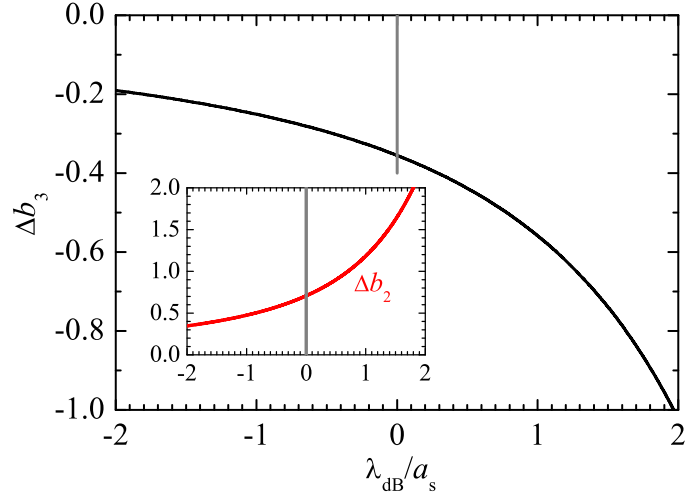} 
\par\end{centering}

\caption{(color online) Third virial coefficient as a function of the dimensionless
parameter $\lambda_{dB}/a_{s}$. The inset shows the second virial
coefficient. Adapted from ref. \cite{LeyronasPreprint}.}

\label{fig:homoB23} 
\end{figure}

The latest field theoretical calculation of the third virial coefficient
was given by Leyronas \cite{LeyronasPreprint}, by using the single-channel
Hamiltonian and Feynman diagrams for $-\partial\Omega^{(3)}/\partial\mu$
or the single-particle Green function. Explicit analytic expressions
of virial coefficient were obtained. To the accuracy of four digits,
it was found that $\Delta b_{3}=-0.3551\pm0.0001$, which is in excellent
agreement with the calculation based on the exact three-particle solutions
in harmonic traps \cite{OurVE,BlumeVE1}, but disagrees slightly with
that obtained by Kaplan and Sun \cite{KaplanPRL2011}. Fig. \ref{fig:homoB23}
shows the prediction by Leyronas \cite{LeyronasPreprint}.

At this stage, we believe that the result of $\Delta b_{3}=-0.3551\pm0.0001$
is robust, as it has been checked independently by two entirely different
methods. The discrepancy between different field theoretic calculations
remains to be understood. We note that it is appealing to calculate
the fourth virial coefficient $\Delta b_{4}$, along the line of Leyronas's
calculation \cite{LeyronasPreprint}, as the four-particle vertex
function is basically known \cite{BrodskyJETP2005}. Together with
an improved calculation with four-fermion solutions in harmonic traps,
$\Delta b_{4}$ could be determined very accurately.

\subsection{Virial equation of state for ultracold Fermi atoms and its comparison
with experimental measurements}

\subsubsection{Virial equation of state}

We are now ready to calculate the virial equations of states in the
high temperature regime, by using the thermodynamic potential 
\begin{equation}
\Omega=\Omega^{(1)}-V\frac{2k_{B}T}{\lambda_{dB}^{3}}\left(\Delta b_{2}z^{2}+\Delta b_{3}z^{3}+\cdots\right)\label{veOmegaHomo}
\end{equation}
 and 
\begin{equation}
\Omega_{T}=\Omega_{T}^{(1)}-\frac{2\left(k_{B}T\right)^{4}}{\left(\hbar\omega_{T}\right)^{3}}\left(\Delta b_{2,T}z^{2}+\Delta b_{3,T}z^{3}+\cdots\right),\label{veOmegaTrap}
\end{equation}
 respectively, for a homogeneous or a harmonically trapped Fermi gas.
Here, the non-interacting thermodynamic potentials are given by Eqs.
(\ref{idealOmegaHomo}) and (\ref{idealOmegaTrap}). All the other
thermodynamic quantities can be derived from the thermodynamic potential
by the standard thermodynamic relations, for example, $N=-\partial\Omega/\partial\mu$,
$S=-\partial\Omega/\partial T$, and then $E=\Omega+TS+\mu N$.

As an concrete example, let us focus on the unitary limit in the thermodynamic
limit, which is of the greatest interest. The equations of states
are easy to calculate because of the temperature independence of virial
coefficients. It is also easy to check the well-known scaling relation
in the unitarity limit: $E=-3\Omega/2$ for a homogeneous Fermi gas
\cite{HoUniversality} and $E=-3\Omega$ for a harmonically trapped
Fermi gas \cite{HLDNJP2010}. The difference of the factor of two
arises from the fact (virial theorem) that in harmonic traps the internal
energy is exactly equal to the trapping potential energy.

To be dimensionless, we take the Fermi temperature $T_{F}$ or Fermi
energy ($E_{F}=k_{B}T_{F}$) as the units for temperature and energy.
For a homogeneous or a harmonically trapped Fermi gas, the Fermi energy
is given by $E_{F}=\hbar^{2}(3\pi^{2}N/V)^{2/3}/2m$ and $E_{F}=(3N)^{1/3}\hbar\omega_{T}$,
respectively. In the actual calculations, we determine the number
of atoms $N$, the total entropy $S$, and the total energy $E$ at
given fugacity and a fixed temperature, and consequently obtain the
Fermi temperature $T_{F}$ and Fermi energy $E_{F}$. We then plot
the energy or energy per particle, $E/(NE_{F})$ and $S/(Nk_{B})$,
as a function of the reduced temperature $T/T_{F}$.

\subsubsection{Experimental measurement of equation of state}

Experimentally, there have been great efforts to measure the thermodynamics
of strongly interacting Fermi gases of $^{6}$Li and $^{40}$K atoms
near a Feshbach resonance \cite{BourdelPRL2003,KinastScience2005,StewardPRL2006,LuoPRL2007,NascimbeneNature2010,HorikoshiScience2010,EoSMIT}.
Initial measurements have focused on trap averaged quantities \cite{BourdelPRL2003,KinastScience2005,StewardPRL2006,LuoPRL2007}.
In the recent development, the \emph{bulk} equation of state of a
homogeneous Fermi gas becomes accessible \cite{NascimbeneNature2010,HorikoshiScience2010,EoSMIT},
following a theoretical proposal by Ho and Zhou \cite{HoNaturePhysics2009}.
Here we focus on the measurements performed by Nascimbène \textit{et
al.} at ENS \cite{NascimbeneNature2010} and by Ku \textit{et al.}
at MIT \cite{EoSMIT}. These two precise measurements allow a \emph{quantitative}
comparison with the virial expansion predictions.

In the ENS experiment, the local pressure $P(\mu(z),T)$ of the trapped
gas was directly probed using{\em \ in-situ} images of the \emph{doubly-integrated}
density profiles along the long $z$-axis (see the theoretical proposal
by Ho and Zhou, ref. \cite{HoNaturePhysics2009}). The temperature
was determined by using a new thermometry approach employing a $^{7}$Li
impurity. The chemical potential could also be determined using the
local density approximation, with $\mu(z)=\mu-V_{T}(z)$ and the central
chemical potential $\mu$ being determined appropriately. By introducing
a universal $h$-function %
\footnote{The universal $h$-function $h(z)$ defined in the experimental paper
\cite{NascimbeneNature2010} is renormalized by the pressure of an
ideal, \emph{single}-component Fermi gas. It is therefore a double
of the universal function defined in Eq. (\ref{hz}).%
} 
\begin{equation}
h\left[z\right]=\frac{P(\mu,T)}{P^{(1)}(\mu,T)},\label{hz}
\end{equation}
 experimentalists were able to determine $h(z)$ with very low noise.
Here, $P(\mu,T)$ is the interacting pressure and $P^{(1)}(\mu,T)$
is the pressure of an ideal two-component Fermi gas. All the other
thermodynamic quantities may then be derived from the universal $h$-function,
i.e., see ref. \cite{HuPRA2011}.

In the MIT experiment \cite{EoSMIT}, instead of the pressure, the
density equation of state $\rho(\mu,T)=\partial P(\mu,T)/\partial\mu$
is measured. Owing to the perfect cylindrical symmetry of the trapping
potential, the 3D density $\rho(\mu,T)$ can be reconstructed from
the measured \emph{column} density, i.e., $\rho_{2D}(x,z)=\int dy\rho(x,y,z)$,
by using an inverse Abel transform \cite{EoSMIT}. The local pressure
and isothermal compressibility $\kappa=[\rho^{2}\partial\mu/\partial\rho]^{-1}$
can then be calculated from the density \cite{EoSMIT}. The crucial
advantage of the MIT experiment is that the temperature $T$ and the
chemical potential $\mu$ can be replaced by the pressure and compressibility.
Thus, the notoriously difficult thermometry of a strongly interacting
Fermi gas may not be required.

\subsubsection{Qualitative comparison between theory and experiment}

\begin{figure}[htp]
\begin{centering}
\includegraphics[clip,width=0.5\textwidth]{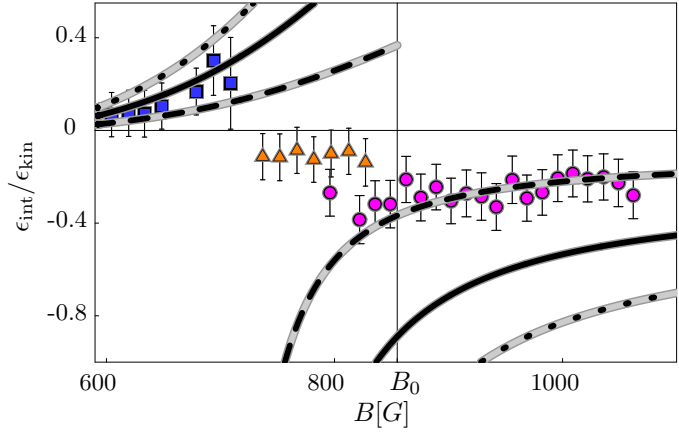} 
\par\end{centering}

\caption{(color online) The second-order virial prediction for the ratio $\epsilon_{int}/\epsilon_{kin}$
of a Fermi gas of $^{6}$Li atoms near the Feshbach resonance. The
dashed, solid, and dotted lines are for $T/T_{F}=1.2$, $0.6$, and
$0.4$, respectively. The symbols show the experimental data from
the ENS group \cite{BourdelPRL2003}, measured at $T=3.5$ $\mu$K
and $T/T_{F}=0.6$. The squares and circles are respectively the results
obtained by approaching the Feshbach resonace ($B_{0}$) from the
BEC side and BCS side. From ref. \cite{HoVE} with permission; copyright
(2004) by APS.}

\label{fig:HoVE} 
\end{figure}

Before quantiative comparing the virial theory with the latest thermodynamics
measurements, we mention briefly the first application of virial expansion
in ultracold atomic Fermi gases, reported by Ho and Mueller in 2004
\cite{HoVE}. This elegant application gave a very good qualitative
explanation for the measured interaction energy at ENS in 2003. 

Experimentally, in a Fermi gas of $^{6}$Li atoms the Feshbach magnetic
field was swept across the resonance from either the positive (BEC)
or negative (BCS) scattering length side. The interaction energy of
the near-resonance Fermi gas was then recorded at different fields.
As shown in Fig. \ref{fig:HoVE} by symbols, crossing the resonance
from the BCS side (the scenario \textbf{A}), the interaction energy
$\epsilon_{int}$ remains negative and continuous across the resonance,
while approaching the resonance from the opposite BEC side (the scenario
\textbf{B}), $\epsilon_{int}$ is positive but drops to a negative
value near the resonance. 

By using the virial expansion theory to the second order, Ho and Mueller
showed conclusively that the different interaction energy is a result
of the different initial state. In the scenario \textbf{A}, the system
is alway in the ground state with strong attractions, while in the
scenario \textbf{B}, the system is initially in the metastable excited
branch, where the interaction between two fermions is repulsive. For
the detailed discussion, see Fig. \ref{fig:2eSpectrum}. To be concrete,
to the second order of virial expansion, the kinetic energy and interaction
energy can be written as \cite{HoVE},
\begin{equation}
\epsilon_{kin}=\frac{3\rho k_{B}T}{2}\left(1+\frac{\rho\lambda_{dB}^{3}}{2^{7/2}}\right)\label{eq:virial2Ekin}
\end{equation}
and
\begin{equation}
\epsilon_{int}=\frac{3\rho k_{B}T}{2}\left(\rho\lambda_{dB}^{3}\right)\left[-\frac{1}{2}\Delta b_{2}+\frac{1}{3}T\frac{\partial\Delta b_{2}}{\partial T}\right].\label{eq:virial2Eint}
\end{equation}
The second-order virial coefficient $\Delta b_{2}$ can be calculated
using the Beth-Uhlenbeck formalism Eq. (\ref{BethUhlenbeck}) and
the usual $s$-wave phase shift. In the metastable excited branch,
the contribution of the bound state to $\Delta b_{2}$ should be removed.
The ratio $\epsilon_{int}/\epsilon_{kin}$, predicted by Eqs. (\ref{eq:virial2Ekin})
and (\ref{eq:virial2Eint}), is compared with the experimental data
in Fig. \ref{fig:HoVE}. The virial prediction at $T/T_{F}=1.2$ agree
well with the experimental results, which were measured at $T/T_{F}=0.6$.
The difference in temperature is understandable, since the fugacity
at $T/T_{F}=0.6$ is already larger than 1 and the agreement must
be affected by the higher order terms in the virial expansion.

\subsubsection{Quantitative comparison: Homogeneous system}

\begin{figure}[htp]
\begin{centering}
\includegraphics[clip,width=0.5\textwidth]{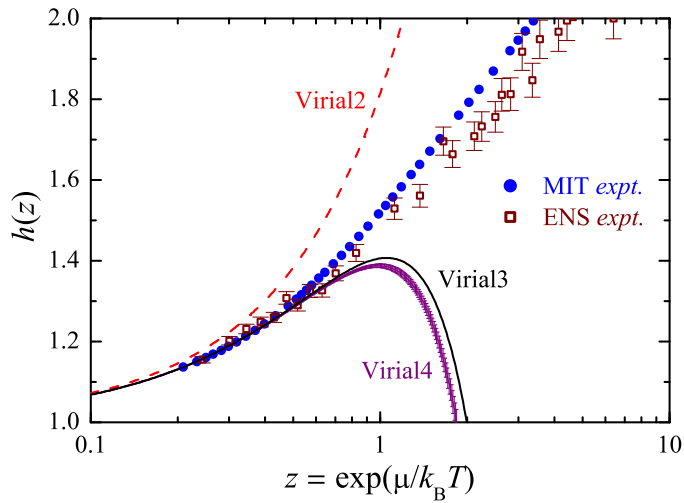} 
\par\end{centering}

\caption{(color online) Virial expansion prediction of the universal function
$h(z)$ up to the second (dashed line), third (solid line) and fourth
order (thin solid line with error bar), compared with the experimental
data (empty squares with error bars). Here, $ $$\Delta b_{4}=-0.016\pm0.004$.
Adapted from ref. \cite{BlumeVE1} with inclusion of experimental
results. The experimental data of the ENS group and of the MIT group
are taken from refs. \cite{NascimbeneNature2010} and \cite{EoSMIT},
respectively.}

\label{fig:hz} 
\end{figure}

We now turn to the quantitative comparison. At high temperature, by
using the virial thermodynamic potential Eq. (\ref{veOmegaHomo}),
the universal function $h(z)$ can be written as, 
\begin{equation}
h(z)=1+\frac{\Delta b_{2}z^{2}+\Delta b_{3}z^{3}+\cdots}{\left(2/\sqrt{\pi}\right)\int_{0}^{\infty}t^{1/2}\ln\left(1+ze^{-t}\right)dt}.\label{hzve}
\end{equation}

In Fig. \ref{fig:hz}, we compare the virial expansion prediction
and the experimental data for the universal function $h(z)$. The
virial results are calculated by using Eq. (\ref{hzve}), with inclusion
virial coefficients $\Delta b_{n}$ up to $\Delta b_{2}=1/\sqrt{2}$
(Virial2), $\Delta b_{3}\simeq-0.35510298$ (Virial3), and $\Delta b_{4}=-0.016\pm0.004$
(Virial4). At small fugacity ($z<0.7$), the experimental data agrees
excellently well with the virial prediction. Using $\Delta b_{3}$
and $\Delta b_{4}$ as independent fitting parameters, experimentally
it was determined that $\Delta b_{3,expt}=-0.35\pm0.02$ and $\Delta b_{4,expt}=0.096\pm0.015$
\cite{NascimbeneNature2010} by the ENS group. The latest measurement
at MIT also showed an excellent agreement with virial expansion and
reported $\Delta b_{4,expt}=0.096\pm0.010$ \cite{EoSMIT}. Thus,
while the theoretical $\Delta b_{3}$ has been confirmed unambiguously
by the experiments, the theoretical prediction for the fourth virial
coefficient $\Delta b_{4}=-0.016\pm0.004$ contradicts with the experimental
observation. This discrepancy remains to be resolved. A possible reason
is the uncertainty of the Feshbach resonance position, which is about
$1.5$ G for $^{6}$Li atoms \cite{EoSMIT}. As we discussed earlier,
this uncertainty will leads to $1\%$ relative error to the second
virial coefficient. When this systematic error passes to the small
fourth virial coefficient, the experimental determination of $\Delta b_{4}$
may become unreliable.

\begin{figure}[htp]
\begin{centering}
\includegraphics[clip,width=0.8\textwidth]{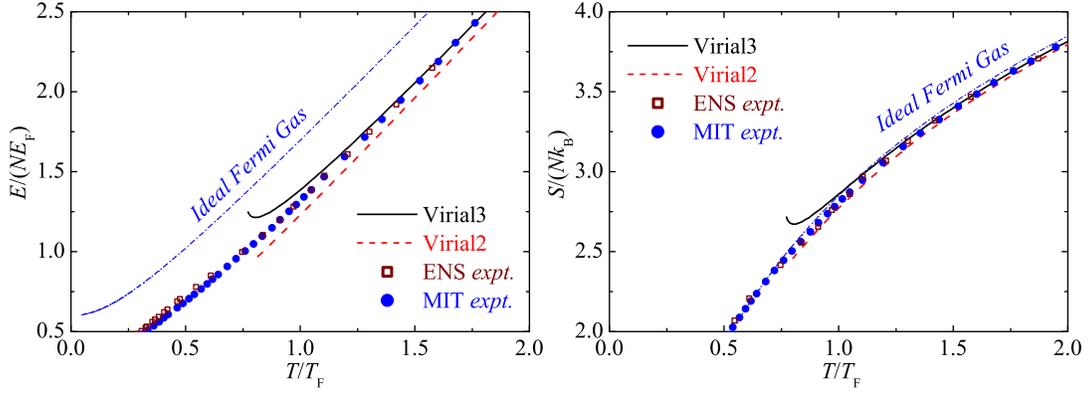} 
\par\end{centering}

\caption{(color online) Energy per particle $E/(NE_{F})$ and the entropy per
particle $S/(NE_{F})$ as a function of reduce temperature $T/T_{F}$
for a homogeneous Fermi gas in the unitary limit. The predictions
of virial expansion up to the second- and third-order are shown by
dashed line and solid line, respectively. For comparison, we plot
the ideal gas result by the dot-dashed line. We show also the experimental
data measured at ENS \cite{NascimbeneNature2010} and MIT \cite{EoSMIT},
which agree extremely well with the prediction from virial expansion.
Adapted from ref. \cite{OurVELongPRA}; copyright (2010) by APS.}

\label{fig:homoES} 
\end{figure}

Let us turn to the other thermodynamic quantities such as energy and
entropy. We report in Fig.\ref{fig:homoES} the temperature dependence
of energy and entropy of a unitary Fermi gas in homogeneous space.
The solid line and dashed line are the predictions of virial expansion
up to the third-order and second-order, respectively. For comparison,
we also show the ideal gas result by the thin dot-dashed line and
the experimental results by symbols. We observe that the virial expansion
is valid down to the degenerate temperature $T_{F}$, where the prediction
up to the second-order or third-order expansion does not differ largely.
The experimental data lie between the two virial expansion predictions,
but clearly agree much better with the third-order expansion, as anticipated.

\subsubsection{Quantitative comparison: Trapped system}

\begin{figure}[htp]
\begin{centering}
\includegraphics[clip,width=0.8\textwidth]{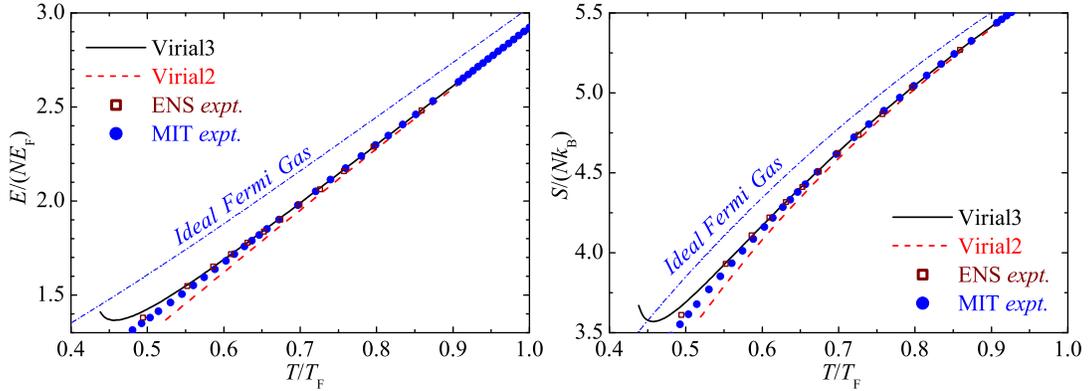} 
\par\end{centering}

\caption{(color online) Energy per particle $E/(NE_{F})$ and the entropy per
particle $S/(NE_{F})$ as a function of reduced temperature $T/T_{F}$
for a trapped Fermi gas in the unitary limit. The predictions of quantum
virial expansion up to the second- and third-order are shown by solid
line and dashed line, respectively. For comparison, we plot the ideal
gas result by the dot-dashed line. The experimental data measured
at ENS \cite{NascimbeneNature2010} and MIT \cite{EoSMIT} are shown
by empty squares and solid circles, respectively. Adapted from ref.
\cite{OurVELongPRA}; copyright (2010) by APS.}

\label{fig:trappedES} 
\end{figure}

Experimentally, the thermodynamics of a harmonically trapped Fermi
gas in the unitary limit can be determined as well, from the measured
universal $h$-function. For the details, see ref. \cite{HLDNJP2010}.
We present in Fig. \ref{fig:trappedES} the high-temperature expansion
predictions for energy and entropy, and compare them with the experimental
measurement. We find a much broader applicability of virial expansion:
it is now quantitatively applicable down to $0.5T_{F}$, as confirmed
by the precise experimental data at ENS \cite{NascimbeneNature2010}
and at MIT \cite{EoSMIT}. This is largely due to the much reduced
higher order virial coefficient in a harmonic traps, i.e., $\Delta b_{n,T}=n^{-3/2}\Delta b_{n}$.
At large $n$, the reduction factor of $n^{-3/2}$ is fairly significant,
implying a better convergence of virial expansion and hence a much
wider applicability.

\subsubsection{Reliability of virial expansion}

\begin{figure}[htp]
\begin{centering}
\includegraphics[clip,width=0.5\textwidth]{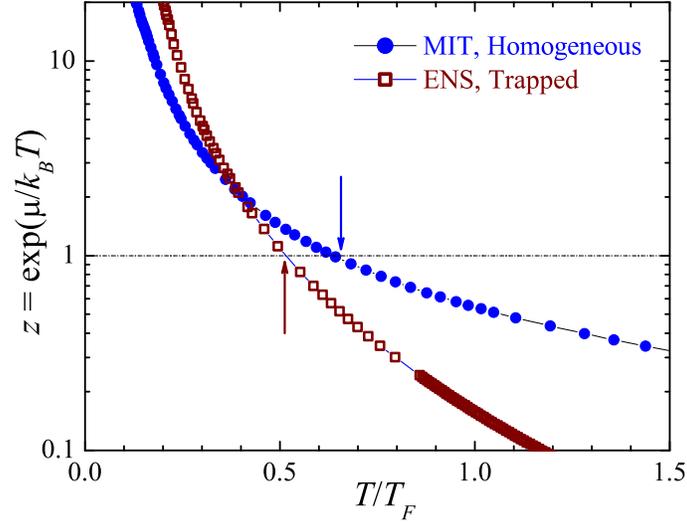} 
\par\end{centering}

\caption{(color online) Fugacity $z$ as a function of reduce temperature $T/T_{F}$,
for a homogeneous unitary Fermi gas (solid circles) and for a trapped
unitary Fermi gas (empty squares). The results, calculated from the
experimental universal function $h(z)$ \cite{NascimbeneNature2010,EoSMIT},
have a relative error at a few percents.}

\label{fig:fugacity} 
\end{figure}

To better understand the reliability of virial expansion, we show
in Fig. \ref{fig:fugacity} the fugacity as a function of temperature,
for a homogeneous or trapped Fermi gas in the unitary limit. These
two curves are determined from the experimental universal function
$h(z)$ measured at ENS and MIT. By setting $z=1$ as the criterion
for qualitative reliability, we find that the virial expansion should
be applicable at $T>0.7T_{F}$ for a homogeneous unitary Fermi gas
and at $T>0.5T_{F}$ for a trapped unitary Fermi gas. As the typical
experimental temperature for a unitary Fermi gas is about $0.5T_{F}$,
we thus demonstrate clearly the virial expansion method is a very
useful tool for understanding the properties of a normal, strongly
interacting Fermi gas.

\subsection{Virial equation of state for a spin-population imbalanced Fermi gas}

We consider so far the balanced Fermi gas with equal mass and spin-populations.
The virial expansion is applicable as well to the imbalanced systems
with either unequal mass \cite{BlumeVE2} or spin-populations \cite{LiuPRA2011}.
Here we focus on the latter case with unequal spin-populations and
use virial expansion to obtain the high-temperature spin susceptibility
of a unitary Fermi gas.

In the presence of spin imbalance, it is necessary to introduce two
fugacities $z_{\uparrow}\equiv\exp(\mu_{\uparrow}/k_{B}T)$ and $z_{\downarrow}\equiv\exp(\mu_{\downarrow}/k_{B}T)$,
and to distinguish different spin-configurations. Quite generally,
we may write the thermodynamic potential as, 
\begin{equation}
\Omega=-k_{B}TQ_{1}\sum_{n=1}^{\infty}\sum_{k=0}^{n}z_{\uparrow}^{n-k}z_{\downarrow}^{k}b_{n,k},
\end{equation}
 where $b_{n,k}$ is the $n$-th (imbalanced) virial coefficient contributed
by the configuration with $n-k$ spin-up fermions and $k$ spin-down
fermions. It is easy to see that the imbalanced virial coefficients
satisfy the relation $b_{n,k}=b_{n,n-k}$ and $\sum_{k=0}^{n}b_{n,k}=b_{n}$.

The calculation of $b_{n,k}$ is straightforward, following the standard
definition of thermodynamic potential. We rewrite the grand partition
function ${\cal Z}\equiv$Tr$\exp[-({\cal H}-\mu_{\uparrow}{\cal N}_{\uparrow}-\mu_{\downarrow}{\cal N}_{\downarrow})/k_{B}T]$
in the form, 
\begin{equation}
{\cal Z}=\sum_{n=0}^{\infty}\sum_{k=0}^{n}z_{\uparrow}^{n-k}z_{\downarrow}^{k}Q_{n,k},
\end{equation}
 where $Q_{n,k}$ is the partition function of a cluster that contains
$n-k$ spin-up fermions and $k$ spin-down fermions. It is apparent
that due to the symmetry in spin configurations we have $Q_{n,k}=Q_{n,n-k}$.
The imbalanced cluster partition functions satisfy as well a sum rule
$\sum_{k=0}^{n}Q_{n,k}=Q_{n}$. By expanding the thermodynamic potential
$\Omega=-k_{B}T\ln{\cal Z}$ into powers of the two fugacities, the
imbalanced virial coefficients can then be expressed in terms of the
cluster partition function $Q_{n,k}$.

\subsubsection{Virial expansion of an imbalanced Fermi gas up to the third order}

To be concrete, let us consider the imbalanced virial expansion up
to the third order. To this order, we may write the grand partition
function as ${\cal Z}=1+x_{1}+x_{2}+x_{3}$, where 
\begin{equation}
x_{1}=z_{\uparrow}Q_{1,0}+z_{\downarrow}Q_{1,1},
\end{equation}
 
\begin{equation}
x_{2}=z_{\uparrow}^{2}Q_{2,0}+z_{\uparrow}z_{\downarrow}Q_{2,1}+z_{\downarrow}^{2}Q_{2,2},
\end{equation}
 and 
\begin{equation}
x_{3}=z_{\uparrow}^{3}Q_{3,0}+z_{\uparrow}^{2}z_{\downarrow}Q_{3,1}+z_{\uparrow}z_{\downarrow}^{2}Q_{3,2}+z_{\downarrow}^{3}Q_{3,3}.
\end{equation}
 By introducing a symmetric cluster partition function $Q_{n}^{s}\equiv Q_{n,0}=Q_{n,n}$
and using the properties of $Q_{n,k}$, it is easy to show that $Q_{1}^{s}=Q_{1}/2$,
$Q_{2,1}=Q_{2}-2Q_{2}^{s}$, and $Q_{3,1}=Q_{3,2}=Q_{3}/2-Q_{3}^{s}$.
Using $\ln(1+x_{1}+x_{2}+x_{3})\simeq(x_{1}+x_{2}+x_{3})-(x_{1}^{2}+2x_{1}x_{2})/2+x_{1}^{3}/3$,
after some algebra we obtain $b_{n,k}$ ($k\leq n/2$), 
\begin{equation}
b_{1,0}=1/2,
\end{equation}
 
\begin{equation}
b_{2,0}=Q_{2}^{s}/Q_{1}-Q_{1}/8,
\end{equation}
 
\begin{equation}
b_{2,1}=Q_{2}/Q_{1}-2Q_{2}^{s}/Q_{1}-Q_{1}/4,
\end{equation}
 
\begin{equation}
b_{3,0}=Q_{3}^{s}/Q_{1}-Q_{2}^{s}/2+Q_{1}^{2}/24,
\end{equation}
 and 
\begin{equation}
b_{3,1}=Q_{3}/(2Q_{1})-Q_{3}^{s}/Q_{1}-Q_{2}/2+Q_{2}^{s}/2+Q_{1}^{2}/8.
\end{equation}
 The virial coefficients with $k\geq n/2$ can be obtained directly
since $b_{n,k}=b_{n,n-k}$.

As before, it is convenient to consider the interaction effect on
the virial coefficients or the differences such as $\Delta Q_{n}=Q_{n}-Q_{n}^{(1)}$,
$\Delta b_{n}=b_{n}-b_{n}^{(1)}$, and $\Delta b_{n,k}=b_{n,k}-b_{n,k}^{(1)}$.
Here, the superscript ``$1$'' denotes an ideal, non-interacting
system with the same fugacities and the operator ``$\Delta$'' removes
the non-interacting contribution. It is clear that the symmetric cluster
partition function $Q_{n}^{s}$ is not affected by interactions since
the interatomic interaction occurs only between fermions with unlike
spins. Thus, we have $\Delta b_{2,0}=\Delta b_{3,0}=0$, $\Delta b_{2,1}=\Delta(b_{2}-2b_{2,0})=\Delta b_{2}$
and $\Delta b_{3,1}=\Delta(b_{3}/2-b_{2,0})=\Delta b_{3}/2$. Accordingly,
we may rewrite the thermodynamic potential into the form (up to the
third order), 
\begin{equation}
\Omega=\Omega^{(1)}-k_{B}TQ_{1}\left[z_{\uparrow}z_{\downarrow}\Delta b_{2}+\frac{z_{\uparrow}^{2}z_{\downarrow}+z_{\uparrow}z_{\downarrow}^{2}}{2}\Delta b_{3}\right],\label{omega}
\end{equation}
 where $\Omega^{(1)}=\Omega^{(1)}(\mu_{\uparrow})+\Omega^{(1)}(\mu_{\downarrow})$
is the thermodynamic potential of a non-interacting Fermi gas. Hereafter,
we consider the homogeneous case, in which $Q_{1}=2V/\lambda_{dB}^{3}$.

With the virial expansion of thermodynamic potential Eq. (\ref{omega}),
we solve the standard thermodynamic relations $N_{\uparrow}=-\partial\Omega/\partial\mu_{\uparrow}$
and $N_{\downarrow}=-\partial\Omega/\partial\mu_{\downarrow}$ for
the two fugacities $z_{\uparrow}$ and $z_{\downarrow}$, at a given
reduced temperature $\tau=T/T_{F}$ and a given spin imbalance $P=(N_{\uparrow}-N_{\downarrow})/N$.
Here, $T_{F}=\hbar^{2}(3\pi^{2}\rho)^{2/3}/(2m)/k_{B}$ is the Fermi
temperature. It is easy to show that we can define a dimensionless
number density $\tilde{\rho}=\rho\lambda^{3}/2=4/(3\sqrt{\pi}\tau^{3/2})$,
$\tilde{\rho}_{\uparrow}=(1+P)\tilde{\rho}$, and $\tilde{\rho}_{\downarrow}=(1-P)\tilde{\rho}$.
We then rewrite the number equations into dimensionless forms, 
\begin{eqnarray}
\tilde{\rho}_{\uparrow} & = & \tilde{\rho}^{(1)}\left(z_{\uparrow}\right)+z_{\uparrow}z_{\downarrow}2\Delta b_{2}+\left(2z_{\uparrow}^{2}z_{\downarrow}+z_{\uparrow}z_{\downarrow}^{2}\right)\Delta b_{3},\\
\tilde{\rho}_{\downarrow} & = & \tilde{\rho}^{(1)}\left(z_{\downarrow}\right)+z_{\uparrow}z_{\downarrow}2\Delta b_{2}+\left(z_{\uparrow}^{2}z_{\downarrow}+2z_{\uparrow}z_{\downarrow}^{2}\right)\Delta b_{3},
\end{eqnarray}
 where $\tilde{\rho}^{(1)}\left(z\right)\equiv(2/\sqrt{\pi})\int\nolimits _{0}^{\infty}\sqrt{t}[ze^{-t}/\left(1+ze^{-t}\right)]dt$.
We can obtain the two fugacities by solving the coupled number equations.

\subsubsection{Virial expansion of spin susceptibility and compressibility}

We now calculate the spin susceptibility $\chi_{S}=(\partial\delta\rho/\partial\delta\mu)$
in the balanced limit of $P=0$. For this purpose, we determine the
two by two susceptibility matrix ${\cal S}=(\partial\rho_{\sigma}/\partial\mu_{\sigma^{\prime}})$
to the third order of fugacity. For a homogeneous unitary Fermi gas,
using the number equation we find that, 
\begin{equation}
{\cal S}\left(P=0\right)=\frac{1}{k_{B}T\lambda_{dB}^{3}}\left[\begin{array}{ll}
A & B\\
B & A
\end{array}\right],
\end{equation}
 where 
\begin{equation}
A=\frac{2}{\sqrt{\pi}}\int\limits _{0}^{\infty}\frac{\sqrt{t}ze^{-t}}{\left(1+ze^{-t}\right)^{2}}dt+2z^{2}\Delta b_{2}+5z^{3}\Delta b_{3}
\end{equation}
 and 
\begin{equation}
B=2z^{2}\Delta b_{2}+4z^{3}\Delta b_{3}.
\end{equation}
 The spin susceptibility $\chi_{S}=2(A-B)/(k_{B}T\lambda_{dB}^{3})$
and compressibility $\kappa=2(A+B)/(k_{B}T\lambda_{dB}^{3})$ are
then given by, 
\begin{equation}
\chi_{S}=\frac{2}{k_{B}T\lambda_{dB}^{3}}\left[\frac{2}{\sqrt{\pi}}\int\limits _{0}^{\infty}\frac{\sqrt{t}ze^{-t}}{\left(1+ze^{-t}\right)^{2}}dt+z^{3}\Delta b_{3}\right],\label{spinkappa}
\end{equation}
 and 
\begin{equation}
\kappa=\frac{2}{k_{B}T\lambda_{dB}^{3}}\left[\frac{2}{\sqrt{\pi}}\int\limits _{0}^{\infty}\frac{\sqrt{t}ze^{-t}}{\left(1+ze^{-t}\right)^{2}}dt+4z^{2}\Delta b_{2}+9z^{3}\Delta b_{3}\right],\label{compkappa}
\end{equation}
 respectively. In the expressions, we see clearly the effect of interactions.
For the spin susceptibility, it appears in the third order of fugacity
only. A high-temperature measurement of spin susceptibility therefore
could be a sensitive way to measure accurately the third virial coefficient.
We note that, for an ideal Fermi gas the spin susceptibility and compressibility
are equal.

\begin{figure}[htp]
\begin{centering}
\includegraphics[clip,width=0.5\textwidth]{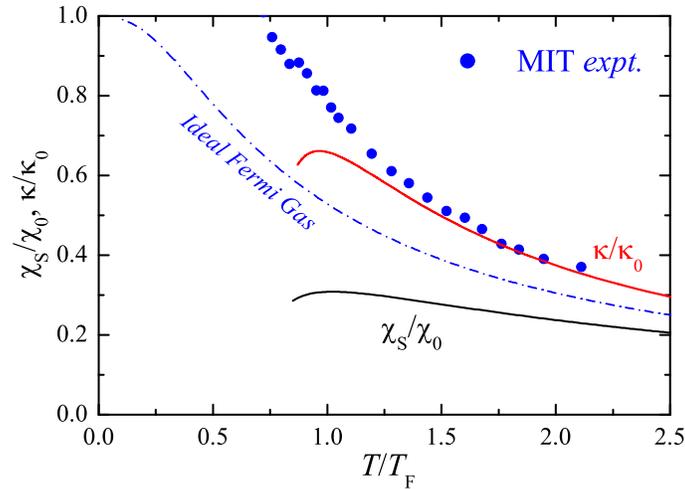} 
\par\end{centering}

\caption{(color online). Spin susceptibility and compressibility of a homogeneous
Fermi gas in the unitary limit, normalized by the $T=0$ ideal Fermi
gas susceptibility value $\chi_{0}=\kappa_{0}=3\rho/(2E_{F})$. Here
$\rho$ is the density and $E_{F}$ the Fermi energy. The experimental
compressibility data are taken from ref. \cite{EoSMIT}. Adapted from
refs. \cite{LiuPRA2011} and \cite{EoSMIT}.}

\label{fig:susceptibility} 
\end{figure}

Fig. \ref{fig:susceptibility} reports the numerical result of Eqs.
(\ref{spinkappa}) and (\ref{compkappa}) for a homogeneous Fermi
gas in the unitary limit, where the fugacity $z$ is solved consistently
to the third order expansion in the number equation. The spin susceptibility
and compressibility are smaller and larger than that of an ideal,
non-interacting Fermi gas, respectively, as expected. Experimentally,
the spin susceptibility at finite temperatures is related to the measurement
of the thermal spin fluctuations: 
\begin{equation}
\frac{\Delta\left(N_{\uparrow}-N_{\downarrow}\right)^{2}}{N}=k_{B}T\frac{\chi_{S}}{\rho}.\label{fluct}
\end{equation}
 A shot noise measurement of the spin fluctuations therefore could
be used as a sensitive thermometry for strongly interacting Fermi
gases \cite{ZhouPRL2011}, provided that the spin susceptibility is
known. Now, this seems to be accessible, since the shot noise measurements
of the density fluctuations in a weakly interacting Fermi gas have
already been demonstrated very recently \cite{MuellerPRL2010,SannerPRL2010}.
We find that the spin susceptibility is strongly suppressed by interactions
with respect to the ideal Fermi gas result, even well above the degenerate
temperature $T_{F}$. At $T=T_{F}$, the reduction is about $40\%$.
For the compressibility, the virial prediction agrees very well with
the latest experimental measurement at $T>T_{F}$ \cite{EoSMIT},
as anticipated.

\section{Virial expansion of Tan's contact}

In this section, we show that the important many-body parameter -
Tan's contact - can be virial expanded in terms of the so-called contact
coefficients \cite{OurVirialContact}. By using few-particle solutions,
we determine the second and third contact coefficients. For a trapped
Fermi gas in the unitary limit, we find that the virial prediction
at $T>0.5T_{F}$ agrees very well with the recent experimental measurements
performed at Swinburne University of Technology \cite{TanSwinburne2,TanSwinburne3}.
The first virial expansion calculation of Tan's contact was given
by Yu, Bruun, and Baym \cite{YuPRA2009}.

\subsection{Virial expansion of Tan's contact}

The virial expansion of the contact follows directly from an alternative
representation of Tan's adiabatic sweep theorem in the {\em grand-canonical}
ensemble, 
\begin{equation}
\left[\frac{\partial\Omega}{\partial\left(-a_{s}^{-1}\right)}\right]_{T,\mu}=\frac{\hbar^{2}}{4\pi m}{\cal I}.
\end{equation}
 This is simply because the adiabatic sweep theorem implies the first
law of thermodynamics, 
\begin{equation}
\Delta E=\hbar^{2}{\cal I}/(4\pi m)\Delta(-a_{s}^{-1})+T\Delta S+\mu\Delta N\,,
\end{equation}
 which can alternatively be written as 
\begin{equation}
\Delta\Omega=\hbar^{2}{\cal I}/(4\pi m)\Delta(-a_{s}^{-1})-S\Delta T-N\Delta\mu\,.
\end{equation}
 Therefore, using virial expansion for $\Omega$ we immediately obtain
a quantum virial expansion for the contact: 
\begin{equation}
{\cal I}=\frac{4\pi mk_{B}T\lambda_{dB}}{\hbar^{2}}Q_{1}\left[c_{2}z^{2}+\cdots+c_{n}z^{n}+\cdots\right],
\end{equation}
 where we have defined the dimensionless contact coefficient, $c_{n}\equiv\partial\Delta b_{n}/\partial(\lambda_{dB}/a_{s})$.
For a homogeneous system, we shall use the contact intensity, ${\cal C}={\cal I}/V$.

In general, the contact coefficient should be a function of $\lambda_{dB}/a_{s}$
and hence is temperature dependent. In the unitarity limit where $\lambda_{dB}/a_{s}=0$,
however, we anticipate a constant, universal contact coefficient,
similar to the universal virial coefficient $\Delta b_{n}$ \cite{HoVE,OurVE}.
This is a manifestation of fermionic universality, shared by all systems
of strongly interacting fermions \cite{HoUniversality,HDLNaturePhysics}.

\subsection{Universal relation between homogeneous and trapped contact coefficients}

In exact analogy with the virial coefficient, fermionic universality
leads to a very simple relation between the trapped and homogeneous
contact coefficients at unitarity. Let us consider the contact of
a harmonically trapped Fermi gas with the trapping potential $V_{T}({\bf x})=m\omega_{T}^{2}(x^{2}+y^{2}+z^{2})/2$.
In the thermodynamic limit of $\omega_{T}\rightarrow0$, as before
we may use the local density approximation and neglect the discrete
energy levels. The whole Fermi system is treated as many cells with
a local chemical potential $\mu({\bf x})=\mu-V_{T}({\bf x})$ and
a local fugacity $z({\bf x})=e^{\mu({\bf x})/k_{B}T}\equiv z\exp[-V_{T}({\bf x})/(k_{B}T)]$.
Due to the constant contact coefficients, the spatial integration
in the total contact ${\cal I}_{T}=\int d{\bf x}[{\cal C}({\bf x})]$
can be easily performed. We find that, 
\begin{equation}
{\cal I}_{T}=\frac{4\pi mk_{B}T\lambda_{dB}}{\hbar^{2}}Q_{1,T}\left[c_{2,T}z^{2}+c_{3,T}z^{3}+\cdots\right],
\end{equation}
 where the trapped contact coefficient is given by a universal relation,
\begin{equation}
c_{n,T}=\frac{c_{n}}{n^{3/2}},\label{UniversalRelationCn}
\end{equation}
 and $Q_{1,T}=2(k_{B}T)^{3}/(\hbar\omega_{T})^{3}$ is the single-particle
partition function in harmonic traps and in the local density approximation.

In the following, using the known solution of two- and three-fermion
problems, we calculate the universal second and third contact coefficients,
in both homogeneous and trapped configurations.

\subsection{Second contact coefficient}

The second contact coefficient of a homogeneous interacting Fermi
gas can be obtained from the well-known Beth-Uhlenbeck formalism for
the second virial coefficient. In the vicinity of the unitary limit,
we have $\Delta b_{2}(a_{s}<0)\simeq1/\sqrt{2}+\lambda_{dB}/(\pi a_{s})$,
giving rise to a homogeneous contact coefficient, 
\begin{equation}
c_{2}=\frac{1}{\pi}.\label{c2}
\end{equation}

To calculate the trapped second contact coefficient, we consider the
second virial sufficient in an isotropic harmonic trap, which is given
by Eq. (\ref{db2}), $\Delta b_{2,T}=(1/2)\sum_{\nu_{n}}[e^{-(2\nu_{n}+3/2)\tilde{\omega}_{T}}-e^{-(2\nu_{n}^{\left(1\right)}+3/2)\tilde{\omega}_{T}}],$
where $\tilde{\omega}_{T}\equiv\hbar\omega_{T}/(k_{B}T)\ll1$ is the
reduced trapping frequency, $\nu_{n}$ satisfies the secular equation
$2\Gamma(-\nu_{n})/\Gamma(-\nu_{n}-1/2)=d/a_{s}$, and $d=\sqrt{2\hbar/(m\omega_{T})}$
is the characteristic length scale of the harmonic trap. In the non-interacting
limit, $\nu_{n}^{\left(1\right)}=n$ ($n=0,1,2,...$), and in the
unitary limit, $\nu_{n}=n-1/2$. It is easy to show that, 
\begin{equation}
\left[\frac{\partial\nu_{n}}{\partial\left(\lambda_{dB}/a_{s}\right)}\right]_{\lambda_{dB}/a_{s}=0}=-\frac{d}{2\pi\lambda_{dB}}\frac{\Gamma\left(n+1/2\right)}{n!}.
\end{equation}
 Thus, we find that in the unitary limit, 
\begin{equation}
c_{2,T}=\frac{d}{2\pi\lambda_{dB}}\tilde{\omega}_{T}\sum_{n=0}^{\infty}\frac{\Gamma\left(n+1/2\right)}{n!}e^{-\left(2n+1/2\right)\tilde{\omega}_{T}}.
\end{equation}
 The sum over $n$ can be exactly performed, leading to, 
\begin{equation}
c_{2,T}=\frac{1}{2\sqrt{2}\pi}\left[\frac{2\tilde{\omega}_{T}}{e^{+\tilde{\omega}_{T}}-e^{-\tilde{\omega}_{T}}}\right]^{1/2}=\frac{1}{2\sqrt{2}\pi}\left[1-\frac{\tilde{\omega}_{T}^{2}}{12}+O\left(\tilde{\omega}_{T}^{4}\right)\right].
\end{equation}
 The leading term in the above equation is universal, satisfying the
universal relation Eq. (\ref{UniversalRelationCn}). The second term
($\propto\tilde{\omega}_{T}^{2}$) is non-universal and is caused
by the length scale of the harmonic trap \cite{OurVE}. It represents
the finite-size correction to the local density approximation that
we have adopted above.

\subsection{Third contact coefficient}

The determination of the third contact coefficient is more cumbersome.
As in the calculation of the third virial coefficient, we can determine
firstly the trapped contact coefficient $c_{3,T}$, and then to use
the universal relations at low trap frequency to obtain the homogeneous
result, $c_{3}=3\sqrt{3}c_{3,T}$.

An estimate of $c_{3,T}$ can already be obtained by the known results
of $\Delta b_{3,T}$ as a function of the coupling constant $1/k_{F}a_{s}$
at different temperatures $T/T_{F}$ and $\tilde{\omega_{T}}\sim0.15$,
as shown in Fig. 8. This is simply because, 
\begin{equation}
c_{3,T}\equiv\frac{1}{k_{F}\lambda_{dB}}\frac{\partial\Delta b_{3,T}}{\partial(1/k_{F}a_{s})}=\sqrt{\frac{T}{4\pi T_{F}}}\frac{\partial\Delta b_{3,T}}{\partial(1/k_{F}a_{s})}.
\end{equation}
 We find that the coefficient $c_{3,T}$ at resonance is indeed nearly
temperature independent and estimate from the slope of $\Delta b_{3,T}$
that, $c_{3,T}($estimate$)\simeq-0.0265$ at $\tilde{\omega}_{T}\sim0.15$.
An accurate determination of $c_{3,T}$ requires a systematic extrapolation
to the limit of $\tilde{\omega_{T}}=0$. For this purpose, we calculate
numerically the derivative $c_{3,T}(\tilde{\omega}_{T})=[\partial\Delta b_{3,T}/\partial(\lambda_{dB}/a_{s})]_{\lambda_{dB}/a_{s}=0}\ $
as a function of $\tilde{\omega}_{T}$. Using the small $\tilde{\omega}_{T}$
data, a numerical extrapolation to $\tilde{\omega}_{T}=0$ gives rise
to the trapped third virial contact coefficient, $c_{3,T}\simeq-0.0271\pm0.0002$.
Thus, we obtain immediately from the universal relation, Eq. (\ref{UniversalRelationCn}),
the homogeneous third contact coefficient, 
\begin{equation}
c_{3}=-0.1408\pm0.0010.\label{c3}
\end{equation}

\begin{figure}[htp]
\begin{centering}
\includegraphics[clip,width=0.5\textwidth]{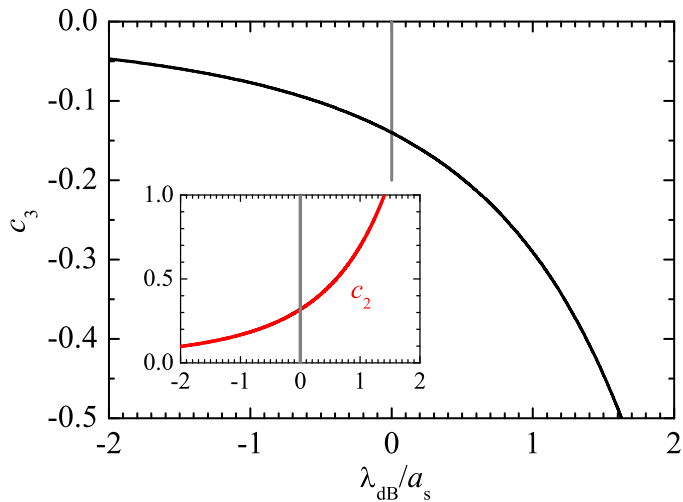} 
\par\end{centering}

\caption{(color online). Third contact coefficient as a function of the dimensionless
parameter $\lambda_{dB}/a_{s}$. The result is calculated using the
third virial coefficient reported by Leyronas \cite{LeyronasPreprint}.
The inset shows the second contact coefficient.}

\label{fig:homoC23} 
\end{figure}

Alternatively, we can determine the third contact coefficient by taking
a numerical derivative of the third virial coefficient $\Delta b_{3}(\lambda_{dB}/a_{s})$,
which was calculated recently by Leyronas \cite{LeyronasPreprint},
by using diagrammatic field theoretic method. As shown in Fig. \ref{fig:homoC23},
in the unitary limit we find that $c_{3}=-0.1399\pm0.0001$, in excellent
agreement with the result Eq. (\ref{c3}) from the exact three-particle
solutions.

\subsection{Large-$T$ contact: the homogeneous case}

We are now ready to calculate the universal contact in the high temperature
regime. For a homogeneous Fermi system, the single-particle partition
function $Q_{1}=2V/\lambda_{dB}^{3}$ and the dimensionless contact
${\cal I}/(Nk_{F})$ is given by, 
\begin{equation}
\frac{{\cal I}}{Nk_{F}}\equiv\frac{{\cal C}}{\rho k_{F}}=3\pi^{2}\left(\frac{T}{T_{F}}\right)^{2}\left[c_{2}z^{2}+c_{3}z^{3}+\cdots\right].\label{homContactVE}
\end{equation}
 Here $N\equiv\rho V$ is the total number of atoms with the homogeneous
density $\rho$.

The fugacity $z$ is determined by the number equation \cite{OurVELongPRA},
\begin{equation}
\tilde{\rho}=\tilde{\rho}^{(1)}\left(z\right)+\left[2\Delta b_{2}z^{2}+3\Delta b_{3}z^{3}+\cdots\right],\label{homNumVE}
\end{equation}
 where we have defined a dimensionless density $\tilde{\rho}\equiv\rho\lambda_{dB}^{3}/2=[4/(3\sqrt{\pi})](T_{F}/T)^{3/2}$
and the density of a non-interacting Fermi gas as $\tilde{\rho}^{(1)}(z)\equiv(2/\sqrt{\pi})\int_{0}^{\infty}dt\sqrt{t}/(1+z^{-1}e^{t})\,$.

In practice, for a given fugacity, we calculate the dimensionless
density using Eq. (\ref{homNumVE}) and hence the reduced temperature
$T/T_{F}$. The dimensionless contact is then obtained from Eq. (\ref{homContactVE}),
as a function of $T/T_{F}$ or the inverse fugacity $z^{-1}$.

\begin{figure}[htp]
\begin{centering}
\includegraphics[clip,width=0.5\textwidth]{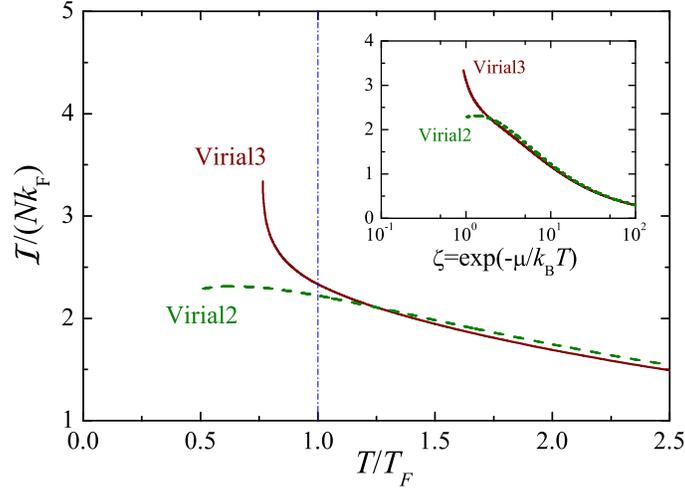} 
\par\end{centering}

\caption{(color online) Universal contact of a homogeneous Fermi gas in the
unitary limit at high temperatures, as predicted by the virial expansion
method up to the second order (dashed line) and the third order (solid
line). Dashed vertical line indicates the Fermi degenerate temperature
$T_{F}$. The inset shows the contact as a function of the inverse
fugacity. From ref. \cite{OurVirialContact}.}

\label{fig:homoContact} 
\end{figure}

Fig. \ref{fig:homoContact} reports the temperature (main figure)
or fugacity (inset) dependence of the homogeneous contact in the unitarity
limit, calculated by virial expanding to the second order (dashed
line) or third order (solid line). The close agreement between the
second and third predictions strongly indicates that the virial expansion
works {\em quantitatively} well down to the Fermi degenerate temperature
$T_{F}$, as indicated by the vertical dashed line. At sufficient
high temperatures, where 
\begin{equation}
z\simeq\tilde{\rho}=[4/(3\sqrt{\pi})](T_{F}/T)^{3/2}\,,\label{eq:homogeneous density}
\end{equation}
 the leading temperature dependence of the contact is given by, 
\begin{equation}
\frac{{\cal I}}{Nk_{F}}\left(T\gg T_{F}\right)=\frac{16}{3}\left(\frac{T}{T_{F}}\right)^{-1},
\end{equation}
 as predicted by Yu and co-workers \cite{YuPRA2009}. We note however
that the pre-factor there is smaller by a factor of $4\pi^{2}$, due
to a different definition for the contact.

\subsection{Large-$T$ contact: the trapped case}

For a trapped Fermi gas at unitarity, the dimensionless contact can
be written as, 
\begin{equation}
\frac{{\cal I}_{T}}{Nk_{F}}=24\pi^{^{3/2}}\left(\frac{T}{T_{F}}\right)^{7/2}\left[c_{2,T}z^{2}+c_{3,T}z^{3}+\cdots\right].\label{trapContactVE}
\end{equation}
 where $c_{2,T}=1/(2\sqrt{2}\pi)$ and $c_{3,T}=-0.02692\pm0.00002$.
The number equation takes the form \cite{OurVELongPRA}, 
\begin{equation}
\tilde{\rho}_{T}=\tilde{\rho}_{T}^{(1)}\left(z\right)+\left[2\Delta b_{2,T}z^{2}+3\Delta b_{3,T}z^{3}+\cdots\right],\label{trapNumVE}
\end{equation}
 where $\tilde{\rho}_{T}\equiv(N/2)(\hbar\omega_{T})^{3}/(k_{B}T)^{3}=(T_{F}/T)^{3}/6$
and the density of a non-interacting trapped Fermi gas $\tilde{\rho}_{T}^{(1)}(z)\equiv(1/2)\int_{0}^{\infty}dtt^{2}/(1+z^{-1}e^{t})$.
In analogy with the homogeneous case, for a given fugacity we determine
the reduced temperature $T/T_{F}$ from the number equation (\ref{trapNumVE})
and then calculate the trapped contact using Eq. (\ref{trapContactVE}).

\begin{figure}[htp]
\begin{centering}
\includegraphics[clip,width=0.5\textwidth]{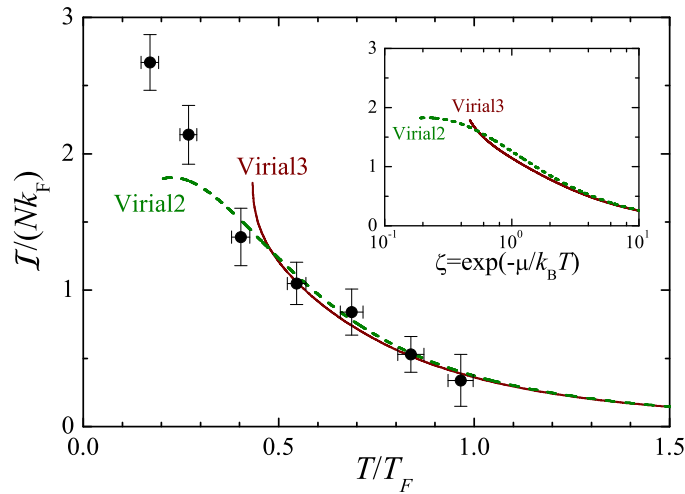} 
\par\end{centering}

\caption{(color online) Universal contact of a trapped Fermi gas in the unitary
limit at high temperatures, obtained by expanding the virial series
to the second order (dashed line) and the third order (solid line).
The inset shows the contact as a function of the inverse fugacity.
Adapted from ref. \cite{TanSwinburne2}; copyright (2011) by APS.}

\label{fig:trappedContact} 
\end{figure}

Fig. \ref{fig:trappedContact} presents the virial expansion prediction
for the trapped universal contact, expanding up to the second order
(dashed line) or third order (solid line). Amazingly, because of the
factor of $n^{-3/2}$ reduction for the $n$-th contact coefficient
in harmonic traps, the convergence of the expansion is much improved.
The expansion now seems to be quantitatively reliable down to $0.5T_{F}$.
The asymptotic behavior of the contact at very high temperatures can
be determined by setting 
\begin{equation}
z\simeq\tilde{\rho}_{T}=(T_{F}/T)^{3}/6\,.\label{eq:trapped density}
\end{equation}
 We find that, 
\begin{equation}
\left(\frac{{\cal I}}{Nk_{F}}\right)_{T}\left(T\gg T_{F}\right)=\frac{\sqrt{2\pi}}{6}\left(\frac{T}{T_{F}}\right)^{-5/2}.
\end{equation}
 Thus, the contact in harmonic traps decays at high temperatures much
faster than in homogeneous space, due to the reduction of the peak
density at the trap center at high temperatures.

The finite-temperature contact of a trapped Fermi gas in the unitary
limit was recently measured at Swinburne University of Technology.
Using the structure factor Tan relation Eq. (\ref{StructureFactorTanRelation}),
the contact was extracted from the static structure factor \cite{TanSwinburne2},
which has been measured by two-photon Bragg spectroscopy. In Fig.
\ref{fig:trappedContact}, the experimental result was shown in solid
circles. At $T>0.5T_{F}$, the data agree well with the virial prediction.

\section{Virial expansion of dynamic structure factor}

So far we consider the virial expansion of static properties of a
strongly correlated Fermi system. In the following, we show that dynamic
properties can be studied as well using virial expansion. This issue
is less explored in the literature. In this section, we consider the
dynamic density response of a strongly correlated Fermi system \cite{OurVirialDSF}.

\subsection{Dynamic structure factor}

The dynamic density response is characterized by the so-called dynamic
structure factor (DSF), which gives the linear response of the many-body
system to an excitation process that couples to density \cite{FetterBook}.
For ultracold atomic gases, it can be conveniently measured by two-photon
Bragg spectroscopy using two laser beams \cite{BraggSwinburne}. Theoretically,
it is difficult to predict DSF in the strongly interacting regime.
Traditional tools such as the perturbative random-phase approximation
(RPA) theory are in principle reliable in the weakly interacting limit
\cite{MinguzziEPJD2001,BruunPRL2001,CombescotEPL2006,ZouPRA2010}.

The DSF $S({\bf q},\omega)$ is the Fourier transform of the density-density
correlation functions at two different space-time points \cite{FetterBook,GriffinBook}.
For a balanced atomic Fermi gas with equal spin populations $N/2$
(referred to as spin-up, $\sigma=\uparrow$, and spin-down, $\sigma=\downarrow$),
$S_{\uparrow\uparrow}({\bf q},\omega)=S_{\downarrow\downarrow}({\bf q},\omega)$
and $S_{\uparrow\downarrow}({\bf q},\omega)=S_{\downarrow\uparrow}({\bf q},\omega)$,
each of which is defined by, 
\begin{equation}
S_{\sigma\sigma^{\prime}}({\bf q},\omega)=Q^{-1}\sum_{nn^{\prime}}e^{-E_{n^{\prime}}/k_{B}T}\left\langle n\left|\delta\rho_{\sigma}\left({\bf q}\right)\right|n^{\prime}\right\rangle \left\langle n^{\prime}\left|\delta\rho_{\sigma^{\prime}}^{\dagger}\left({\bf q}\right)\right|n\right\rangle \delta\left(\hbar\omega-E_{nn^{\prime}}\right).
\end{equation}
 Here $\left|n\right\rangle $ and $E_{nn^{\prime}}=E_{n}-E_{n^{\prime}}$
are, respectively, the eigenstate and eigenvalue of the many-body
system, while $Q=\sum_{n}\exp(-E_{n}/k_{B}T)$ is the partition function.
The density operator $\delta\hat{\rho}_{\sigma}({\bf q})=\sum_{i\sigma}e^{-i{\bf q\cdot x}_{i}}$
is the Fourier transform of the atomic density operator $\delta\hat{\rho}_{\sigma}\left({\bf r}\right)$
for spin-$\sigma$ atoms. The total DSF is given by $S({\bf q},\omega)\equiv2[S_{\uparrow\uparrow}({\bf q},\omega)+S_{\uparrow\downarrow}({\bf q},\omega)]$.
The DSF satisfies two remarkable $f$-sum rules \cite{GuoPRL2010},
\begin{equation}
\int_{-\infty}^{+\infty}S({\bf q},\omega)\omega d\omega=N\frac{\hbar{\bf q}^{2}}{2m}
\end{equation}
 and 
\begin{equation}
\int_{-\infty}^{+\infty}S_{\uparrow\downarrow}({\bf q},\omega)\omega d\omega=0,
\end{equation}
 which hold irrespective of interactions and temperatures.

According to the finite-temperature quantum field theory \cite{GriffinBook},
it is convenient to calculate DSF from dynamic susceptibility, $\chi_{\sigma\sigma^{\prime}}\left({\bf q},\tau\right)\equiv-\left\langle T_{\tau}\hat{\rho}_{\sigma}\left({\bf q},\tau\right)\hat{\rho}_{\sigma^{\prime}}\left({\bf q},0\right)\right\rangle $,
where $\tau$ is an imaginary time in the interval $0<\tau\leq\beta=1/k_{B}T$.
The Fourier component $\chi_{\sigma\sigma^{\prime}}\left({\bf q},i\omega_{n}\right)$
at discrete Matsubara imaginary frequencies $i\omega_{n}=i2n\pi k_{B}T$
($n=0,\pm1,...$) gives directly the DSF, after taking analytic continuation
and using the fluctuation-dissipation theorem: 
\begin{equation}
S_{\sigma\sigma^{\prime}}\left({\bf q,}\omega\right)=-\frac{\mathop{\rm Im}\chi_{\sigma\sigma^{\prime}}\left({\bf q};i\omega_{n}\rightarrow\omega+i0^{+}\right)}{\pi(1-e^{-\hbar\omega/k_{B}T})}\,\,\,.\label{fluct-dissi-theorem}
\end{equation}

The frequency integral of the DSF defines the so-called static structure
factor (SSF). For different spin components, we have, 
\begin{equation}
S_{\sigma\sigma^{\prime}}\left({\bf q}\right)=\frac{2}{N}\int_{-\infty}^{+\infty}S_{\sigma\sigma^{\prime}}\left({\bf q,}\omega\right)d\omega.
\end{equation}
 The total SSF is given by, $S\left({\bf q}\right)=(1/N)\int_{-\infty}^{+\infty}d\omega S\left({\bf q,}\omega\right)=S_{\uparrow\uparrow}({\bf q})+S_{\uparrow\downarrow}({\bf q})$.
As we mentioned earlier, the SSF is related to the two-body pair correlation
function $g_{\sigma\sigma^{\prime}}\left({\bf r}\right)$ \cite{FetterBook}
through a Fourier transform.

Experimentally, the DSF is measured by inelastic scattering experiments
of two-photon Bragg spectroscopy \cite{BraggSwinburne}. The atoms
are exposed to two laser beams with differences in wave-vector and
frequency. In a two-photon scattering event, atoms absorb a photon
from one of the beams and emit a photo into the other. Therefore,
the difference in the wave-vectors of the beams defines the momentum
transfer $\hbar{\bf q}$, while the frequency difference defines the
energy transfer $\hbar\omega$. In the regime of large transferred
momentum, which is exactly the case in current experiments for the
crossover Fermi gas \cite{BraggSwinburne}, the single-particle response
is dominant and peaks at the quasi-elastic resonance frequency $\omega_{res}=\hbar{\bf q}^{2}/(2M)$,
where $M$ is the mass of the elementary constituents of the system.
Therefore, we may anticipate that the Bragg response peaks at $\omega_{R}=\hbar{\bf q}^{2}/(2m)$
in the BCS limit and peaks at $\omega_{R,mol}=\hbar{\bf q}^{2}/(4m)=\omega_{R}/2$
in the BEC limit, since the underlying particles are respectively
free atoms ($M=m$)\ and molecules ($M=2m$).

\begin{figure}[htp]
\begin{centering}
\includegraphics[clip,width=0.8\textwidth]{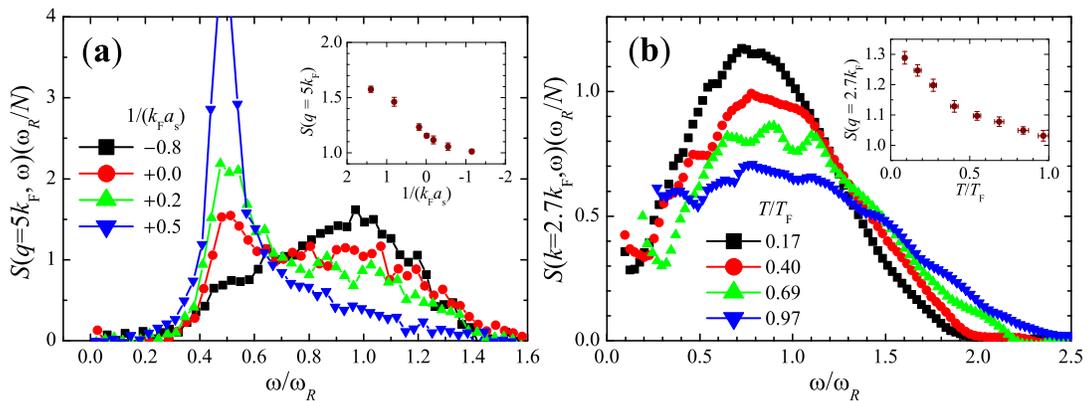} 
\par\end{centering}

\caption{(color online) (\textbf{a}) Measured dynamic structure factor of a
harmonically trapped Fermi gas in the BEC-BCS crossover at the lowest
attainable temperature ($T<0.1T_{F}$) and at a large transferred
wave-vector $q=5k_{F}$. The inset shows the static structure factor
as a function of the dimensionless interaction parameter. Adopted
from ref. \cite{BraggSwinburne} with permission; copyright (2008)
by APS. (\textbf{b}) Temperature dependence of dynamic structure factor
of a trapped Fermi gas in the unitary limit, measured at $q=2.7k_{F}$.
The inset shows the temperature dependence of static structure factor.
Adapted from ref. \cite{TanSwinburne2}; copyright (2011) by APS.}

\label{fig:BraggExperiments} 
\end{figure}

In Fig. \ref{fig:BraggExperiments}, we summarize the main experimental
results for a harmonically trapped Fermi gas in the BEC-BCS crossover
\cite{BraggSwinburne,TanSwinburne1,TanSwinburne2,TanSwinburne3}.
Fig. \ref{fig:BraggExperiments}a shows the DSF (main panel) and SSF
(inset) at several dimensionless interaction strengths and at the
lowest experimentally attainable temperature (i.e., $T<0.1T_{F}$,
where $T_{F}$ is the Fermi temperature) \cite{BraggSwinburne}, while
Fig. \ref{fig:BraggExperiments}b presents the temperature dependence
of structure factors in the most interesting unitary limit \cite{TanSwinburne2,TanSwinburne3}.
As anticipated, in Fig. \ref{fig:BraggExperiments}a the DSF peaks
at $\omega_{R}/2$ and $\omega_{R}$ on the BEC side (i.e., $1/(k_{F}a_{s})=+0.5$)
and on the BCS side ($1/(k_{F}a_{s})=-0.8$), respectively. In the
unitary limit, where the statistics of the elementary excitations
is not well defined, we observe a two-peak structure with responses
from both molecules and free-atoms. As the temperature increases (Fig.
\ref{fig:BraggExperiments}b), however, these two peaks merge. The
resultant broad peak shifts eventually to $\omega_{R}$ at high temperatures.

\subsection{Virial expansion of dynamic structure factor}

We construct first the virial expansion for the dynamic susceptibility
$\chi_{\sigma\sigma^{\prime}}\left({\bf x},{\bf x}^{\prime};\tau>0\right)$,
which is given by, 
\begin{equation}
\chi_{\sigma\sigma^{\prime}}\equiv-\frac{\text{Tr}\left[e^{-\left({\cal H}-\mu{\cal N}\right)/k_{B}T}e^{{\cal H}\tau}\hat{n}_{\sigma}\left({\bf x}\right)e^{-{\cal H}\tau}\hat{n}_{\sigma^{\prime}}\left({\bf x}^{\prime}\right)\right]}{\text{Tr}e^{-\left({\cal H}-\mu{\cal N}\right)/k_{B}T}}.
\end{equation}
 At high temperatures, Taylor-expanding in terms of the powers of
small fugacity $z\equiv\exp(\mu/k_{B}T)\ll1$ leads to $\chi_{\sigma\sigma^{\prime}}\left({\bf x},{\bf x}^{\prime};\tau\right)=(zX_{1}+z^{2}X_{2}+\cdots)/(1+zQ_{1}+z^{2}Q_{2}+\cdots)=zX_{1}+z^{2}\left(X_{2}-X_{1}Q_{1}\right)+\cdots$,
where we have introduced the cluster functions $X_{n}=-$ Tr$_{n}[e^{-{\cal H}/k_{B}T}e^{\tau{\cal H}}\hat{n}_{\sigma}({\bf x)}e^{-\tau{\cal H}}\hat{n}_{\sigma^{\prime}}({\bf x}^{\prime})]$
and $Q_{n}=$Tr$_{n}[e^{-{\cal H}/k_{B}T}]$, with $n$ denoting the
number of particles in the cluster and Tr$_{n}$ denoting the trace
over $n$-particle states of proper symmetry. We shall refer to the
above expansion as the virial expansion of dynamic susceptibilities,
$\chi_{\sigma\sigma^{\prime}}\left({\bf x},{\bf x}^{\prime};\tau\right)=z\chi_{\sigma\sigma^{\prime},1}\left({\bf x},{\bf r}^{\prime};\tau\right)+z^{2}\chi_{\sigma\sigma^{\prime},2}\left({\bf x},{\bf x}^{\prime};\tau\right)+\cdots,$
where, 
\begin{eqnarray}
\chi_{\sigma\sigma^{\prime},1}\left({\bf x},{\bf x}^{\prime};\tau\right) & = & X_{1},\nonumber \\
\chi_{\sigma\sigma^{\prime},2}\left({\bf x},{\bf x}^{\prime};\tau\right) & = & X_{2}-X_{1}Q_{1},\ \text{etc}.
\end{eqnarray}
 Accordingly, we shall write for the dynamic structure factors, 
\begin{equation}
S_{\sigma\sigma^{\prime}}\left({\bf q},\omega\right)=zS_{\sigma\sigma^{\prime},1}\left({\bf q},\omega\right)+z^{2}S_{\sigma\sigma^{\prime},2}\left({\bf q},\omega\right)+\cdots.
\end{equation}

\subsection{Trapped virial dynamic structure factor up to the second order}

The calculation of the $n$-th expansion coefficient requires the
knowledge of all solutions up to $n$-body, including both the eigenvalues
and eigenstates \cite{OurVE,OurVELongPRA}. Here we aim to calculate
the leading effect of interactions, which contribute to the 2nd-order
expansion function \cite{OurVirialDSF}. For this purpose, it is convenient
to define $\Delta\chi_{\sigma\sigma^{\prime},2}\equiv\left\{ \chi_{\sigma\sigma^{\prime},2}\right\} ^{(I)}=\left\{ X_{2}\right\} ^{(I)}$
and $\Delta S_{\sigma\sigma^{\prime},2}\equiv\left\{ S_{\sigma\sigma^{\prime},2}\right\} ^{(I)}$.
The notation $\left\{ {}\right\} ^{(I)}$ means the contribution due
to interactions inside the bracketed term, so that $\left\{ X_{2}\right\} ^{(I)}=X_{2}-X_{2}^{(1)}$,
where the superscript ``1'' in $X_{2}^{(1)}$ denotes quantities
for a noninteracting system. We note that the inclusion of the 3rd-order
expansion function is straightforward, though involving more numerical
effort.

It is easy to see that $\Delta\chi_{\sigma\sigma^{\prime},1}=0$,
according to the definition of notation $\left\{ {}\right\} ^{(I)}$.
To calculate the 2nd-order expansion function for the dynamic susceptibility,
$\Delta\chi_{\sigma\sigma^{\prime},2}=-\left\{ Tr_{\uparrow\downarrow}\left[e^{-{\cal H}/k_{B}T}e^{\tau{\cal H}}\hat{n}_{\sigma}\left({\bf x}\right)e^{-\tau{\cal H}}\hat{n}_{\sigma^{\prime}}\left({\bf x}^{\prime}\right)\right]\right\} ^{(I)}$,
we insert the identity $\sum_{Q}\left|Q\right\rangle \left\langle Q\right|={\bf \hat{1}}$
and take the trace over the state $P$, i.e., 
\begin{equation}
\Delta\chi_{\sigma\sigma^{\prime},2}=-\sum_{P,Q}\left\{ e^{-E_{P}/k_{B}T+\tau(E_{P}-E_{Q})}\left\langle P\left|\hat{n}_{\sigma}\right|Q\right\rangle \left\langle Q\left|\hat{n}_{\sigma^{\prime}}\right|P\right\rangle \right\} ^{(I)}.
\end{equation}
 Here, $P$ and $Q$ are the two-atom eigenstates with energies $E_{P}$
and $E_{Q}$, respectively. Expressing the density operator in the
first quantization: $\hat{n}_{\uparrow}\left({\bf x}\right)=\sum_{i}\delta\left({\bf x}-{\bf x}_{i\uparrow}\right)$
and $\hat{n}_{\downarrow}\left({\bf x}\right)=\sum_{j}\delta({\bf x}-{\bf x}_{j\downarrow})$,
it is straightforward to show that, 
\begin{equation}
\Delta\chi_{\sigma\sigma^{\prime},2}=-\sum_{P,Q}\left\{ e^{-E_{P}/k_{B}T+\tau\left(E_{P}-E_{Q}\right)}C_{\sigma\sigma^{\prime}}^{PQ}\left({\bf x},{\bf x}^{\prime}\right)\right\} ^{(I)},
\end{equation}
 where 
\begin{equation}
C_{\uparrow\uparrow}^{PQ}\equiv\int d{\bf x}_{2}d{\bf x}_{2}^{\prime}\left[\Phi_{P}^{*}\Phi_{Q}\right]\left({\bf x},{\bf x}_{2}\right)[\Phi_{Q}^{*}\Phi_{P}]\left({\bf x}^{\prime},{\bf x}_{2}^{\prime}\right)
\end{equation}
 and 
\begin{equation}
C_{\uparrow\downarrow}^{PQ}\equiv\int d{\bf x}_{1}d{\bf x}_{2}[\Phi_{P}^{*}\Phi_{Q}]\left({\bf x},{\bf x}_{2}\right)[\Phi_{Q}^{*}\Phi_{P}]\left({\bf x}_{1},{\bf x}^{\prime}\right).
\end{equation}
 The dynamic structure factor can be obtained by taking the analytic
continuation. This result is $\Delta S_{\sigma\sigma^{\prime},2}\left({\bf x},{\bf x}^{\prime};\omega\right)=\sum_{P,Q}\left\{ \delta\left(\omega+E_{P}-E_{Q}\right)e^{-\beta E_{P}}C_{\sigma\sigma^{\prime}}^{PQ}\left({\bf x},{\bf x}^{\prime}\right)\right\} ^{(I)}$.
Applying a further Fourier transform with respect to ${\bf r}={\bf x}-{\bf x}^{\prime}$
and integrating over ${\bf X}=({\bf x}+{\bf x}^{\prime})/2$, we obtain
the response $\Delta S_{\sigma\sigma^{\prime},2}\left({\bf q},\omega\right)$,
\begin{equation}
\Delta S_{\sigma\sigma^{\prime},2}=\sum_{P,Q}\left\{ \delta\left(\omega+E_{P}-E_{Q}\right)e^{-\beta E_{P}}\tilde{C}_{\sigma\sigma^{\prime}}^{PQ}\left({\bf q}\right)\right\} ^{(I)},
\end{equation}
 where $\tilde{C}_{\sigma\sigma^{\prime}}^{PQ}\left({\bf q}\right)=\int d{\bf x}d{\bf x}^{\prime}e^{-i{\bf q\cdot}({\bf x}-{\bf x}^{\prime})}C_{\sigma\sigma^{\prime}}^{PQ}\left({\bf x},{\bf x}^{\prime}\right)$.

The calculation of $C_{\sigma\sigma^{\prime}}^{PQ}\left({\bf x},{\bf x}^{\prime}\right)$
or $\tilde{C}_{\sigma\sigma^{\prime}}^{PQ}\left({\bf q}\right)$ is
straightforward but tedious, by using the two-atom solutions in an
isotropic harmonic trap $m\omega_{T}^{2}x^{2}/2$. We refer to ref.
\cite{OurVirialDSF} for details. The final result is given by, 
\begin{equation}
\Delta S_{\sigma\sigma^{\prime},2}=B\sqrt{\frac{m}{\pi}}\sum_{p2q2}\left\{ e^{-\frac{\left(\omega-\omega_{R}/2+\epsilon_{p2}-\epsilon_{q2}\right)^{2}}{2\omega_{R}k_{B}T}}e^{-\frac{\epsilon_{p2}}{k_{B}T}}A_{p2q2}^{\sigma\sigma^{\prime}}\right\} ^{(I)}\,\,,\label{ddsf2}
\end{equation}
 where $B\equiv(k_{B}T)^{5/2}/(q\hbar^{4}\omega_{0}^{3})$, and 
\begin{equation}
A_{p2q2}^{\sigma\sigma^{\prime}}=(-1)^{l(1-\delta_{\sigma\sigma^{\prime}})}(2l+1)\left[\int_{0}^{\infty}drr^{2}j_{l}\left(\frac{qr}{2}\right)\phi_{n_{p}l_{p}}\left(r\right)\phi_{n_{q}l_{q}}\left(r\right)\right]^{2}.\label{Arel}
\end{equation}
 In Eq. (\ref{Arel}), we specify $p2=\{n_{p}l_{p}\}$ and $q2=\{n_{q}l_{q}\}$,
and $l=\max\{l_{p},l_{q}\}$ for the two-atom relative radial wave
functions $\phi(r)$ with energy $\epsilon$. We require that either
$l_{p}$ or $l_{q}$ should be zero (i.e., $\min\{l_{p},l_{q}\}=0$),
otherwise $A_{p2q2}^{\sigma\sigma^{\prime}}$ will be cancelled exactly
by the non-interacting terms.

Together with the non-interacting DSF $S_{\sigma\sigma^{\prime}}^{\left(1\right)}$,
we calculate directly the interacting structure factor, 
\begin{equation}
S_{\sigma\sigma^{\prime}}({\bf q},\omega)=S_{\sigma\sigma^{\prime}}^{\left(1\right)}({\bf q},\omega)+z^{2}\Delta S_{\sigma\sigma^{\prime},2},
\end{equation}
 once the fugacity $z$ is determined by the virial expansion for
equation of state.

\subsubsection{Comparison of theory with the Swinburne experiment}

Considerable insight into the dynamic structure factor of a strongly
correlated Fermi gas can already be seen from Eq. (\ref{ddsf2}),
in which the spectrum is peaked roughly at $\omega_{R,mol}=\omega_{R}/2$,
the resonant frequency for molecules. Therefore, the peak is related
to the response of molecules with mass $M=2m$. Eq. (\ref{ddsf2})
shows clearly how the molecular response develops with the modified
two-fermion energies and wave functions as the interaction strength
increases. In the BCS limit where $\Delta S_{\sigma\sigma^{\prime},2}$
is small, the response is determined by the non-interacting background
$S_{\sigma\sigma^{\prime}}^{\left(1\right)}$ that peaks at $\omega_{R}$.
In the extreme BEC limit ($a\rightarrow0^{+}$), however, $\Delta S_{\sigma\sigma^{\prime},2}$
dominates. The sum in $\Delta S_{\sigma\sigma^{\prime},2}$ is exhausted
by the (lowest) tightly bound state $\phi_{rel}(r)\simeq\sqrt{2/a_{s}}e^{-r/a_{s}}$
with energy $\epsilon_{rel}\simeq E_{B}\equiv-\hbar^{2}/(ma_{s}^{2})$.
The chemical potential of molecules is given by $\mu_{m}=2\mu-E_{B}$.
Therefore, the DSF of fermions takes the form, 
\begin{equation}
S_{\sigma\sigma^{\prime}}^{BEC}\simeq z_{m}B\sqrt{\frac{M}{\pi}}\exp\left[-\frac{(\omega-\omega_{R,mol})^{2}}{4\omega_{R,mol}k_{B}T}\right],\label{dsfbec}
\end{equation}
 where $z_{m}=e^{\mu_{m}/k_{B}T}$ is the molecular fugacity. This
peaks at the molecular resonant energy. As anticipated, Eq. (\ref{dsfbec})
is exactly the leading virial expansion term in the DSF of non-interacting
molecules. It is clear that $S_{\uparrow\uparrow}({\bf q},\omega)\simeq S_{\uparrow\downarrow}({\bf q},\omega)$
in the BEC limit, since the spin structure in a single molecule can
no longer be resolved.

\begin{figure}[htp]
\begin{centering}
\includegraphics[clip,width=0.8\textwidth]{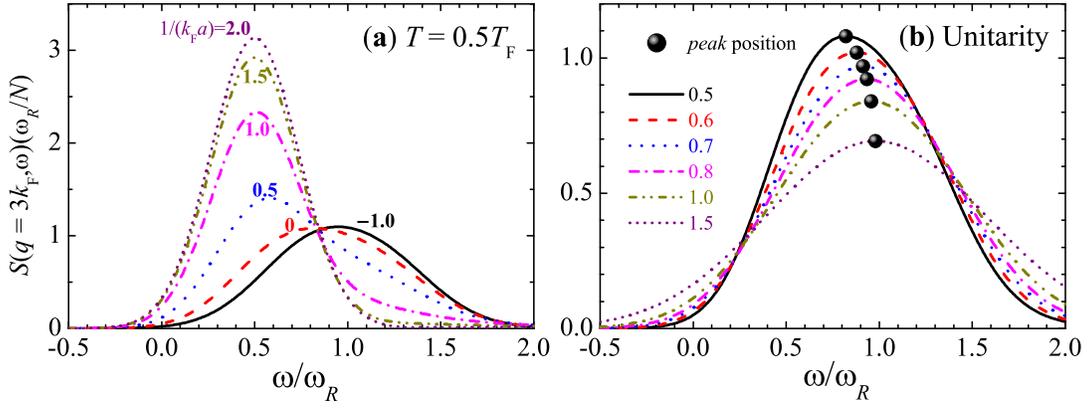} 
\par\end{centering}

\caption{(color online) (\textbf{a}) Evolution of dynamic structure factor
of a trapped Fermi gas in the BEC-BCS crossover with increasing interaction
strength $1/(k_{F}a_{s})$ at $T=0.5T_{F}$. (\textbf{b}) Temperature
dependence of dynamic structure factor of a trapped unitary Fermi
gas. The dark circles indicate the peak position of spectra. The transferred
wave-vector is $q=3k_{F}$. Adapted from ref. \cite{OurVirialDSF};
copyright (2010) by APS.}

\label{fig:virialDSF} 
\end{figure}

To understand the intermediate regime, in Fig. \ref{fig:virialDSF}a
we report numerical results for the DSF as the interaction strength
increases from the BCS to BEC regimes at $T=0.5T_{F}$ \cite{OurVirialDSF}.
The temperature dependence of the DSF in the unitary limit is shown
in Fig. \ref{fig:virialDSF}b \cite{OurVirialDSF}. In a trapped gas
with total number of fermions $N$, we use the zero temperature Thomas-Fermi
wave vector $k_{F}=(24N)^{1/6}/a_{ho}$ and temperature $T_{F}=(3N)^{1/3}\hbar\omega_{T}/k_{B}$
as characteristic units. In accord with the experiment \cite{BraggSwinburne,TanSwinburne2},
we take a large transferred momentum of $q=3k_{F}$. At $T=0.5T_{F}$,
A smooth transition from atomic to molecular responses is evident
as the interaction parameter $1/(k_{F}a_{s})$ increases, in qualitative
agreement with the experimental observation (c.f. Fig. \ref{fig:BraggExperiments}a).
In the unitary limit, the peak of total DSF shifts towards the molecular
recoil frequency, as indicated by the dark circles. This red-shift
is again in qualitative agreement with experiment (c.f. Fig. \ref{fig:BraggExperiments}b).

\begin{figure}[htp]
\begin{centering}
\includegraphics[clip,width=0.4\textwidth]{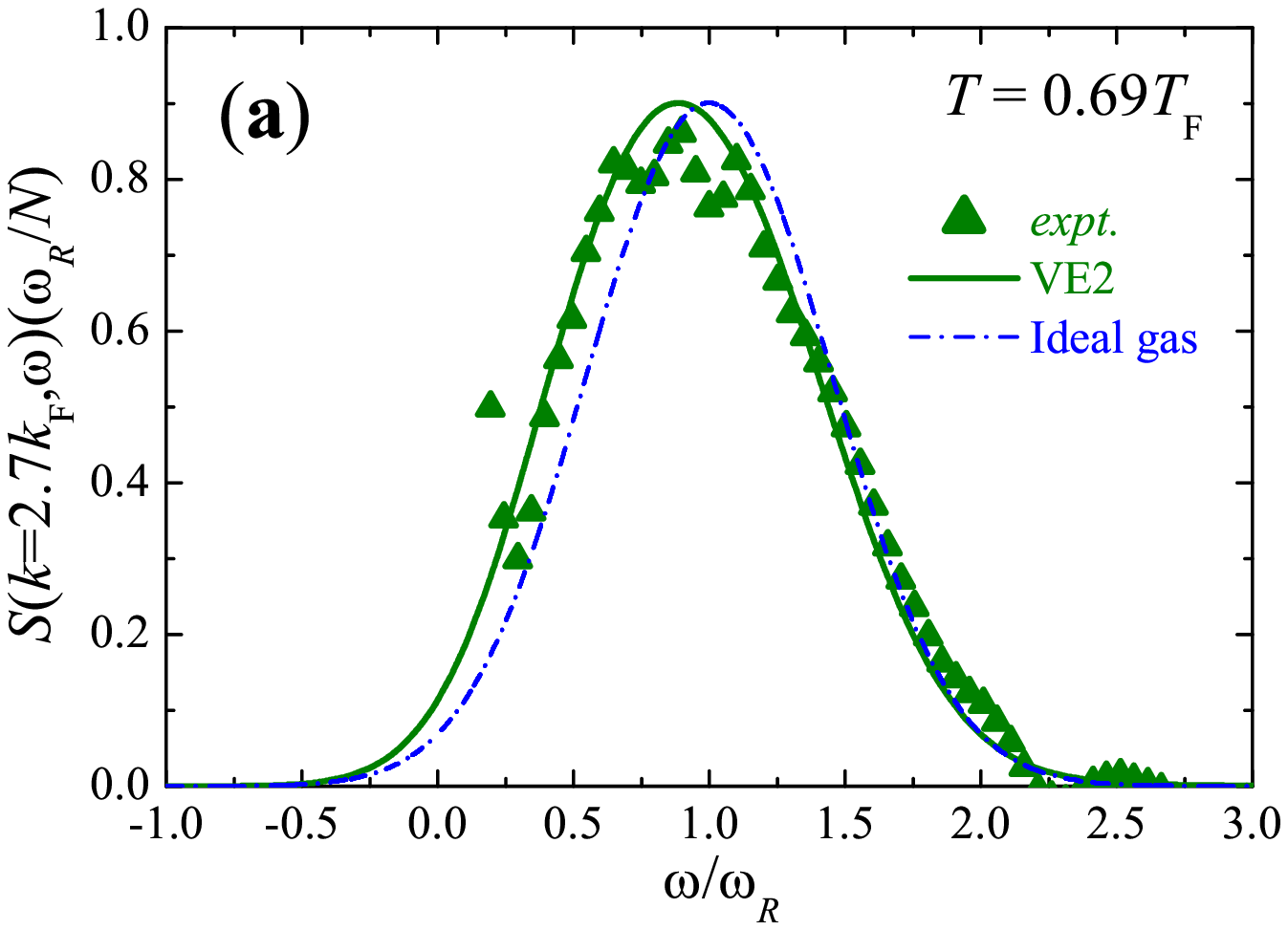}
\includegraphics[clip,width=0.4\textwidth]{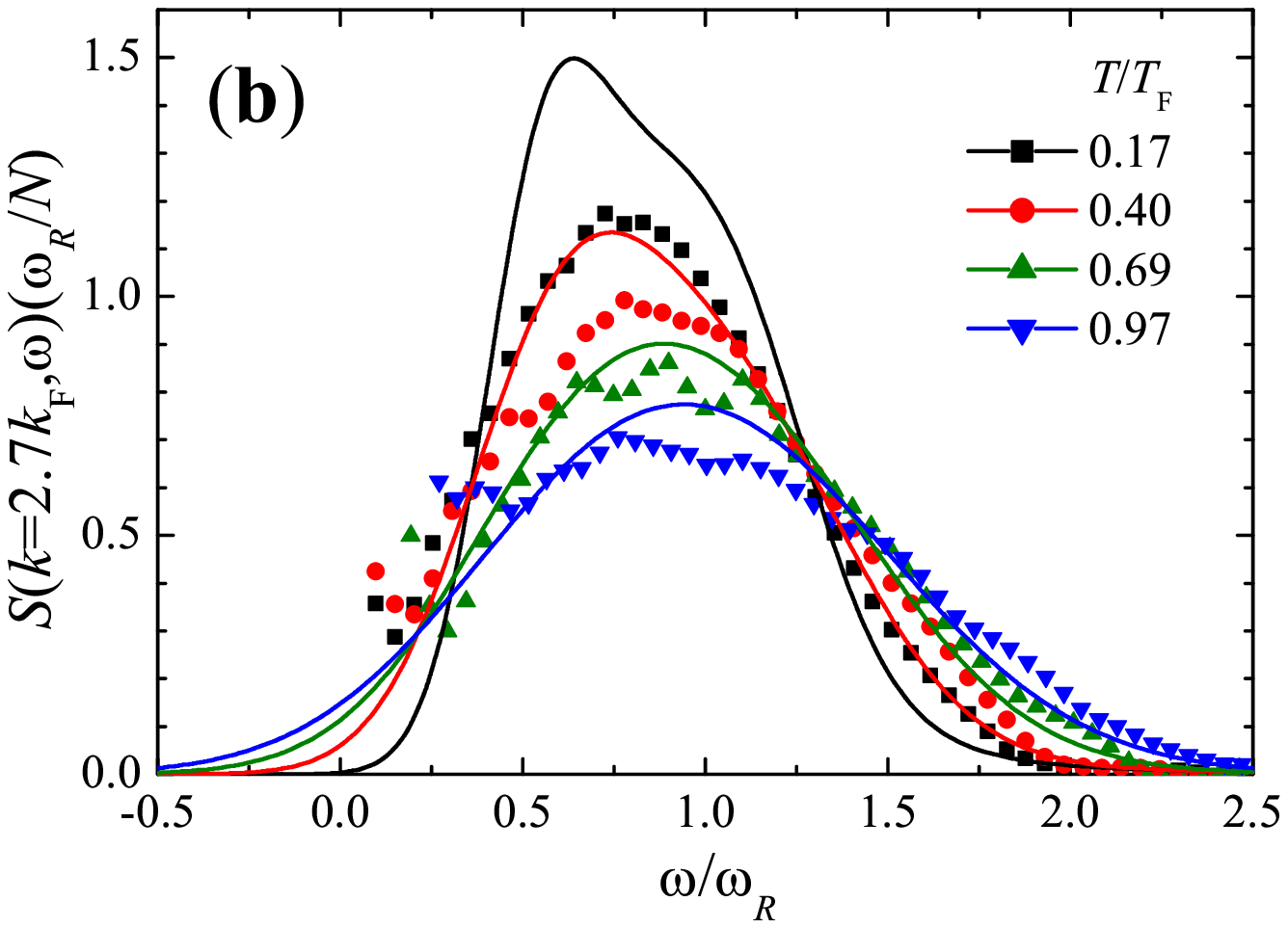}
\par\end{centering}

\caption{(color online) Comparison between theory and experiment for the dynamic
structure factor of a trapped unitary Fermi gas at finite temperatures.
Here, the transferred wave-vector is $q=2.7k_{F}$. Adapted from ref.
\cite{TanSwinburne3}; copyright (2011) by NJP.}

\label{fig:virialDSFComparison} 
\end{figure}

For a close comparison, we plot in Fig. \ref{fig:virialDSFComparison}
the virial expansion prediction and experimental data for the DSF
at several temperatures in the unitary limit. The theory is in very
good agreement with experimental data at high temperatures (see, for
example, the case of $T=0.69T_{F}$ in Fig. \ref{fig:virialDSFComparison}a)
\cite{TanSwinburne2,TanSwinburne3}, where the fugacity $z$ is less
than $1$. Towards low temperatures, the agreement becomes worse.
However, the virial expansion does capture the qualitative feature
of the DSF, for temperature down to the onset of superfluid transition,
$T_{c}\sim0.2T_{F}$.

\subsection{Homogeneous virial dynamic structure factor up to the second order}

Let us now consider the expansion functions of a {\em homogeneous}
Fermi gas in the unitarity limit. This can be extracted from the trapped
expansion function because of fermionic universality in the unitary
limit. As the scattering length diverges, all microscopic scales of
the interaction are absent \cite{HoUniversality}. For this few-body
problem, the only energy scale is $k_{B}T$ and length scale is the
thermal de Broglie wavelength $\lambda_{dB}$. Dimensional analysis
leads to, 
\begin{equation}
\Delta S_{\sigma\sigma^{\prime},n}({\bf q},\omega,T)=\frac{V}{k_{B}T\lambda_{dB}^{3}}\Delta\tilde{S}_{\sigma\sigma^{\prime},n}(\tilde{q},\tilde{\omega}),
\end{equation}
 where $V$ is the volume, $\tilde{q}=[\hbar^{2}{\bf q}^{2}/(2mk_{B}T)]^{1/2}$,
$\tilde{\omega}=\hbar\omega/(k_{B}T)$, and $\Delta\tilde{S}_{\sigma\sigma^{\prime},n}$
is a dimensionless expansion function. The temperature $T$ is now
implicit in the variables $\tilde{q}$ and $\tilde{\omega}$. This
universal form implies a simple relation between the trapped and homogeneous
expansion function. In a shallow harmonic trap, $V_{T}({\bf x})=m\omega_{T}^{2}(x^{2}+y^{2}+z^{2})/2$,
where $\omega_{T}\rightarrow0$, the system may be viewed as a collection
of many cells with a local chemical potential $\mu({\bf r})=\mu-V_{T}({\bf r})$
and fugacity $z(r)=z\exp[-V_{T}({\bf r})/k_{B}T]$, so that the trapped
DSF is given by $\Delta S_{\sigma\sigma^{\prime},T}\left({\bf q},\omega,T\right)=\int d{\bf r[}\Delta S_{\sigma\sigma^{\prime}}\left({\bf q},\omega,T,{\bf r}\right)/V]$.
Owing to the universal $\tilde{q}$- and $\tilde{\omega}$-dependence
in the expansion functions, the spatial integration can be easily
performed, giving rise to 
\begin{equation}
\Delta\tilde{S}_{\sigma\sigma^{\prime},n}(\tilde{q},\tilde{\omega})=n^{3/2}\frac{\left(\hbar\omega_{T}\right)^{3}}{\left(k_{B}T\right)^{2}}\Delta S_{\sigma\sigma^{\prime},n,T}({\bf q},\omega,T).\label{UniversalRelationDSF}
\end{equation}
 The (non-universal) correction to the above local density approximation
is at the order of $O[(\hbar\omega_{T})^{2}/(k_{B}T)^{2}]$. Eq. (\ref{UniversalRelationDSF})
is vitally important because the calculation of expansion functions
in harmonic traps is much easier than in free space.

\begin{figure}[htp]
\begin{centering}
\includegraphics[clip,width=0.5\textwidth]{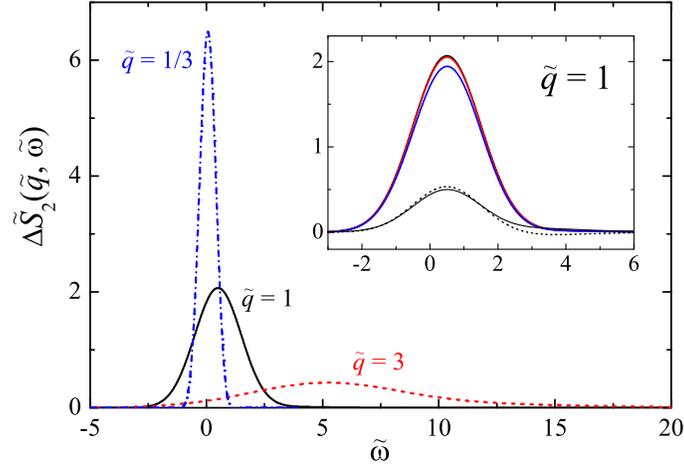} 
\par\end{centering}

\caption{(color online) Universal second order expansion function of DSF at
$\tilde{q}=1/3$, $1$, and $3$. The inset shows the rapid convergence
of $\Delta\tilde{S}_{2}(\tilde{q},\tilde{\omega})$ at small $\hbar\omega_{T}/k_{B}T$
(thick lines) and, $\Delta\tilde{S}_{\uparrow\uparrow,2}$ (thin solid
line) and $\Delta\tilde{S}_{\uparrow\downarrow,2}$ (thin dashed line)
at $\tilde{q}=1$. From ref. \cite{HLPreprint}.}

\label{fig:homoDeltaDSF} 
\end{figure}

\begin{figure}[htp]
\begin{centering}
\includegraphics[clip,width=0.6\textwidth]{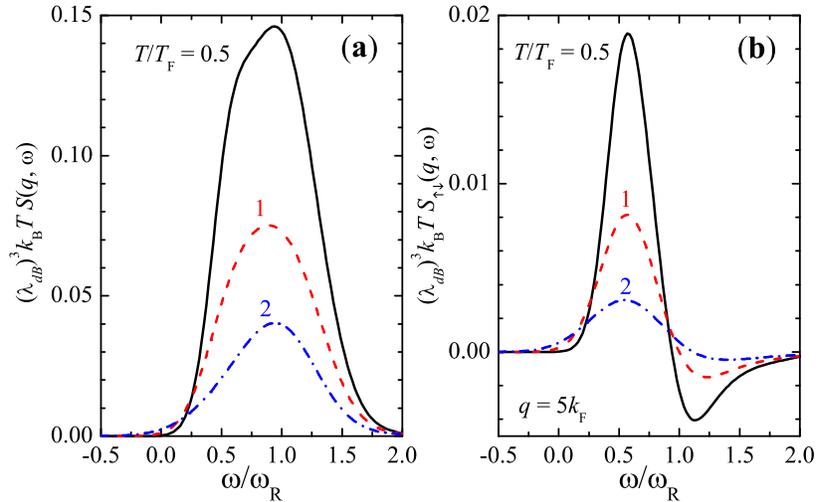} 
\par\end{centering}

\caption{(color online) Homogeneous dynamic structure factors $S(q,\omega)$
and $S_{\uparrow\downarrow}(q,\omega)$ at $q=5k_{F}$ and different
temperatures, calculated up to the second order.}

\label{fig:homoDSF} 
\end{figure}

Fig. \ref{fig:homoDeltaDSF} reports the homogeneous expansion function
$\Delta\tilde{S}_{2}=2[\Delta\tilde{S}_{\uparrow\uparrow,2}+\Delta\tilde{S}_{\uparrow\downarrow,2}]$
at three different momenta, using $\Delta S_{\sigma\sigma^{\prime},2,T}$
in Ref. \cite{OurVirialDSF} as the input. One observes a quasielastic
peak at $\tilde{\omega}=\tilde{q}^{2}/2$ or $\omega=\hbar{\bf q}^{2}/(4m)$,
as a result of the formation of fermionic pairs. In Fig. \ref{fig:homoDSF},
we show the total homogeneous dynamic structure factor at $q=5k_{F}$,
calculated up to the second order.

\subsubsection{The f-sum rules}

We may derive sum rules that constrain the expansion functions, using
the well-known \textit{f}-sum rules satisfied by DSF. Using the virial
expansion of the total number of fermions $N$, we shall have $f$-sum
relations 
\begin{equation}
\int\nolimits _{-\infty}^{+\infty}\tilde{\omega}\Delta\tilde{S}_{\uparrow\uparrow,n}(\tilde{q},\tilde{\omega})d\tilde{\omega}=n\tilde{q}^{2}\Delta b_{n}
\end{equation}
 and 
\begin{equation}
\int\nolimits _{-\infty}^{+\infty}\tilde{\omega}\Delta\tilde{S}_{\uparrow\downarrow,n}(\tilde{q},\tilde{\omega})d\tilde{\omega}=0,
\end{equation}
 which hold for {\em arbitrary} transferred momentum.

\subsubsection{Virial and contact coefficients from the large-$q$ expansion functions}

At large momentum, the spin-antiparallel static structure factor satisfies
the structure factor Tan relation Eq. (\ref{StructureFactorTanRelation}),
$\int S_{\uparrow\downarrow}({\bf q},\omega,T)d\omega\simeq{\cal I}/(8\hbar q)$.
By virial expanding both sides of the equation, we find that, 
\begin{equation}
\Delta\tilde{S}_{\uparrow\downarrow,n}(\tilde{q}\gg1)\equiv\int\nolimits _{-\infty}^{+\infty}\Delta\tilde{S}_{\uparrow\downarrow,n}(\tilde{q},\tilde{\omega})d\tilde{\omega}=\frac{\pi^{3/2}c_{n}}{\tilde{q}}.\label{contactSumRule}
\end{equation}
 On the other hand, in the same limit of large momentum, the spin-parallel
static structure factor is nearly unity so that $\int S_{\uparrow\uparrow}({\bf q},\omega,T)d\omega\simeq N/(2\hbar)$
\cite{OurVirialDSF,TanSwinburne1}. This leads to 
\begin{equation}
\Delta\tilde{S}_{\uparrow\uparrow,n}(\tilde{q}\gg1)\equiv\int\nolimits _{-\infty}^{+\infty}\Delta\tilde{S}_{\uparrow\uparrow,n}(\tilde{q},\tilde{\omega})d\tilde{\omega}=n\Delta b_{n}.
\end{equation}
 For the second expansion function, $\Delta\tilde{S}_{\sigma\sigma^{\prime},2}$,
we have checked numerically that all the above mentioned sum rules
are strictly satisfied.

\section{Virial expansion of single-particle spectral function}

In this section, we present the virial expansion of single-particle
spectral function, a quantity that plays a key role in understanding
the nature of pairing in strongly correlated Fermi gases. It has been
argued that there might be a small window for pseudogap - the precursor
of fermionic pairing in the normal state above the superfluid transition
temperature - in analogy with high-$T_{c}$ superconductors \cite{ChenPhysicsReport2005}.
However, its unambiguous identification is still under debate. Some
of strong-coupling theories predict a pseudogap \cite{ChenPRL2009,TsuchiyaPRA2009,WatanabePRA2010,DomanskiPRA2011,MuellerPRA2011},
while some others claim no such effects \cite{HaussmannPRA2009}.
Ab-initio quantum Monte Carlo simulations of the spectral function
have been performed \cite{MagierskiPRL2009,MagierskiPRL2011}, but
the accuracy is yet to be improved. To date, the experimental measurements,
through the momentum-resolved rf spectroscopy \cite{rfJILANature},
were not conclusive, though there is a weak indication of pseudogap
\cite{rfJILANaturePhysics}. Here, we show that one can use the virial
expansion up to the second order to qualitatively understand the experimental
results \cite{OurVirialAkw}. Further improvements of virial expansion
might be useful to solve the delicate pseudogap problem.

\subsection{Virial expansion of single-particle spectral function}

To virial expand the single-particle spectral function, let us consider
the related finite-temperature Green function at different space-time
points \cite{OurVirialAkw}, 
\begin{equation}
G_{\sigma\sigma^{\prime}}\left({\bf x},{\bf x}^{\prime};\tau\right)\equiv-\frac{\text{Tr}\left[e^{-\left({\cal H}-\mu{\cal N}\right)/k_{B}T}\hat{\Psi}_{\sigma}\left({\bf x,}\tau\right)\hat{\Psi}_{\sigma^{\prime}}^{\dagger}\left({\bf x}^{\prime}\right)\right]}{\text{Tr}e^{-\left({\cal H}-\mu{\cal N}\right)/k_{B}T}},
\end{equation}
 where at finite temperatures we are working with an imaginary time
$\tau$ in the interval $0<\tau\leq\beta=1/k_{B}T$. At high temperatures,
both numerator and denominator may be expanded into the powers of
$z\ll1$, leading to $G_{\sigma\sigma^{\prime}}\left({\bf r},{\bf r}^{\prime};\tau\right)=(X_{0}+zX_{1}+\cdots)/(1+zQ_{1}+\cdots)=X_{0}+z\left(X_{1}-X_{0}Q_{1}\right)+\cdots$,
where $X_{n}=-$ Tr$_{n}[e^{-{\cal H}/k_{B}T}\hat{\Psi}_{\sigma}\left({\bf x,}\tau\right)\hat{\Psi}_{\sigma^{\prime}}^{\dagger}({\bf x}^{\prime})]$
is the expansion function and $Q_{n}=$Tr$_{n}[e^{-{\cal H}/k_{B}T}]$
is the cluster partition function. The above expansion is to be referred
to as the virial expansion of Green function, $G_{\sigma\sigma^{\prime}}\left({\bf x},{\bf x}^{\prime};\tau\right)=G_{\sigma\sigma^{\prime},0}\left({\bf x},{\bf x}^{\prime};\tau\right)+zG_{\sigma\sigma^{\prime},1}\left({\bf x},{\bf x}^{\prime};\tau\right)+\cdots,$
where, 
\begin{equation}
G_{\sigma\sigma^{\prime},0}\left({\bf x},{\bf x}^{\prime};\tau\right)=X_{0},\quad G_{\sigma\sigma^{\prime},1}\left({\bf x},{\bf x}^{\prime};\tau\right)=X_{1}-X_{0}Q_{1},\ \text{etc}.
\end{equation}
 We then take the Fourier transformation with respect to $\tau$,
to obtain $G_{\sigma\sigma^{\prime}}({\bf x,x}^{\prime};i\omega_{n})$.
The experimentally measured spectral function $A\left({\bf k},\omega\right)$
can be calculated from the finite-temperature Green function via analytic
continuation, 
\begin{equation}
A_{\sigma\sigma^{\prime}}\left({\bf x},{\bf x}^{\prime};\omega\right)=-\frac{1}{\pi}\mathop{\rm Im}G_{\sigma\sigma^{\prime}}\left({\bf x},{\bf x}^{\prime};i\omega_{n}\rightarrow\omega+i0^{+}\right).
\end{equation}
 A final Fourier transform on ${\bf x}-{\bf x}^{\prime}$ leads to
$A_{\sigma\sigma^{\prime}}({\bf k},\omega)$, as measured experimentally.
For a normal, balanced Fermi gas, $A_{\uparrow\uparrow}=A_{\downarrow\downarrow}$
$\equiv A({\bf k},\omega)$ and $A_{\uparrow\downarrow}=0$. In accord
with the virial expansion of Green function, we may write the spectral
function, 
\begin{equation}
A\left({\bf k},\omega\right)=A_{0}\left({\bf k},\omega\right)+zA_{1}\left({\bf k},\omega\right)+\cdots.
\end{equation}
 The calculation of the $n$-th expansion function $G_{n}\left({\bf x},{\bf x}^{\prime};\tau\right)$
or $A_{n}\left({\bf k},\omega\right)$ requires the knowledge of solutions
up to the $n$-body problem, including both energy levels and wavefunctions.

As before, in the calculations of the Green function or spectral function,
it is convenient to separate out the contribution arising from interactions.
To this aim, for any physical quantity ${\cal Q}$ we may write ${\cal Q}=\{{\cal Q}\}^{(I)}+{\cal Q}^{(1)}$,
where the superscript ``1'' in ${\cal Q}^{(1)}$ denotes the part
of a non-interacting system having the {\em same} fugacity. The
operator $\{\}^{(I)}$ then picks up the residues due to interactions.
We then may write, 
\begin{equation}
G\left({\bf x},{\bf x}^{\prime};\tau\right)=\left\{ G\left({\bf x},{\bf x}^{\prime};\tau\right)\right\} ^{\left(I\right)}+G^{\left(1\right)}\left({\bf x},{\bf x}^{\prime};\tau\right),
\end{equation}
 where $\{G\}^{(I)}$ can be expanded in terms of $\{X_{n}\}^{(I)}$.

\subsection{Trapped virial spectral function up to the second order}

We now calculate the second-order expansion function, which accounts
for the leading interaction effect. The next-order expansion function,
could be treated straightforward using exact three-fermion solutions
\cite{OurVELongPRA}. The leading term of $\{G_{\uparrow\uparrow}({\bf x},{\bf x}^{\prime};\tau)\}^{(I)}$
takes the form, 
\begin{equation}
-ze^{\mu\tau}\left\{ \text{Tr}_{1}\left[e^{-{\cal H}/k_{B}T}e^{\tau{\cal H}}\hat{\Psi}_{\uparrow}\left({\bf x}\right)e^{-\tau{\cal H}}\hat{\Psi}_{\uparrow}^{+}\left({\bf x}^{\prime}\right)\right]\right\} ^{\left(I\right)}.
\end{equation}
 The trace has to be taken over all the single-particle states (i.e.,
$\psi_{p}$ with energy $\epsilon_{p}$) for a spin-down fermion.
We insert in the bracket an identity $\sum_{Q}\left|Q\right\rangle \left\langle Q\right|={\bf \hat{1}}$,
where $Q$ refers to the ``paired'' state (i.e., $\Phi_{Q}$ with
energy $E_{Q}$) for two fermions with {\em unlike} spins. It is
straightforward to show that, at the leading order, 
\begin{equation}
\{G_{\uparrow\uparrow}\}^{(I)}=-ze^{\mu\tau}\sum_{p,Q}\left\{ e^{-\epsilon_{p}/k_{B}T+\tau\left(\epsilon_{p}-E_{Q}\right)}F_{pQ}\left({\bf x,x}^{\prime}\right)\right\} ^{\left(I\right)},
\end{equation}
 where $F_{pQ}\equiv\int d{\bf x}_{1}d{\bf x}_{2}\psi_{p}^{*}({\bf x}_{1})\Phi_{Q}\left({\bf x},{\bf x}_{1}\right)\Phi_{Q}^{*}\left({\bf x}^{\prime},{\bf x}_{2}\right)\psi_{p}({\bf x}_{2})$.
Accordingly, the leading interaction correction to the spectral function,
$\{A\left({\bf k},\omega\right)\}^{(I)}$, is given by, 
\begin{equation}
z\left(1+e^{-\hbar\omega/k_{B}T}\right)\sum_{p,Q}\left\{ \delta\left(\omega+\epsilon_{p}-E_{Q}+\mu\right)e^{-\epsilon_{p}/k_{B}T}\left|\tilde{F}_{pQ}\right|^{2}\right\} ^{(I)},
\end{equation}
 where $\tilde{F}_{pQ}({\bf k})\equiv\int d{\bf x}d{\bf x}_{1}e^{-i{\bf k}\cdot{\bf r}}\psi_{p}^{*}({\bf x}_{1})\Phi_{Q}\left({\bf x},{\bf x}_{1}\right)$.

In an isotropic harmonic trap with frequency $\omega_{T}$, we can
solve exactly the two-fermion problem for relative wavefunctions \cite{OurVELongPRA}
and obtain $\{A\left({\bf k},\omega\right)\}^{(I)}$ using the above
procedure. In the end, we calculate 
\begin{equation}
I({\bf k},\omega)=\frac{k^{2}}{2\pi^{2}}\left[\left\{ A\right\} ^{(I)}f_{F}\left(\omega\right)+A^{(1)}({\bf k},\omega)f_{F}\left(\omega\right)\right],
\end{equation}
 as measured experimentally \cite{rfJILANature,rfJILANaturePhysics}.
Here, $f_{F}\left(\omega\right)=1/(e^{\hbar\omega/k_{B}T}+1)$ is
the Fermi distribution function and 
\begin{equation}
A^{(1)}=4\sqrt{2}\pi/(m^{3/2}\omega_{T}^{3})\left(\omega+\mu-\epsilon_{{\bf k}}\right)^{1/2},
\end{equation}
is the spectral function of an ideal, non-interacting Fermi gas. To
account for the experimental resolution, we may further convolute
$I({\bf k},\omega)$ with a gaussian broadening curve.

In the BEC limit, we may show analytically that, 
\begin{equation}
\left\{ A\right\} ^{(I)}f_{F}\propto\exp\left[-\beta\left(\sqrt{\epsilon_{{\bf k}}-\omega-\mu+E_{B}}-\sqrt{\epsilon_{{\bf k}}}\right)^{2}\right],
\end{equation}
 where $E_{B}=-\hbar^{2}/(ma_{s}^{2})$ is the binding energy. Thus,
at large momentum $k$ the intensity due to interactions peaks at
$\omega+\mu=-\epsilon_{{\bf k}}+E_{B}$, with a width $\sim\sqrt{k_{B}T\epsilon_{{\bf k}}}$.
At low temperatures, the width should be replaced by $\sqrt{E_{F}\epsilon_{{\bf k}}}$,
where the Fermi energy $E_{F}$ provides a cut-off to the thermal
energy $k_{B}T$.

We may also calculate the momentum distribution $\rho_{\sigma}({\bf k})=\int_{-\infty}^{+\infty}d\omega A({\bf k},\omega)f_{F}(\omega)$.
At large momentum, we confirm the Tan relation \cite{TanRelations},
$\rho_{\sigma}({\bf k})\simeq{\cal I}/k^{4}$, where the contact ${\cal I}$
is given by, 
\begin{equation}
{\cal I}=4\pi z^{2}\left(\frac{k_{B}T}{\hslash\omega_{T}}\right)^{3}\sum_{n}e^{-\epsilon_{rel,n}/k_{B}T}\phi_{rel,n}^{2}\left(0\right).
\end{equation}
 Here, $\phi_{rel}$ is the relative radial wavefunction of the paired
state with energy $\epsilon_{rel}$ \cite{OurVELongPRA}. At low temperatures,
a finite contact therefore implies a finite spectral weight below
the chemical potential.

In the calculation, consistent with the leading order expansion in
$A\left({\bf k},\omega\right)$, we determine the fugacity from the
number equation $N=-\partial[\Delta\Omega+\Omega^{(1)}]/\partial\mu$,
by expanding the interacting part of thermodynamic potential, $\Delta\Omega,$
up to the second-order virial coefficient.

\begin{figure}[htp]
\begin{centering}
\includegraphics[clip,width=0.6\textwidth]{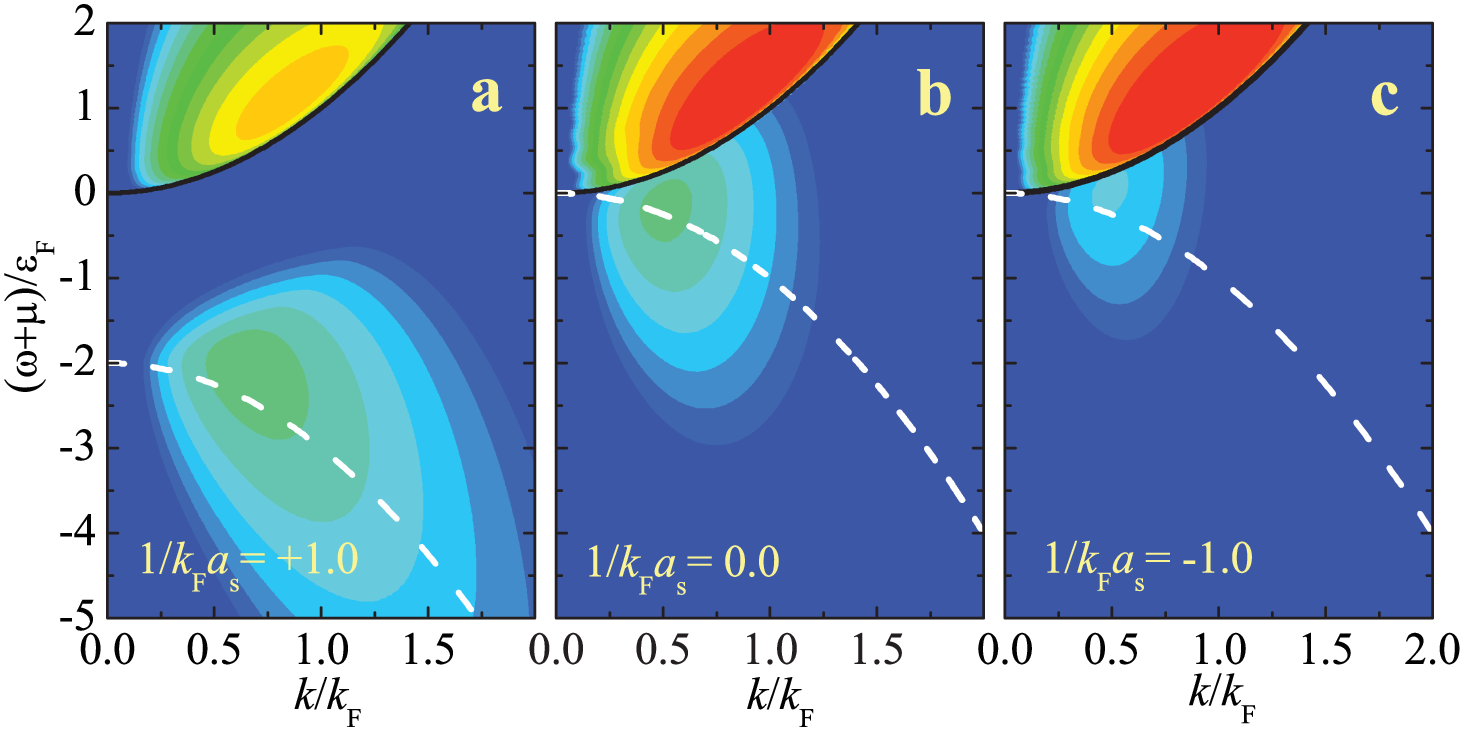} 
\par\end{centering}

\caption{(color online) Contour plots of the occupied spectral intensity at
crossover. The intensity $I(\omega)=A({\bf k},\omega)f_{F}(\omega)k^{2}/(2\pi^{2})$
increases from blue ($10^{-3}I_{max}$) to red ($I_{max}$) in a logarithmic
scale. The calculations were performed with harmonic traps at $T=0.7T_{F}$
and $1/(k_{F}a_{s})=+1$, $0$, $-1$, with a resulting fugacity at
the trap center of $z\simeq0.14$, $0.42$, and $0.48$, respectively.
From ref. \cite{OurVirialAkw}; copyright (2010) by APS.}

\label{fig:virialAkw} 
\end{figure}

Fig. \ref{fig:virialAkw} shows contour plots of the occupied spectral
intensity of a trapped Fermi gas in the crossover at $T=0.7T_{F}$.
At this temperature, our results are {\em quantitatively} reliable.
We observe that, in addition to the response from coherent Landau
quasiparticles (black lines), there is a broad incoherent spectral
weight centered about $\omega+\mu=-\epsilon_{{\bf k}}+E_{B}$ (white
dashed lines), where $\epsilon_{{\bf k}}=\hbar^{2}k^{2}/(2m)$ and
$E_{B}=-\hbar^{2}/(ma_{s}^{2})$ is the binding energy. Thus, the
spectra clearly exhibit a gap-like double peak structure in the normal
state. This is a remarkable feature: the dispersion at negative energies
seems to follow the BCS-like dispersion curve, $\omega=-\sqrt{(\epsilon_{{\bf k}}-\mu)^{2}+\Delta^{2}}$,
and behaves as if the gas was superconducting, even though we are
above the critical temperature $T_{c}$. Therefore, the incoherent
spectral weight indicates the tendency of pseudogap: the precursor
of fermionic pairing due to strong attractions, i.e., it arises from
the atoms in the paired state or ``molecules''. The pairing response
is very broad in energy and bends down towards lower energy for increasing
$k$. At large $1/(k_{F}a_{s})$, the width is of order $\sqrt{\mbox{max}\{k_{B}T,E_{F}\}\epsilon_{{\bf k}}}$.
The incoherent spectral weight found by our leading cluster expansion
is a universal feature of interacting Fermi gases. At large momentum
$k\gg k_{F}$, it is related to the universal $1/k^{4}$ tail of momentum
distribution \cite{TanRelations,SchneiderPRA2010}.

\subsection{Comparison of theory with the JILA experiment}

\begin{figure}[htp]
\begin{centering}
\includegraphics[clip,width=0.4\textwidth]{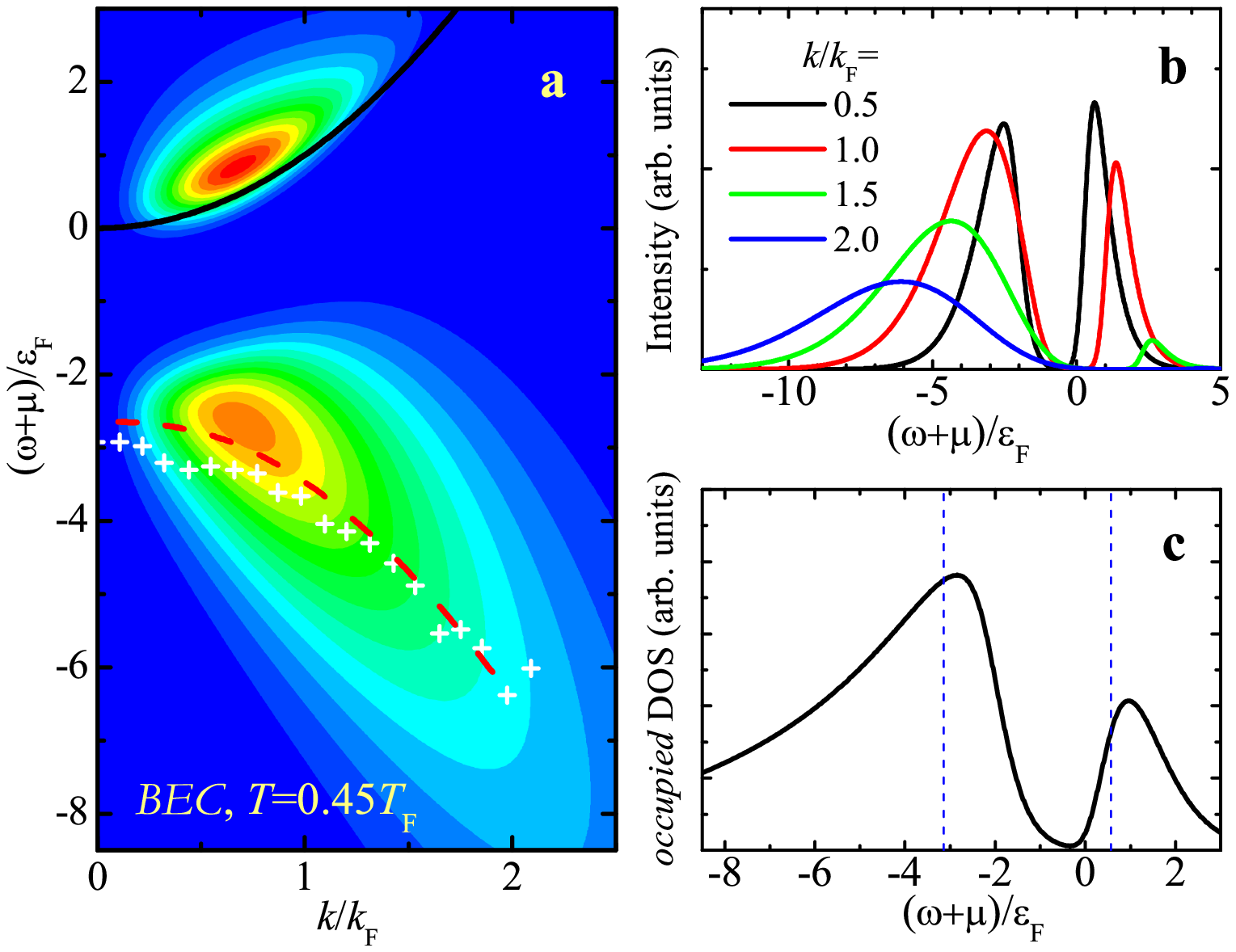}\includegraphics[clip,width=0.4\textwidth]{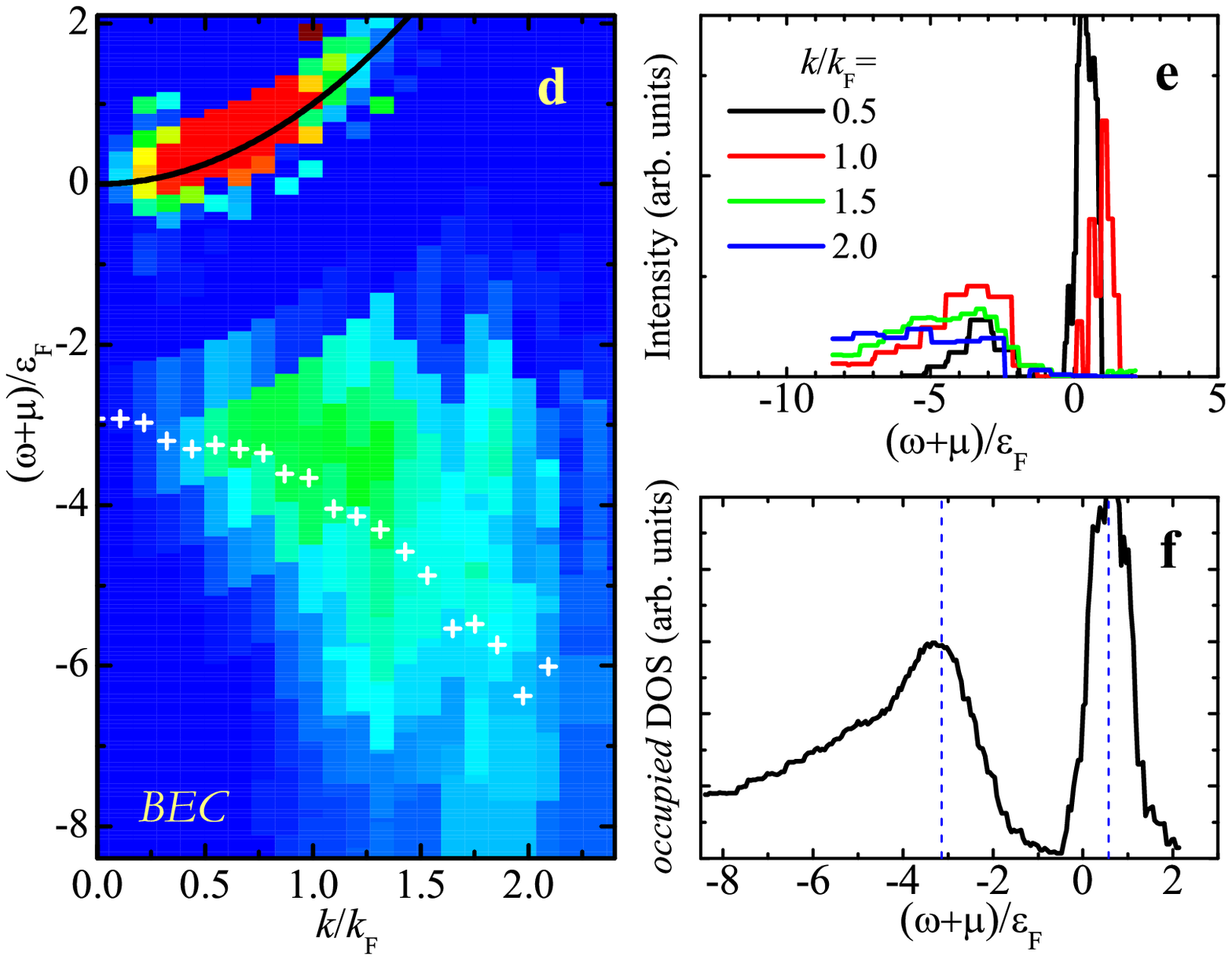}
\par\end{centering}

\caption{(color online) Single-particle excitation spectra on the BEC side
of crossover. \textbf{a-c}, Cluster expansion predictions ($z\simeq0.1$
and $\mu\simeq-1.08E_{F}$). \textbf{d-e}, Corresponding experimental
data \cite{rfJILANature}. \textbf{a}, The linear-scale intensity
map. Our results were convoluted with a gaussian broadening curve
of width $\sigma=0.22E_{F}$, to account for the measurement resolution
\cite{rfJILANature}. The black line shows upper free-atom dispersion.
The red dashed line is the lower dispersion curve of molecules, obtained
via fitting each fixed-$k$ energy distribution curve (in \textbf{b})
with a two gaussian distribution. It agree fairly well with the experimental
result (white symbols). \textbf{b}, Energy distribution curves for
selected values of $k$. \textbf{c}, The occupied density of state
(DOS). Blue dashed lines show the experimental peak positions. Adapted
from ref. \cite{OurVirialAkw}; copyright (2010) by APS.}

\label{fig:virialAkwComparisonBEC} 
\end{figure}

For a close comparison with experiment \cite{rfJILANature}, we perform
calculations using realistic experimental parameters, including the
measurement resolution. Fig. \ref{fig:virialAkwComparisonBEC} presents
the results on the BEC side of crossover with $1/(k_{F}a_{s})=1.1$.
The temperature $T=0.45T_{F}$ is estimated from an initial temperature
$T_{i}=0.17T_{F}$ obtained before the field sweep to the BEC side
\cite{rfJILANature}. The experimentally observed upper and lower
features, caused respectively by unpaired atoms and molecules, are
faithfully reproduced. In particular, the experimental data for the
quasiparticle dispersion of molecules, marked by white symbols, agrees
with our theory (lower red dashed line). There is also a qualitative
agreement for the energy distribution curves (Figs. \ref{fig:virialAkwComparisonBEC}b
and \ref{fig:virialAkwComparisonBEC}e) and the occupied density of
states (Figs. \ref{fig:virialAkwComparisonBEC}c and \ref{fig:virialAkwComparisonBEC}f).
A narrow peak due to free atoms and a broader feature due to molecules
are reproduced theoretically with very similar width at nearly the
same position. It is impressive that the simple quantum cluster expansion
is able to capture the main feature of the experimental spectra.

\begin{figure}[htp]
\begin{centering}
\includegraphics[clip,width=0.4\textwidth]{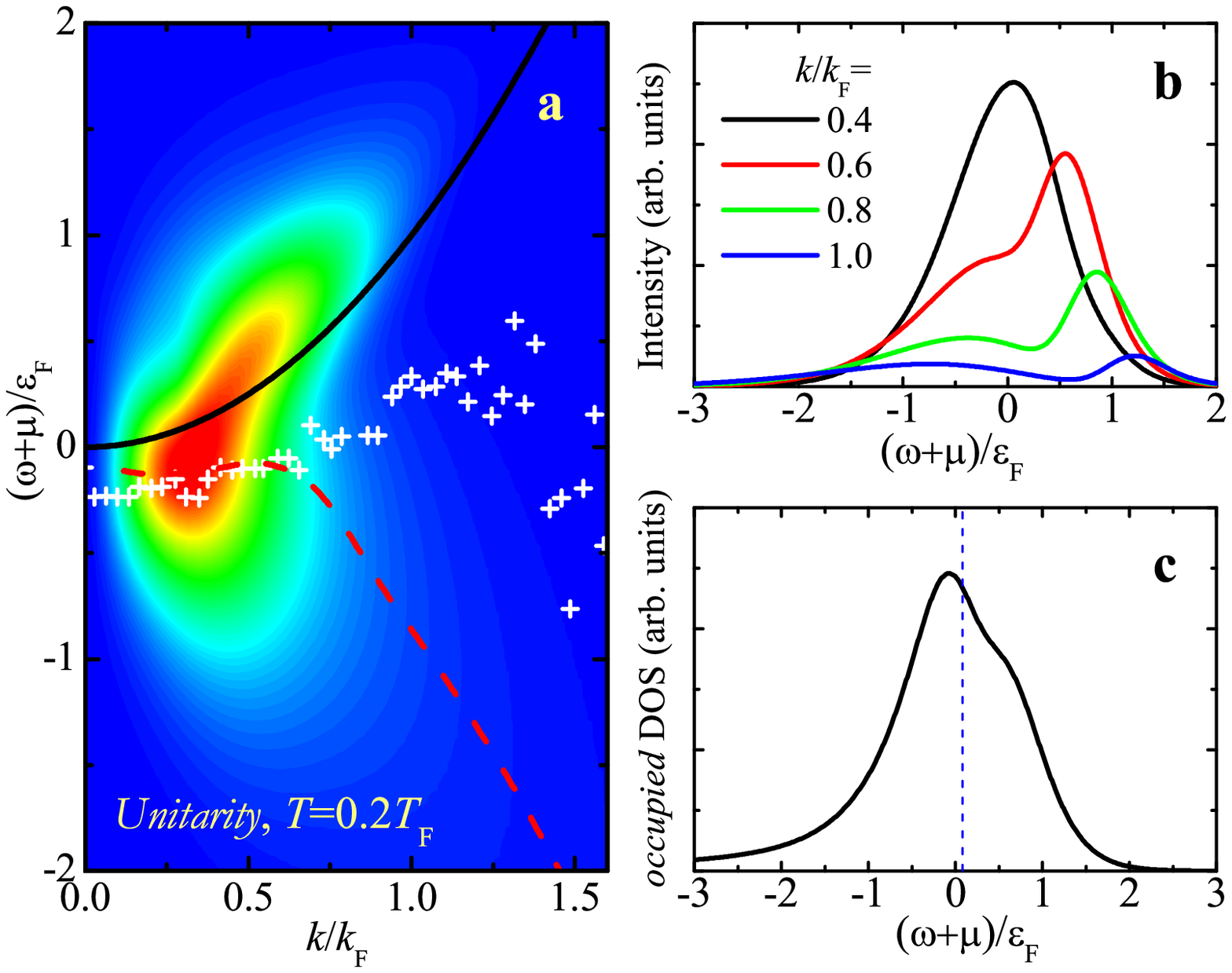}
\includegraphics[clip,width=0.4\textwidth]{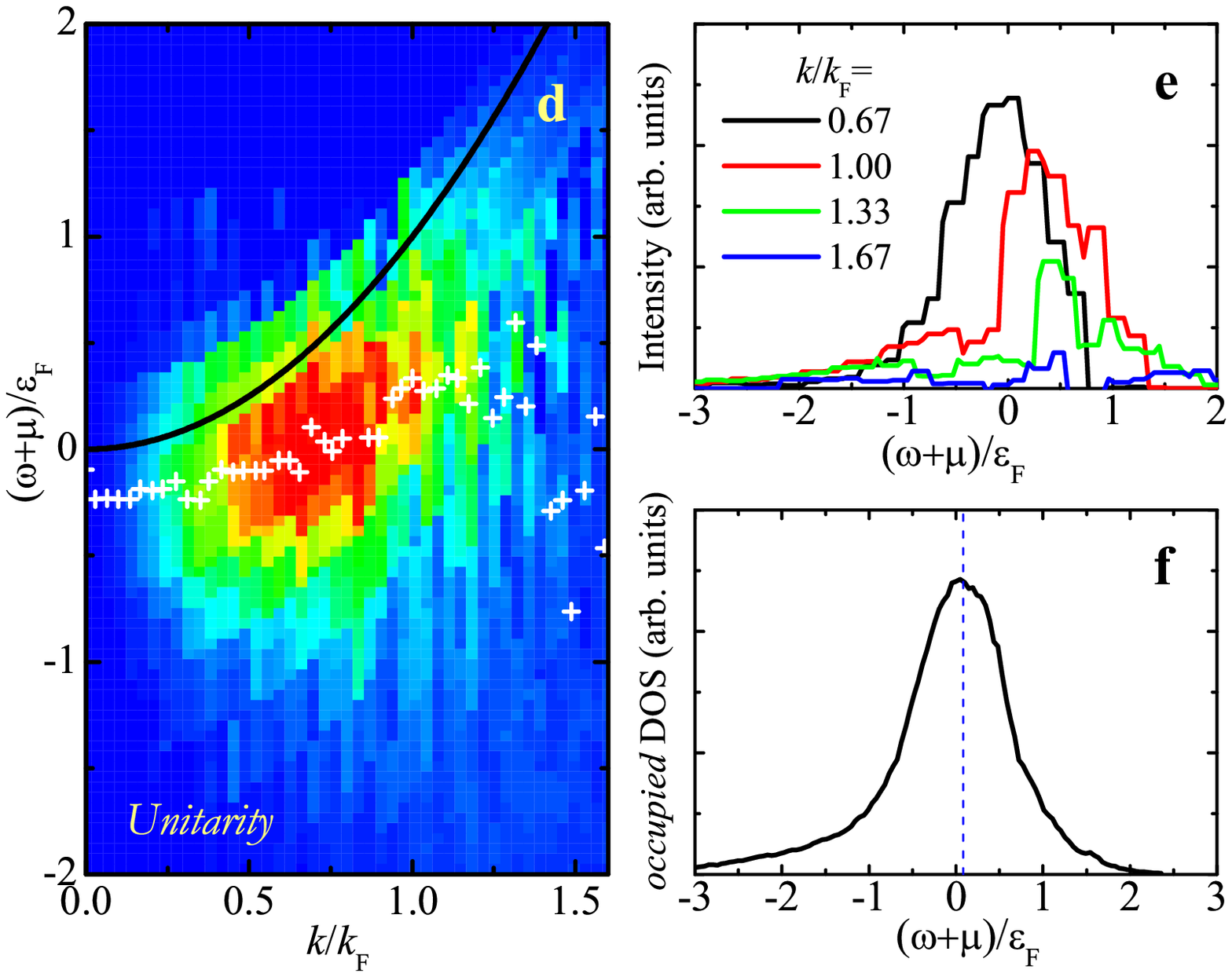}
\par\end{centering}

\caption{(color online) Single-particle excitation spectra of a strongly interacting
Fermi gas. \textbf{a-c}, Cluster expansion predictions ($z\simeq6$
and $\mu\simeq0.37E_{F}$). \textbf{d-e}, Corresponding experimental
data \cite{rfJILANature}. In\textbf{\ e}, for the experimental energy
distribution curves, we use a larger value of $k$ (i.e., enlarged
by a factor of 5/3) to account for a scaling discrepancy due to many-body
correlations. Adapted from ref. \cite{OurVirialAkw}; copyright (2010)
by APS.}

\label{fig:virialAkwComparisonUnitary} 
\end{figure}

Fig. \ref{fig:virialAkwComparisonUnitary} reports the spectra in
the unitarity limit at the critical temperature $T_{c}\simeq0.2T_{F}$.
At such low temperatures, the use of a cluster expansion becomes highly
questionable as the fugacity at the center $z\simeq6>>1$. Nevertheless,
we find that the dispersion curve is lowered by the attractions by
an amount comparable to the Fermi energy $\epsilon_{F}$, as shown
clearly by the red dashed line in Fig. \ref{fig:virialAkwComparisonUnitary}a.
The calculated energy distribution curves bifurcates from a single
peak with increasing $k$ and becomes dominated by the lower molecular
branch (Fig. \ref{fig:virialAkwComparisonUnitary}b), which eventually
leads to the bending back of the dispersion curve to negative energy.
This picture may be view as an indication of the existence of a pseudogap,
which is consistent with the experimental findings (Fig. \ref{fig:virialAkwComparisonUnitary}e).
This surprisingly good agreement merits further investigation. We
conjecture that even at these relatively low temperatures the virial
expansion captures the dominant two-body correlations measured in
these experiments, apart from a possible overall scaling factor due
to the missing higher-order terms.

\section{Virial expansion function and Wilson coefficient}

In this section, we discuss briefly the relation between virial expansion
and Tan relations, both of which provides useful insights to the challenging
many-body problem. The virial expansion is a natural tool to bridge
few-body and many-body physics, while the exact Tan relations give
perspective from the point of view of short-distance and/or short-time
scale. It has been shown by Braaten and Platter that Tan's relations
can be understood using the short-distance and/or short-time operator
product expansion (OPE) method \cite{BraatenPRL2008,BraatenPRL2010},
in which the few-body and many-body scales are separated. At this
point, there should be a close relation between virial expansion and
Tan relations. Here, we show that the Wilson coefficient appearing
in the OPE equations is given by the virial expansion function \cite{HLPreprint}.

\subsection{Operator product expansion method}

The OPE gives a powerful tool to understand the strongly correlated
many-body system in the short-distance/short-time limit. It is a {\em
hypothesis} independently conjectured by Wilson, Kadanoff, and Polyakov
in 1969 \cite{OPE}. The OPE expands the product of local operators
at different space-time points in local operators with coefficients
that are functions of the separation in space and time. For density
correlation, it takes the form, 
\begin{equation}
\hat{\rho}_{\sigma}\left({\bf x},\tau\right)\hat{\rho}_{\sigma^{\prime}}\left({\bf x}^{\prime},\tau^{\prime}\right)=\sum_{C}W_{\sigma\sigma^{\prime}}^{C}\left({\bf x}-{\bf x}^{\prime},\tau-\tau^{\prime}\right){\cal O}_{C},\label{eq:OPE-definition}
\end{equation}
 where the sum is over infinitely many local operators ${\cal O_{C}}[({\bf x}+{\bf x}^{\prime})/2,(\tau+\tau^{\prime})/2]$
and $W_{\sigma\sigma^{\prime}}^{C}\left({\bf r}-{\bf r}^{\prime},\tau\right)$
are called Wilson coefficients. The original hypothesis concerns the
real time $t$ \cite{OPE}. Here, we generalize it to an imaginary
time $\tau$ via the analytical continuation, $t=-i\tau$. As a result,
the Wilson coefficients defined in this way are amenable for calculations
at both zero and finite temperatures. The Wilson coefficients rely
only on few-body physics. Hence, in order to determine $W_{\sigma\sigma^{\prime}}^{C}$
of a local operator ${\cal O}_{C}$ at zero temperature, one may choose
a simple few-body state for which $\left\langle {\cal O}_{C}\right\rangle \neq0$
and match the expectation values on both sides of Eq. (\ref{eq:OPE-definition}).
At finite temperatures, however, this matching procedure may be considerably
complicated.

In the short-distance/short-time limit, only a few terms in the sum
of Eq. (\ref{eq:OPE-definition}) contribute. By neglecting the un-important
single-particle contribution, it was shown \cite{BraatenPRL2010,SonPRA2010}
that after a Fourier transform ($q\rightarrow\infty$ and $\omega\rightarrow\infty$),

\begin{equation}
S_{\sigma\sigma^{\prime}}\left({\bf q},\omega,T\right)-S_{\sigma\sigma^{\prime}}^{(1)}\left({\bf q},\omega,T\right)\simeq W_{\sigma\sigma^{\prime}}\left({\bf q},\omega,T\right){\cal I},\label{OPE}
\end{equation}
 where ${\cal I}$ is the Tan's contact. At zero temperature, the
Wilson coefficient of the DSF has been determined by Son and Thompson
\cite{SonPRA2010}.

\subsection{Wilson coefficient from the virial expansion function}

The relation between the virial expansion function and the Wilson
coefficient becomes evident, if we expand both sides of Eq. (\ref{OPE})
in fugacity. As $W_{\sigma\sigma^{\prime}}$ involves only the few-body
physics and hence does not contain the fugacity $z$, a count of the
term $z^{n}$ on both sides of Eq. (\ref{OPE}) leads to 
\begin{equation}
W_{\sigma\sigma^{\prime}}\left({\bf q},\omega,T\right)=\frac{z^{2}}{{\cal I}_{2}}\Delta S_{\sigma\sigma^{\prime},2}\left({\bf q},\omega,T\right)\label{WilsonCoefficient}
\end{equation}
 and 
\begin{equation}
\Delta S_{\sigma\sigma^{\prime},n}\left({\bf q},\omega,T\right)=\frac{c_{n}}{c_{2}}\Delta S_{\sigma\sigma^{\prime},2}\left({\bf q},\omega,T\right),\label{HigherOrderExpansion}
\end{equation}
 where ${\cal I}_{2}=z^{2}16\pi^{2}Vc_{2}/\lambda_{dB}^{4}$ is the
contact up to the second order expansion. Therefore, the Wilson coefficient
is given by the second expansion function, in the case of two-body
contact interactions. This result is obtained by applying the OPE
and virial expansion method. As a result, in principle it should be
valid at temperatures above the superfluid transition. However, we
may expect that it holds at all temperatures, as both the Wilson coefficient
and second expansion function are irrelevant to the many-body pairing
in the superfluid phase. The many-body effect enters through the many-body
parameter of contact only. As shown by Eq. (\ref{HigherOrderExpansion}),
in the limits of $q\rightarrow\infty$ and $\omega\rightarrow\infty$,
the virial expansion functions becomes proportional to the contact
coefficients, as a direct result of the OPE hypothesis.

\begin{figure}[htp]
\begin{centering}
\includegraphics[clip,width=0.6\textwidth]{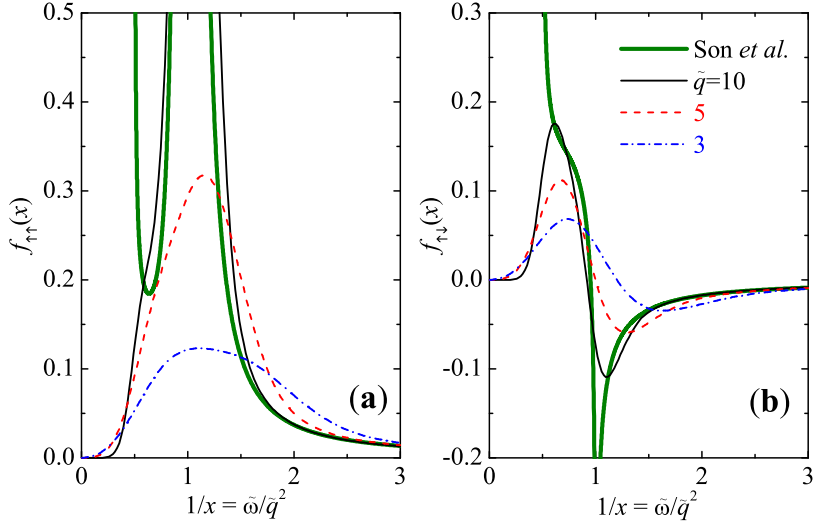} 
\par\end{centering}

\caption{(color online) Wilson coefficients $f_{\sigma\sigma^{\prime}}=\sqrt{m\hbar}\omega^{3/2}(z^{2}/{\cal I}_{2})\Delta S_{\sigma\sigma^{\prime},2}$
at $\tilde{q}=3$, $5$, and $10$. With increasing momentum and/or
frequency, $f_{\sigma\sigma^{\prime}}$ approaches smoothly to the
$T=0$ result by Son and Thompson \cite{SonPRA2010}. Adapted from
ref. \cite{HLPreprint}.}

\label{fig:WilsonCoefficient} 
\end{figure}

At zero temperature, the Wilson coefficient of the DSF of a unitary
Fermi gas can be analytically calculated, by using the matching procedure
using diagrammatic theory. It is given by \cite{HLPreprint,SonPRA2010},
$W_{\uparrow\uparrow}^{T=0}=f_{\uparrow\uparrow}/(\sqrt{m\hbar}\omega^{3/2})$
and $W_{\uparrow\downarrow}^{T=0}=f_{\uparrow\downarrow}/(\sqrt{m\hbar}\omega^{3/2})$,
where, 
\begin{equation}
f_{\uparrow\uparrow}=\frac{1}{4\pi^{2}}\frac{\sqrt{1-x/2}}{\left(1-x\right)^{2}}-\frac{1}{4\pi^{2}}\frac{1}{2x\sqrt{1-x/2}}\left[\ln^{2}\frac{1+\sqrt{2x-x^{2}}}{\left|1-x\right|}-\pi^{2}\Theta(x-1)\right],
\end{equation}
 and 
\begin{equation}
f_{\uparrow\downarrow}=\frac{1}{4\pi^{2}}\frac{1}{\sqrt{2x}}\ln\frac{1+\sqrt{2x-x^{2}}}{\left|1-x\right|}-\frac{1}{4\pi^{2}}\frac{1}{2x\sqrt{1-x/2}}\left[\ln^{2}\frac{1+\sqrt{2x-x^{2}}}{\left|1-x\right|}-\pi^{2}\Theta(x-1)\right],
\end{equation}
 $x\equiv\hbar^{2}{\bf q}^{2}/(2m\hbar\omega)$, and $\Theta$ is
the step function. On the other hand, the second virial expansion
function of the DSF at zero temperature can be calculated from the
trapped results in the limit of large $\tilde{q}\equiv[\hbar^{2}{\bf q}^{2}/(2mk_{B}T)]^{1/2}$.

In Fig. \ref{fig:WilsonCoefficient} we check the validity of Eq.
(\ref{WilsonCoefficient}) at zero temperature, by calculating $(m\hbar)^{1/2}\omega^{3/2}(z^{2}/{\cal I}_{2})\Delta S_{\sigma\sigma^{\prime},2}$
at different momenta. With decreasing temperature $T$ or increasing
$\tilde{q}$, it approaches gradually to $(m\hbar)^{1/2}\omega^{3/2}W_{\sigma\sigma^{\prime}}^{T=0}$
when $\tilde{\omega}>\tilde{q}^{2}/2$. This confirms numerically
that Eq. (\ref{WilsonCoefficient}) holds at zero temperature at large
momentum and frequency. For small frequency (i.e., $\tilde{\omega}\rightarrow0$),
the Wilson coefficient becomes divergent. The confirmation of equivalence
in this limit is stringet and requires a large value of $\tilde{q}$.
Our virial expansion function at $\tilde{q}$ up to 10 is unable to
approach the Wilson coefficient at $\tilde{\omega}<\tilde{q}^{2}/2$.
We also note that, in the limit of large frequency, $W_{\uparrow\uparrow}^{T=0}$
and $W_{\uparrow\downarrow}^{T=0}$ have an interesting high-frequency
power-law tail $\omega^{-5/2}$ \cite{SonPRA2010,TaylorPRA2010},
\begin{equation}
W_{\uparrow\uparrow}^{T=0}=-W_{\uparrow\downarrow}^{T=0}=\frac{\hbar^{1/2}{\bf q}^{2}}{12\pi^{2}m^{3/2}\omega^{5/2}}.\label{WilsonTail}
\end{equation}
 This is fairly evident in the second order virial expansion functions.

The identification of the Wilson coefficient as the virial expansion
function is very useful. For example, in the system where the three-body
interactions dominate, we anticipate that the third virial expansion
function would give the Wilson coefficient. At this point, we note
that, for identical bosons with a large scattering length in which
three-body Efimov physics occurs, the Wilson coefficient and new universal
relation have been derived very recently \cite{BrattenPRL2011,CastinPRA2011}.

\section{Outlook}

In this review, we have demonstrated that virial expansion provides
a powerful tool to understand a normal, strongly correlated atomic
Fermi gas at temperature down to a half of the Fermi degenerate temperature.
The virial predictions generally agree well with the experimental
measurements.

\begin{table}
$\begin{array}{ccccc}
\\
\\
\\
\\
\\
\end{array}$

\begin{tabular}{|c|c|c|c|}
\hline 
$n$ & $\Delta b_{n}$ (theory) & $\Delta b_{n}$ (experiment) & $c_{n}$ (theory)\tabularnewline
\hline 
\hline 
2 & $1/\sqrt{2}$ \cite{BethPhysica1937} &  & $1/\pi$ \cite{YuPRA2009,OurVirialContact}\tabularnewline
\hline 
3 & $\begin{array}{c}
+1.05\pm0.01^{a}\\
-0.35510298^{b}\\
-0.3551030264897^{c}\\
-0.3573\pm0.0005^{d}\\
-0.3551\pm0.0001^{e}
\end{array}$ & $-0.35\pm0.02$ \cite{NascimbeneNature2010} & $\begin{array}{c}
-0.1408\pm0.0010^{f}\\
-0.1399\pm0.0001^{g}
\end{array}$\tabularnewline
\hline 
4 & $-0.016\pm0.004$ \cite{BlumeVE1} & $\begin{array}{c}
0.096\pm0.015^{h}\\
0.096\pm0.010^{i}
\end{array}$ & \tabularnewline
\hline 
5 & $0.0017\leq\Delta b_{5}\leq0.101$ \cite{BlumeVE1} &  & \tabularnewline
\hline 
\end{tabular}

\caption{List of the theoretical predictions and experimental measurements
for the virial coefficients of a homogeneous Fermi gas in the unitary
limit. The last column shows the contact coefficients. For the superscripts
(a)-(i), the references are: (a) \cite{RupakPRL2007}, (b) \cite{OurVE},
(c) \cite{BlumeVE1}, (d) \cite{KaplanPRL2011}, (e) \cite{LeyronasPreprint},
(f) \cite{OurVirialContact}, (g) \cite{LeyronasPreprint}, (h) \cite{NascimbeneNature2010},
and (i) \cite{EoSMIT}. }
\end{table}

\subsection{Successes}

These remarkable results cover both static and dynamic properties.
\begin{itemize}
\item For thermodynamics, in the calculation of virial coefficients in the
strongly-interacting regime, a convenient way is proposed, based on
the few-particle solutions in harmonic traps. The exact three-fermion
solution leads to a very accurate determination of the long-sought
third virial coefficient in the unitary limit: $\Delta b_{3}=-0.3551030264897$.
The resulting virial equation of state serves as an important benchmark
for accurate experimental measurements (see Sec. II). It also provides
a possible thermometry for strongly interacting Fermi gases \cite{EoSMIT}.
The calculation of the fourth virial coefficient in the unitary limit
has been attempted, by solving numerically the four-fermion problem.
In Table 1, we summarize the past theoretical and experimental efforts
in determining the virial coefficients of a strongly interacting homogeneous
Fermi gas. 
\item For the universal Tan's contact ${\cal I}$ that governs the short-range/short-time
physics, thanks to the adiabatic relation, we can virial expand it
in terms of contact coefficients. In the unitary limit, the second
and third contact coefficients, $\Delta c_{2}=1/\pi$ and $\Delta c_{3}=-0.1399\pm0.0001$
(see Table 1), provide a good explanation for the recent measurement
in harmonic traps (Sec. III). 
\item For dynamic properties, the dynamic structure factor and single-particle
spectral function can be virial expanded as well, in terms of virial
expansion functions. The second order expansion (in the leading order
of interactions) gives a good qualitative understanding of recent
experimental measurements on two-photon Bragg spectroscopy (Sec. IV)
and momentum-resolved rf-spectroscopy (Sec. V), at temperature down
to the onset of superfluid transition.
\end{itemize}

\subsection{Future developments}

Encouraged by these remarkable achievements, we may foresee a number
of potential developments and applications of virial expansion in
the near future.

\subsubsection{Higher-order expansions and new applications}

It is technically straightforward to calculate higher-order virial
coefficients and expansion functions. However, much heavier numerical
efforts would be involved. Owing to the ever-growing power in computation,
we anticipate optimistically that the fourth and fifth virial coefficients
could be calculated accurately. Accordingly, the third to fifth virial
expansion functions for the single-particle spectral function may
be determined. These results will clearly bring in-depth understanding
of the existing measurements on thermodynamics and spectral function.

In the novel atomic systems such as multi-component Fermi gases or
strongly interacting Bose gases, where the three-body or four-body
physics becomes important, virial expansion would be particularly
useful. Using the third or fourth expansion functions, we anticipate
to address the many-body consequence of the multi-component (i.e.,
triplet) pairing and Efimov physics. New universal relations may be
predicted.

With these in-mind, we note that the methodology of virial expansion
is very general. It can be used as well to study many other interesting
properties of strongly-correlated atomic Fermi gases, which now become
available with current experimental techniques. Important examples
includes the universal transport coefficient (i.e. the shear viscosity)
of a unitary Fermi gas, which has been investigated already by the
damping rate in collective excitations and by the hydrodynamic expansion
\cite{CaoScience2011}, fermionic pairing in low-dimensions, which
can be probed by the rf-spectroscopy \cite{FrohlichPRL2011,FeldPreprint,SommerPreprint},
and non-$s$-wave fermionic pairing \cite{HoPreprint,PengPRA2011}.

\subsubsection{Insights for reliable low-temperature strong-coupling theories}

An important motivation of the virial expansion study is to gain insights
for developing reliable low-temperature strong-coupling theories.
Ideally, we wish to apply in a {\em quantitative} manner the virial
expansion down to the superfluid transition temperature. However,
in the deep quantum degenerate regime, where the fugacity is larger
than unity, we may not anticipate convergence of the virial series,
evaluated up to certain order. To extract the infinite-order result,
it is necessary to apply some resummation techniques \cite{HouckePreprint}.

\begin{figure}[htp]
\begin{centering}
\includegraphics[clip,width=0.5\textwidth]{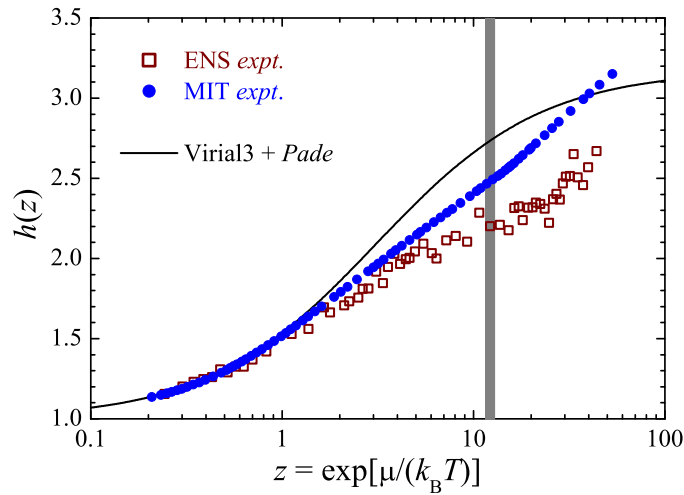} 
\par\end{centering}

\caption{(color online) Universal $h$-function as a function of fugacity.
The Padé {[}$2/2${]} approximant is compared with the two experimental
data sets, from Salomon's group at ENS (empty squares) \cite{NascimbeneNature2010}
and from Zwierlein group at MIT (solid circles) \cite{EoSMIT}. The
vertical shaded line shows the critical fugacity for the onset of
superfluid transition. }

\label{fig:hzPade} 
\end{figure}

\begin{figure}[htp]
\begin{centering}
\includegraphics[clip,width=0.6\textwidth]{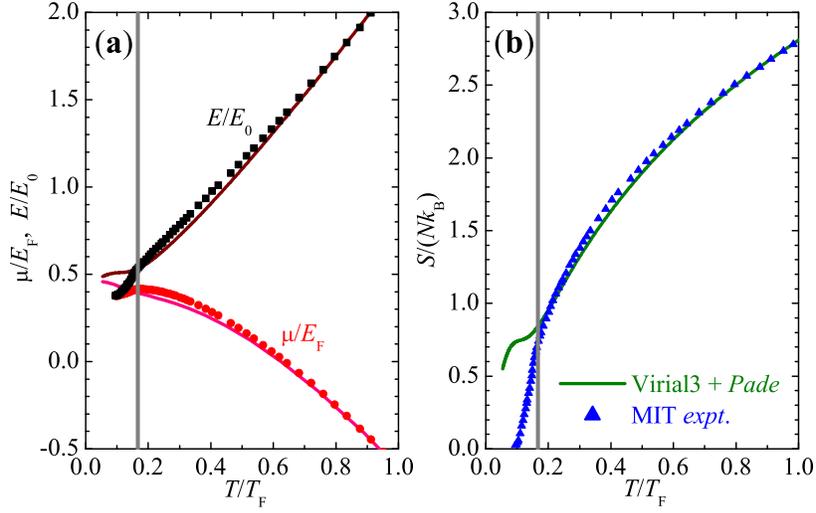} 
\par\end{centering}

\caption{(color online) Universal equation of state obtained by using the third-order
virial expansion within the {[}$1/1${]} Padé approximant. The result
is contrasted with the experimental data from MIT \cite{EoSMIT}.
Here $E_{0}=(3/5)NE_{F}$ is the ground state energy of an ideal,
non-interacting Fermi gas. The vertical shaded line indicates the
critical temperature for the onset of superfluid transition. }

\label{fig:eosPade} 
\end{figure}

As an interesting example, here we discuss briefly the Padé resummation
method, in which a virial series, i.e., the universal $h$-function
Eq. (\ref{hzve}), is written into the form, 
\begin{equation}
h_{_{\text{Pade}}}(z)=\frac{p_{0}+p_{1}z+p_{2}z^{2}+\cdots+p_{m}z^{m}}{1+q_{1}z+q_{2}z^{2}+\cdots+q_{n}z^{n}}.
\end{equation}
 This is the so-called Padé approximant of order {[}$m/n${]}. To
the order {[}$1/1${]}, the three Padé coefficients $p_{0}$, $p_{1}$,
and $q_{1}$ can be uniquely determined using the three virial coefficients,
i.e., 
\begin{equation}
h_{_{\text{Pade}}}^{[1/1]}(z)=\frac{1+\left[b_{2}^{(1)}+\Delta b_{2}-\Delta b_{3}/\Delta b_{2}\right]z}{1+\left[b_{2}^{(1)}-\Delta b_{3}/\Delta b_{2}\right]z}.
\end{equation}
 In Fig. \ref{fig:hzPade}, we compare the universal $h$-function
in the Padé {[}$1/1${]} form with the experimental data. It agrees
very well with the latest measurement (the MIT data set) reported
by Ziwerlein's group at $all$ temperatures. The relative discrepancy
is about $10\%$ in maximum, comparable with the discrepancy of the
two experimental data sets. In Fig. \ref{fig:eosPade}, we compare
the virial equation of state, calculated using $h_{_{\text{Pade}}}^{[1/1]}$,
with the MIT data set. The third-order virial expansion within Padé
approximant works extremely well, for temperatures down to the onset
of superfluid phase transition, $T_{c}\sim0.16T_{F}$. This remarkable
agreement, over a wide parameter window in fugacity, is entirely unexpected,
since the Padé approximant is not controllable and therefore its application
can not be justified {\em a prior}. We anticipate that the accuracy
of the virial equation of state could be improved by the inclusion
of more Padé terms in $h_{_{\text{Pade}}}(z)$, such as the $z^{2}$
term. This is straightforward once the fourth and fifth virial coefficients
are accurately calculated.

It is reasonable to anticipate that the similar Padé approximant may
work for the single-particle spectral function. In this respect, the
virial spectral function within the Padé {[}$1/1${]} or {[}$2/2${]}
approximant could be useful to clarify the delicate pseudogap puzzle
in a unitary Fermi gas.

\section*{Acknowledgments}

We have benefited from discussions and collaborations with many physicists:
here we would like to especially thank Hui Hu, Peter D. Drummond,
Peter Hannaford, Chris J. Vale, and Eva D. Kuhnle for valuable interactions
in recent years, and Tin-Lun Ho for his continuous encouragement.
We also thank Xavier Leyronas for sending his data file of $\Delta b_{3}$
in ref. \cite{LeyronasPreprint} , Sylvain Nascimbène and Christophe
Salomon for providing us the experimental data of the universal $h$-function
in ref. \cite{NascimbeneNature2010}, and Martin W. Zwierlein and
Mark J.-H. Ku for providing us the experimental data in ref. \cite{EoSMIT}
. This research was supported by the Australian Research Council Discovery
Project (Grant No. DP0984637) and NFRP-China (Grant No. 2011CB921502).

\appendix

\section{Calculation of $C_{nn^{\prime}}$}

In this appendix, we outline the details of how to construct the matrix
element $C_{nn^{\prime}}$ in Eq. (\ref{amat}), which is given by,
\begin{equation}
C_{nn^{\prime}}\equiv\int\limits _{0}^{\infty}\rho^{2}d\rho R_{nl}\left(\rho\right)R_{n^{\prime}l}\left(\frac{\rho}{2}\right)\psi_{2b}^{rel}(\frac{\sqrt{3}}{2}\rho;\nu_{l,n^{\prime}}),
\end{equation}
 where 
\begin{equation}
R_{nl}\left(\rho\right)=\sqrt{\frac{2n!}{\Gamma\left(n+l+3/2\right)}}\rho^{l}e^{-\rho^{2}/2}L_{n}^{\left(l+1/2\right)}\left(\rho^{2}\right),
\end{equation}
 is the radial wave function of an isotropic 3D harmonic oscillator
and the two-body relative wave function is 
\begin{equation}
\psi_{2b}^{rel}=\Gamma(-\nu_{l,n^{\prime}})U(-\nu_{l,n^{\prime}},\frac{3}{2},\frac{3}{4}\rho^{2})\exp(-\frac{3}{8}\rho^{2}).
\end{equation}
 Here, for convenience we have set $d=1$ as the unit of length. $L_{n}^{\left(l+1/2\right)}$
is the generalized Laguerre polynomial and $U$ is the second Kummer
confluent hypergeometric function. A direct integration for $C_{nn^{\prime}}$
is difficult, since the second Kummer function has a singularity at
the origin. The need to integrate for different values of $\nu_{l,n^{\prime}}$
also causes additional complications.

It turns out that a better strategy for the numerical calculations
is to write, 
\begin{equation}
\psi_{2b}^{rel}=\sum_{k=0}^{\infty}\frac{1}{k-\nu_{l,n^{\prime}}}\sqrt{\frac{\Gamma\left(k+3/2\right)}{2k!}}R_{k0}\left(\frac{\sqrt{3}}{2}\rho\right),
\end{equation}
 by using the exact identity, 
\begin{equation}
\Gamma(-\nu)U(-\nu,\frac{3}{2},x^{2})=\sum_{k=0}^{\infty}\frac{L_{k}^{1/2}\left(x^{2}\right)}{k-\nu}.
\end{equation}
 Therefore, we find that 
\begin{equation}
C_{nn^{\prime}}=\sum_{k=0}^{\infty}\frac{1}{k-\nu_{l,n^{\prime}}}\sqrt{\frac{\Gamma\left(k+3/2\right)}{2k!}}C_{nn^{\prime}k}^{l},
\end{equation}
 where 
\begin{equation}
C_{nn^{\prime}k}^{l}\equiv\int\limits _{0}^{\infty}\rho^{2}d\rho R_{nl}\left(\rho\right)R_{n^{\prime}l}\left(\frac{\rho}{2}\right)R_{k0}\left(\frac{\sqrt{3}}{2}\rho\right)
\end{equation}
 can be calculated to high accuracy with an appropriate integration
algorithm. In checking convergence of the summation over $k$, we
find numerically that for a cut-off $n_{\max}$ (i.e., $n,n^{\prime}<n_{\max}$),
$C_{nn^{\prime}k}^{l}$ vanishes for a sufficient large $k>k_{\max}\sim4n_{\max}$.

In practical calculations, we tabulate $C_{nn^{\prime}k}^{l}$ for
a given total relative angular momentum. The calculation of $C_{nn^{\prime}}$
for different values of $\nu_{l,n^{\prime}}$ then reduces to a simple
summation over $k$, which is very efficient. Numerically, we have
confirmed that the matrix $C_{nn^{\prime}}$ is symmetric, i.e., $C_{nn^{\prime}}=C_{n^{\prime}n}$.

\section{Calculation of $s_{l,n}$}

The calculation of $s_{l,n}$ seems straightforward by using the Bethe-Peierls
boundary condition in hyperspherical coordinates (\ref{BP3eHyper}).
However, we find that numerical accuracy is low for large $n$ and
$l$ due to the difficulty of calculating the hypergeometric function
$_{2}F_{1}$ accurately using IEEE standard precision arithmetic.
We have therefore utilized MATHEMATICA software that can perform analytical
calculations with unlimited accuracy. For this purpose, we introduce
$\Delta s_{l,n}=s_{l,n}-\bar{s}_{l,n}$. After some algebra, we find
the following boundary condition for $t\equiv\Delta s_{l,n}/2$, 
\begin{equation}
\sin\left(\pi t\right)=\sqrt{\frac{\pi}{3}}\frac{\left(-1\right)^{n+l}\Gamma\left(n+l+1+t\right)}{2^{l}\Gamma\left(l+\frac{3}{2}\right)\Gamma\left(n+1+t\right)}f\left(t\right),\label{dsln}
\end{equation}
 where we have defined a function 
\begin{equation}
f\left(t\right)\equiv\text{ }_{2}F_{1}\left(-n-t,n+l+1+t,l+\frac{3}{2};\frac{1}{4}\right).
\end{equation}
 The above equation can be solved using the MATHEMATICA routine ``FindRoot'',
by seeking a solution around $t=0$. It is also easy to write a short
program to solve Eq. (\ref{dsln}) continuously for $n<n_{\max}=512$
and $l<l_{\max}=512$. In a typical current PC, this takes several
days. The results can be tabulated and stored in a file for further
use.

\end{document}